\documentclass[12pt,english]{book}
\usepackage{psfig,epsfig,babel,epsf,here,
amsthm,amsfonts,amssymb,wrapfig}

\newcommand{\beq}{\begin{equation}} 
\newcommand{\eeq}{\end{equation}} 
\newcommand{\bea}{\begin{eqnarray}} 
\newcommand{ \eea}{\end{eqnarray}}
\def \bq{\begin{quote}} 
\def \eq{\end{quote}}
\def \nn{\nonumber}
\def\dd{\displaystyle}
\def\sls#1{/ \hskip -6pt #1}
\def\slsh#1{/ \hskip -8pt #1}
\def\sq{\frac{1}{\sqrt{2}}}
\def\ep{\epsilon}

\newcommand{\clearemptydoublepage}{\newpage{\pagestyle{empty}\cleardoublepage}}

\begin{document}

%\pagestyle{empty}

%\begin{titlepage}

%\begin{flushleft}
%{\large{Sede Amministrativa: Universit\`a degli Studi di Padova}}

%\vspace{0.2cm}
%{\large{Dipartimento di Fisica ``Galileo Galilei''}}

%\vspace{1.5cm}

%{\scshape{\large{Dottorato di Ricerca in Fisica}}}
%\vspace{0.5cm}

%{\scshape{\large{Ciclo XIII}}}
%\end{flushleft}

%\vspace{\fill}

%\begin{center}
%{\Huge{\bf{
%The Problem of Neutrino Masses 
%}}}
%\\
%\end{center}
%\begin{center}
%{\Huge{\bf{ 
%in the Extensions 
%}}}
%\\ 
%\end{center}
%\begin{center}
%{\Huge{\bf{   
%of the Standard Model
%}}}
%\\ 

%\end{center}

%\vspace{\fill}
%\begin{flushleft}
%{\large{Coordinatore: Ch.mo Prof.~Attilio Stella}}

%\vspace{1.5cm}
%{\large{Supervisore: Ch.mo Prof.~Ferruccio Feruglio}}
%\end{flushleft}

%\begin{flushright}
%\vspace{1.5cm}
%{\large{Dottoranda: Dott.ssa~Isabella Masina}}
%\end{flushright}

%\begin{center}
%\vspace{3.5cm}
%\vspace{\fill}
%31 Dicembre 2000
%\end{center}
%\end{titlepage}

%\pagebreak
%\mbox{}
%\clearemptydoublepage

%\maketitle
%\clearemptydoublepage

%%%%%%%%%%%%%%%%%%%%%%%%%%%%%%%%%%%%%%%%%%%%%%%%%%%
%\newpage \newpage

\thispagestyle{empty}

\begin{flushright} 
{DFPD 01/TH/30}
\vspace{0.8cm}
\end{flushright}

\centerline{ \large \bf THE PROBLEM OF NEUTRINO MASSES}
\vspace*{0.2 cm}
\centerline{ \large \bf IN EXTENSIONS OF THE STANDARD MODEL
\footnote{Based on PhD thesis work, February 2001}}
\vspace*{1cm}
\centerline{ISABELLA MASINA}
\vspace*{0.5 cm}
\centerline{\it{Department of Physics G.Galilei,
Univ. of Padova}}
\centerline{\it{Via Marzolo 8, I-35131 Padova, Italy
\footnote{Now at Service de Physique Th\'eorique, CEA-Saclay, F-91191
Gif-sur-Yvette, France \\  
e-mail: masina@spht.saclay.cea.fr}
}}
\vspace{3cm}
\noindent{We review the problem of neutrino masses and mixings in 
the context of Grand Unified Theories. After a brief summary 
of the present experimental status of neutrino physics, we describe 
how the see-saw mechanism can automatically account 
for the large atmospheric mixing angle. We provide two specific 
examples where this possibility is realized by means of a 
flavour symmetry. We then review in some detail the various 
severe problems which plague minimal GUT models (like the doublet-triplet 
splitting and proton-decay) and which force to investigate the possibility
of constructing more elaborate but realistic 
models. We then show an example of a quasi-realistic SUSY $SU(5)$ 
model which, by exploiting the crucial presence of an abelian 
flavour symmetry, does not require any fine-tuning and 
predicts a satisfactory phenomenology with respect to coupling unification,
fermion masses and mixings and bounds from proton decay.}

\pagestyle{headings}
\pagenumbering{roman}

\tableofcontents
\clearemptydoublepage

\addcontentsline{toc}{chapter}{Introduction}
\pagestyle{plain} 

\chapter*{Introduction} 

\pagenumbering{arabic}

It is quite obvious that the Standard Model must be extended.
Various arguments bring along the important conceptual implication that
the Standard Model should be seen as {\it a low energy effective theory}
valid up to some physical cut-off scale.

It is thus compelling to speculate about the underlying more fundamental
theory. Various ideas have been proposed and developed to answer this
question. For instance, recently a lot of interest and efforts have been
devoted to the idea of extra-dimensions, where the different behaviour of the
gauge and gravitational interactions may provide a solution to the hierarchy
problem. 
But the attractive proposal of Grand Unification has not lost its charm.

The concept that all particle interactions unify at very high energies is
indeed not very recent.  The first paper on this subject was published 26
years ago
and from then on this idea captured the interest of physicists. 
Many fundamental papers that explore the various aspects and 
consequences of Grand Unification have been written. 
What is then inside this proposal which renders it
still so interesting to collect the general interest?
The main point is that it received {\it phenomenological support} by: \\
A) the success of gauge coupling unification;\\
B) recent developments on neutrinos.

A) Gauge coupling unification works terribly well. The
original proposal has however notably evolved because now we know that
Grand Unification needs Supersymmetry. Only then gauge coupling
unification successfully happens at the scale $M_{GUT} \sim 10^{16}$ GeV.
In addition, in non supersymmetric Grand Unified theories the prediction of
proton lifetime for the process $p\rightarrow e^+ \pi^0$ is in conflict with lower experimental 
limits, but including Supersymmetry it turns out to be close to the safe side.

B) Recent data coming from neutrino physics are much more
robust and less fragmented then they were just few years ago.
There is definitive evidence for a neutrino mass associated with 
atmospheric oscillations in the range $0.05 \div 1$ eV.
There is no room in the Standard Model for neutrinos to have a mass,
so that neutrinos represent the first true signal and actually give 
some hints on possible extensions of the Standard Model.
Baryon and lepton numbers are expected to be violated in Grand Unified Theories,
leading to the fundamental consequences of proton instability and
possibility of Majorana mass terms. 
The previously quoted mass range points in favour of lepton number
violation at a scale $\sim 10^{(14 \div 15)}$ GeV.

Thus both gauge coupling unification and neutrino masses independently
strongly suggest that beyond the electroweak scale there is another
crucial scale where some new physics really arises and which lies just two
orders of magnitude below the Planck scale.

Neutrinos are curious also because, in contrast with the observed smallness of
left handed quark mixings, it is now experimentally well established by
Super-Kamiokande that they possess a large left handed mixing associated 
with the
neutrino flavours involved in atmospheric oscillations.
{\it To explain from the theoretical point of view the possible causes
which determine the atmospheric mixing to be large and the 
possible consequences
is the thread of the present review.}

Grand Unification represents an excellent framework for the study of neutrino
masses and mixings, because it predicts fermion mass matrices to be related. 
Several Grand Unified models have been constructed which show that the presence of the
large atmospheric mixing may naturally arise via the see-saw. 
These models can be distinguished into two classes, according to the basic
mechanism exploited.

Most models are based on the possible presence of a large right handed down 
quark mixing, not directly observable in weak transitions, which is strictly 
correlated to a large left handed mixing for charged leptons. 
The latter can be transferred to the Dirac neutrino mass matrix and 
the desired large mixing is achieved if the right handed Majorana neutrinos
are not widely split. 

But there is also a particularly interesting alternative mechanism. 
This mechanism is based on the observation that for a sufficiently hierarchical
right handed neutrino spectrum, a large atmospheric mixing
automatically arises if the left handed $\nu_{\mu}$ and $\nu_{\tau}$ 
couple with comparable
strength to the right handed $\nu^c_{\mu}$. 
Thus, remarkably, this situation does not require the presence of large mixings
in the Dirac and/or Majorana mass matrices. As opposite to the first mechanism,
this second one can then be realized even if all left and right mixings of quarks and leptons 
are very small. It is then a possibility for the large mixing to actually 
arise from nothing, that is only as an effect of the see-saw mechanism.
This situation is particularly attractive from the theoretical point of
view because it is compatible with left right symmetric scenarios. 

The first two chapters of this review 
contain a summary of the present experimental status and future perspectives 
about neutrinos, 
show the two basic mechanisms providing an explanation of the large atmospheric mixing 
and present explicit Grand Unified models which realize these two possibilities.

The two independent supports in favour of the physical
relevance of $M_{GUT}$, i.e. supersymmetric coupling unification and 
neutrino masses, can be hardly interpreted as just a 
coincidence.
Armed with the results obtained from the study of the neutrino sector,
we face the more ambitious task of {\it constructing a sufficiently complete
and realistic Grand Unified model which could embed one of the two
mechanisms for the generation of the large atmospheric mixing}.
After 26 years, one could wonder if it is still possible and
eventually what kind of informations could be added to further
increase our knowledge of Grand Unification.
There are mainly two new facts:
\begin{enumerate}
\item much stronger limits on proton decay rates, in particular
on the channels which are expected to be dominant in supersymmetric Grand Unified
theories. Super-Kamiokande gives a lower limit for proton lifetime of about 
$2\cdot 10^{33}$ yr for the channel $p\rightarrow K^+ \bar{\nu}$);
\item the approach itself. Since Grand Unification seems so successful, 
it is desirable to use it as a framework where to address also the flavour
problem, that is to explain the observed hierarchical spectra
and mixings of all fermions, including neutrinos.   
\end{enumerate}

The flavour problem is independent of Grand Unification, in the sense that 
Grand Unification alone is not able to provide any light on this. 
However it is really an optimal framework in which to embed possible solutions 
of the flavour problem, because Grand Unification predicts more or less strict group 
theoretical relations between fermion mass matrices.
One interesting way to address the flavour problem is by introducing
an horizontal flavour symmetry, namely a symmetry under which
the three copies of each fermion transform differently.
 
Thus the modern point of view is to construct not too
complicated, sufficiently realistic and relatively complete
models. These models have to be enough detailed to actually
carry out the calculation of as much low energy observables as
possible. We are not satisfied with too rough and incomplete models. 
It is clear that only those models which provide predictions    
in agreement with experiment survive.
 
In this respect, minimal versions of Grand Unified Theories 
(e.g. $SU(5)$ with only $24$, $5$ and $\overline 5$ of Higgs fields) are not realistic. 
They suffer from serious problems:
\begin{itemize}
\item doublet-triplet splitting;
\item predictions of proton decay rates;
\item masses and mixings.
%\item the prediction of $\alpha_3(m_Z)$ assuming coupling unification.
\end{itemize}
To address some of these points, an enormous fine-tuning is
required, e.g. of about 14 orders of magnitude for the doublet-triplet
splitting. But the serious fact is that, even if one tolerates to do this
unpleasant tunings, minimal models give wrong predictions. 
In fact, minimal models
predict protons decaying at a too fast rate for the channel $p \rightarrow K^+ \bar \nu$,
in conflict with experiment (point 1.) and
give wrong relations between fermion masses (point 2.).

It is clear that something is lacking in minimal models.
In particular it would be of crucial importance to extend
the Higgs sector in order to: 
\begin{itemize}
\item give an explanation to the doublet-triplet splitting; 
\item enhance the prediction of proton lifetime;
\item adjust down quarks and charged lepton spectra and correctly 
describe neutrino masses.
\end{itemize}
However, the addition of new fields in the Higgs sector 
must be done in such a way not to destroy the only firm starting point,
that is coupling unification.
Here troubles come because a realistic model requires so many improvements
over the minimal versions that one could think
that the beautiful idea of Grand Unification is actually not realizable.
Note that the effort does not go in the direction of making
precise predictions on poorly measured observables, 
but rather of doing predictions which are in not rough conflict
with experiment.

The problems and the technical difficulties which are typical 
of minimal models 
are reviewed in chapter 3, together with a review of the main ideas proposed 
to overcome them.

Many attempts of constructing realistic models have been done in the context
of $SO(10)$ but in general they lack of predictivity on unification, which
should then occur in an accidental way.  
This is due to the fact that despite $SO(10)$ is interesting because
it contains right handed neutrinos, it requires a lot of
assumptions on the Higgs content in order to reduce its rank, 
thus reaching $SU(5)$ or $SU(4)\otimes SU(2) \otimes SU(2)$.

We instead focus the attention directly on Supersymmetric $SU(5)$,
which already possesses the same rank as the Standard Model,
trying here to address as many problems as possible. 
We present a model which successfully overcomes fine-tunings
problems and which gives predictions in agreement with
experiments. This model exploits the crucial presence of an additional
$U(1)$ flavour symmetry. 

The fine-tunings linked to the doublet-triplet splitting are avoided by adopting
the Missing Doublet Mechanism and by using the flavour symmetry to stabilize the Higgs 
doublet mass against large mass corrections due to non renormalizable
operators. Fermion masses and mixings are
reproduced, at the level of the correct order of magnitude, by the same flavour symmetry. 
In the neutrino sector the preferred solution is one with nearly maximal mixing both for
atmospheric and solar neutrinos, which turn out to be compatible with the
LOW solution.    
The large atmospheric mixing follows from the presence of large
left mixings both in the Dirac neutrino and charged lepton mass matrices. The latter
is correlated to an unobservable large right mixing of the down quark mass matrix. 
A third effect of the flavour symmetry is to adequately suppress 
non renormalizable operators inducing proton decay at a too fast rate.
Remarkably, the presence of the large representations demanded by the 
Missing Doublet Mechanism for the solution of the doublet-triplet splitting problem, 
makes possible a decrease of the predicted value of $\alpha_3(m_Z)$   
assuming coupling unification thanks to an increase of the effective mass
that mediates proton decay induced by dimension 5 operators.
As a consequence the value of the strong coupling is in
better agreement with the experimental value and
the proton decay rate is still compatible with present bounds, even if it
is difficult to avoid the conclusion that proton decay must occur with a rate 
at reach of current and next generation experiments.
Chapter 4 contains a detailed description of this model.
We think that this model
is interesting because it proves that a Supersymmetric $SU(5)$ Grand
Unification is not excluded and
offers a benchmark for comparison with experiment.

%%%%%%%%%%%%%%%%%%%%%%%%%%%%%%%

\clearemptydoublepage
\pagestyle{headings}
\chapter{Neutrino Masses and Mixings}

\label{numem}

Recent data from Super-Kamiokande \cite{SKatm} have provided a more solid experimental 
basis for neutrino oscillations as an explanation of the atmospheric neutrino anomaly. 
Also the solar neutrino deficit, observed by several experiments, is 
probably an indication of a different sort of neutrino oscillations. 
Neutrino oscillations imply neutrino masses, 
so that it is compelling to look for viable extensions of the Standard Model. 
In order to fulfill such a program, it is important to recognize that the extreme 
smallness of neutrino masses in comparison with quark and charged lepton masses 
indicates that the formers have different nature, linked to lepton number violation.
 
Thus, neutrino masses provide some insight on the very large energy scale where 
lepton number is violated and on Grand Unified Theories (GUTs). Indeed, GUTs 
constitute a very natural framework for the generation of neutrino masses, because they 
generically predict lepton and baryon number violation.
Neutrino masses could also give an important feedback on the problem 
of quark and charged lepton masses, as all these masses are related in GUTs. 
In particular the observation of a nearly maximal mixing angle for 
$\nu_{\mu}\rightarrow \nu_{\tau}$ is particularly interesting. 
At present, a large solar neutrino mixings seems to be slightly more preferable than
a small one. Large mixings in the neutrino sector are very interesting 
because a first guess was in favour of small mixings, in analogy to 
what is observed for left mixings in the quark sector. 
If confirmed, single or double maximal mixings can provide 
an important hint on the mechanism that generate neutrino masses. 
With neutrino masses settled, observation of proton decay will be
the next decisive challenge remained to support or eventually 
put in crisis GUTs.

After a concise summary of the present experimental status 
of neutrino masses and mixings and its future perspective, in this chapter 
we will 
review the most promising mechanisms  which reproduce 
a large atmospheric mixing.
 
\section[Present Experimental Status and Future Perspective]
{Neutrino Oscillations: Present Experimental Status and Future Perspective}

Despite the remarkable progress reached during the last few years, 
the experimental status of neutrino oscillations is still very preliminary. 
While the evidence for the existence of neutrino oscillations from solar and atmospheric 
neutrino data is rather convincing by now, the values of the mass 
squared differences $|\Delta m^2|$ and mixing angles are not firmly established. 
However, there is a general confidence that future experiments will 
not only succeed in clarifying such a picture but hopefully will also find one of the 
CP violating phases of the neutrino sector and establish the sign of the $\Delta m^2$.

\subsection{Atmospheric}

\subsubsection{Present}

Data with impressive high statistics are coming from Super-Kamiokande \cite{SKatm} 
which are in perfect agreement with the hypothesis of flavour oscillations \cite{bp, mns}.
Such hypothesis is consistent with all Super-Kamiokande data and is also corroborated
by independent 
atmospheric neutrino results from the Soudan-2 \cite{s2atm} and MACRO \cite{macroatm}
experiments, as well as by the pioneering Kamiokande experiment \cite{Katm}.

Zenith-angle distributions, up-down asymmetries and so on provide plenty of smoking guns
for the fact that something actually happens to atmospheric $\nu_{\mu}$ as a function
of $L E^n$ with $n \approx -1$, where $L$ is the baseline length and $E$ the energy.

In the simplest $2\nu$ case of  $\nu_{\mu} \rightarrow \nu_{\tau}$, a Super-Kamiokande
combined fit indicates that: \\
(i) the preferred value of
$\Delta m^2_{atm}$ lies in the range $(1.5 \div 5) \cdot 10^{-3}$ eV$^2$; \\
(ii) a large mixing angle appears to be well established: 
$sin^2 2\theta_{atm} > 0.88$ at 90\% confidence level.\\
Interpreted in terms of $2\nu$ oscillations, Super-Kamiokande favours $\nu_{\mu}
\rightarrow \nu_{\tau}$ over $\nu_{\mu} \rightarrow \nu_{s}$ at 99\% confidence level,
based on separate analyses of the zenith-angle distributions.
The same conclusions also come from the rate of $\pi^0$ production,
but in this case systematic uncertainties are still large. 
Pure $\nu_{\mu}\rightarrow \nu_e$ transitions do not provide a good fit to the data,
and are independently excluded by the negative $\nu_e$ disappearance searches 
in the CHOOZ \cite{chooz} and Palo Verde \cite{palov} reactors.

However, these results have to be taken {\it cum grano salis}. 
Additional oscillation channels
may be open, as naturally expected in $3\nu$ and $4\nu$ schemes accommodating
the current phenomenology. 
In fact, it has been realized that the dominant $\nu_{\mu} \rightarrow \nu_{\tau}$ 
oscillations plus subdominant $\nu_{\mu} \rightarrow \nu_e$ oscillations are also
consistent \cite{3nu} with combined Super-Kamiokande and CHOOZ data, leading to a much
richer three-flavour phenomenology, as will be discussed in the following.
On the other hand, when the data are analyzed 
\cite{bklww,4nu} for $\nu_{\mu}$ oscillating into a linear combination of  
$\nu_{\tau}$ and $\nu_{s}$, one finds such a scenario to be still viable 
(and also compatible with solar neutrino data) for a sizeable region of parameter space.
These analysis are very important to determine to what extent the $4\nu$ 
scenario motivated by the LSND experiment can be accommodated.

It is clear that the definite evidence would be direct $\tau$ production, but this is 
not an easy task for Super-Kamiokande. Actually, there is not yet direct evidence 
for an oscillation pattern.

Thus, there is still room to speculate on  other interpretations, 
like neutrino decay \cite{blpw}, decoherence scenarios \cite{lmm} and $\nu_{\mu}$
mixing with neutrino states propagating in large extra dimensions \cite{bcs}, 
which appear indistinguishable with present data with respect to the
$L/E$ distribution. Detailed analysis together with high statistics data could
test these non standard explanations. 
For instance, the exotic scenario of neutrino decay seems 
to fail \cite{3nu,flms} in reproducing the zenith-angle distributions of sub-GeV, multi-GeV and
up-going muons observed by Super-Kamiokande.

\subsubsection{Future}

So, future experiments are highly needed, among which 
long-baseline laboratory experiments, which require a muon decay source for the neutrino 
beam, seem to be very promising in order to: \\
(i) confirm the existence of atmospheric neutrino oscillations by using a controllable
beam and measuring an oscillation pattern;\\
(ii) precisely measure $|\Delta m^2_{atm}|$ and $sin^2 2\theta_{atm}$;\\
(iii) make new measurements possible.

The K2K experiment \cite{k2k} (L=250 km, $<E_{\nu}> =1.4$ GeV,
from KEK to Super-Kamiokande, in operation), 
finds an encouraging deficit of events with respect to the 
non oscillation hypothesis. The rate is perfectly compatible
with $\Delta m^2_{atm}=3 \cdot10^{-3}$ eV$^2$ and $sin^2 2\theta_{atm} =1$, 
but the statistics is still low and the significance only about $2 \sigma$. 

Long-baseline investigations will be pursued
by MINOS \cite{minos} (L=732 km, from Fermilab to Soudan, begin in 2002).
In addition to (i) and (ii), it should be able to discriminate between $\nu_{\mu}
\rightarrow \nu_{\tau}$ and $\nu_{\mu} \rightarrow \nu_{s}$ and
to measure $|U_{e3}|$ from $\nu_e \rightarrow \nu_{\mu}$ appearance. 

OPERA and ICANOE \cite{opic}
(L=743 km, from CERN to Gran Sasso, approved),
have the main purpose of looking for $\tau$ detection as the smoking gun for 
$\nu_{\mu} \rightarrow \nu_{\tau}$ oscillations. They also should be sensitive to
$|U_{e3}|$.

With respect to long-term future,
neutrino factories together with very long baselines ($L \sim 3000$ km)
appear \cite{nufact} to offer the possibility of
determining the sign of $\Delta m^2_{atm}$ from matter effects on the Earth crust,
measuring with an accuracy of tenths of degrees the angle $\theta_{13}$, down to
values of  $\theta_{13}=1\raisebox{1ex}{\scriptsize o}$.
The CP violating phase $\delta$ could also be measured if the solar neutrino
parameters lie in the range of the MSW-LMA solution.

\subsection{Solar}

\subsubsection{Present}

For solar neutrinos the experimental situation and the interpretation in terms of
frequency and mixing angles is less clear.

The solar neutrino experiments \cite{solexp} sample different $\nu_e$ energy ranges
and find different flux deficits compared to the Standard Solar Model (SSM) \cite{ssm},
as shown in Table \ref{defflux},
so that the $\nu_e$ survival probability is inferred to be energy dependent. 
\begin{table} [h]
\begin{center}
\begin{tabular}{|c|c|}
\hline 
& \\ & flux$^{\rm{(exp)}}$ / flux$^{\rm{(SSM)}}$ \\ &\\ \hline
&\\ GALLEX & $0.60\pm 0.06$ \\ 
SAGE & $0.52 \pm 0.06$ \\ & \\ Homestake & $0.33 \pm 0.03$ \\
& \\ Super-Kamiokande & $0.47 \pm 0.02$ \\ & \\ \hline
\end{tabular}
\end{center}
\caption{For the energy range specific of each type of experiment, 
the ratio between the experimentally measured neutrino flux and the 
flux predicted by the SSM is shown.}
\label{defflux}
\end{table}

Global oscillation fits have been made which include: total rates (the different suppression
ratio for Homestake plays a vital role), day-night asymmetry, seasonal dependence 
beyond $1/r^2$.
Typical candidate solar solutions are given in Table \ref{solsol}.
\begin{table} [h]
\begin{center}
\begin{tabular}{|c|c|c|}
\hline & &\\ solution &$\Delta m^2_{sol} (eV^2)$ & $sin^2 2\theta_{sol}$\\ & &\\ 
\hline & &\\ SMA-MSW & $\sim 5 \cdot 10^{-6} $ & $\sim 5 \cdot 10^{-3}$ \\ & &\\
LMA-MSW & $ \sim 2 \cdot 10^{-5}$ & $\sim 1$ \\ & &\\ LOW &$\sim 10^{-7}$ & $\sim 1$ \\
& &\\ VO & $\sim 10^{-(9 \div 10)}$&$ \sim 1$\\ & &\\ \hline
\end{tabular}
\end{center}
\caption{Typical solar solutions from global fits.}
\label{solsol}
\end{table}

Interesting recent developments by Super-Kamiokande (with reduced background and 
threshold of 5 MeV) show that the flux deficit is 
confirmed but no other direct signal of oscillations emerges:
the spectrum is compatible with flat, the day-night asymmetry is
reduced at the 1.3 $\sigma$ level and the seasonal variation is compatible
with the flux variation from the orbit eccentricity. These last two facts
disfavour respectively the SMA-MSW and the VO at the 95\% confidence level.
We are left with the LMA-MSW and the LOW, both with large mixing.
The pure transition $\nu_e \rightarrow \nu_s$ is disfavoured at 95\%.
It is important to keep in mind that the $\Delta m^2_{sol}$ values of the above solutions
are determined by the experimental result that the flux suppression is energy
dependent. If the Homestake indication were disregarded, new energy independent
solutions would emerge, with much more extended ranges for $\Delta m^2_{sol}$.

All these conclusions need to be confirmed by updated global analysis,
without forgetting the region with $\theta_{sol} > \pi/4$, as well as matter and
vacuum effects.
It is also important to analyze the data in the $3 \nu$ and $4 \nu$ schemes, to understand
to what extent a component $\nu_e \rightarrow \nu_s$ is tolerated \cite{bklww}. 

\subsubsection{Future}

The experimental situation will be hopefully clarified with the results from future 
experiments like SNO \cite{sno}, BOREXINO \cite{borex} and KamLAND \cite{kaml}.

In the present Phase I,
SNO clearly sees the signal from charged current (CC) reactions 
$\nu_e +d \rightarrow p+p+e^-$, with an energy spectrum which seems to be 
proportional to that of the SSM.
It also sees a clear signal from elastic scattering (ES) 
$\sum_{i} \nu_i +e \rightarrow \sum_{i} \nu_i +e$.  The ratio CC/ES is expected to be 
measured in a sufficiently accurate way to reveal at 3$\sigma$ if high energy neutrinos
with $i \not=  e$ are coming from the Sun at a rate consistent with
the observed deficit of $\nu_e$ with respect to the SSM expectation. 
In the following Phase II, the neutral current (NC) cross section $\sum_{i} \nu_i +d
\rightarrow p+p+ \sum_{i} \nu_i$ will be measured. With also the ratio of NC/CC, 
SNO should be able to discriminate between different solar neutrino scenarios.

BOREXINO can measure in real time the $^{7}Be$ flux and is sensitive to the
VO solution.

KamLAND, starting in 2001, should be able to test definitively the LMA-MSW and LOW 
solutions, using neutrinos from nuclear power reactors. It should also measure the 
day-night and seasonal effects for $^{7}Be$ neutrinos. 

Thus it seems possible that the solar neutrino problem will be better
clarified in the near future.

\subsection{LSND}

The issue of the additional claim for $\nu_e \rightarrow \nu_{\mu}$ neutrino oscillations
from the LSND experiment \cite{LSND} is still open, because it is neither confirmed nor 
completely excluded by KARMEN-2 \cite{Karmen}. 
The range still surviving is at small mixing angle
and with $\Delta m^2 = (0.2 \div 2)$ eV$^2$. This range is much above the solar and 
atmospheric neutrino frequencies, so that the need to accommodate a third frequency
would imply the existence of a fourth light sterile neutrino, in addition to the three 
established weak interacting flavours, which should mix with solar and/or atmospheric
neutrinos. 
MiniBOONE, starting in 2002, is an experiment designed to solve this problem.
It will be possibly followed by BOONE, if the signal is confirmed.

\subsection{Three Flavour Analysis}

\label{threeflan}
Given the present experimental uncertainties the theorist has to
make some assumptions on how the data will finally look like in the future. 
Here and in all the following sections, we tentatively assume that the LSND 
evidence will disappear. 
If so then we only have two oscillations frequencies, which can be given in terms of 
the three known species of light neutrinos without additional sterile kinds.

As pointed out, Super-Kamiokande data are compatible
with nearly maximal $\nu_{\mu}\rightarrow\nu_{\tau}$ two flavour oscillations,
but, among the possible deviations from this standard picture, three flavour oscillations
appear to be perfectly viable. In fact, Super-Kamiokande data do not exclude a 
non-vanishing $U_{e3}$ element. 

We then take for granted that the frequency of atmospheric neutrino oscillations 
will remain well separated from the solar neutrino frequency, even for the MSW solutions:
$\Delta m^2_{atm} > \Delta m^2_{sol}$.
Assuming that two out of three active neutrinos are almost degenerate, say 
$m_1\approx m_2$, one can show \cite{3nu} that atmospheric neutrinos probe only 
$\Delta m^2_{atm}= m_3^2-m^2_{1,2}$ and the mixing matrix elements $U_{\alpha 3}$,
which satisfy the unitarity constraint $U^2_{e 3}+U^2_{\mu 3}+U^2_{\tau 3}=1$.
The unitarity constraint can be conveniently embedded in a triangle plot.
The best fit for Super-Kamiokande data (70.5 kTy) is found at \cite{flmwinprog}
$(\Delta m^2_{atm}, U^2_{e 3}, U^2_{\mu 3}, U^2_{\tau 3})$~=
$(3.5 \cdot 10^{-3} eV^2,0.07,0.57,0.36)$.
The tolerated $\nu_\mu \rightarrow \nu_e$ mixing is constrained by $U^2_{e 3} < 0.31$ at
90\% confidence level.

The weak preference for $U^2_{e 3} \not=0$ is suppressed by including CHOOZ data.
The best fit for combined Super-Kamiokande and CHOOZ data gives 
$(\Delta m^2_{atm}, U^2_{e 3}$, $U^2_{\mu 3}, U^2_{\tau 3})$~=$(3.0 \cdot 10^{-3} eV^2,0.,0.5,0.5)$,
with the further constraint $U^2_{e 3} < 0.04$ at 90\% confidence level.

It appears very difficult to probe, with present atmospheric data, values of $U^2_{e 3}$
as small as a few \%. Constraining $U^2_{e 3}$ is one of the major tasks for future reactor
and accelerator neutrino experiments.

\section{Beyond the Standard Model}

Assuming just three flavours, neutrino oscillations are due to a misalignment 
between the flavour basis 
$\nu'\equiv(\nu_e,\nu_{\mu},\nu_{\tau})$, where $\nu_e$ is the partner of the mass and 
flavour eigenstate $e^-$ in a left-handed weak isospin $SU(2)$ doublet (similarly for 
$\nu_{\mu}$ and $\nu_{\tau}$), and the neutrino mass eigenstates 
$\nu\equiv(\nu_1, \nu_2,\nu_3)$: 
\beq
\vert\nu'\rangle =U\vert\nu\rangle,\label{U}
\eeq 
where $U$ is the MNS 3 by 3 mixing matrix \cite{mns}. 
The presence of mixing implies that neutrinos cannot be all massless and actually the
presence of two different oscillation frequencies implies at least two different non zero 
masses. 
Neutrino oscillations are practically only sensitive to differences $\Delta m^2$ so that 
the absolute scale of squared masses is not fixed by the observed frequencies.

\subsection{Neutrino Masses and Lepton Number Violation}

Neutrino oscillations imply neutrino masses which in turn demand either the 
existence of right handed neutrinos (Dirac masses) or lepton number (L) violation 
(Majorana masses) or both. Anyway, one is forced to go beyond the Standard
Model (SM). 
In fact, referring to three neutrino generations, 
in the SM Lagrangian density only the term 
$i\overline{\nu_L}^T \slsh D \nu_L$ appears, where
$D_{\mu}$ is the $SU(2) \otimes U(1)$ gauge covariant derivative and
$\nu_{L}=({\nu_e} ~~{\nu_\mu} ~~{\nu_\tau})^T$ is a column of three
Weyl spinors in a four-component notation (a Weyl spinor has only two 
independent components) satisfying
$\nu_{L}=1/2 ~(1+\gamma_5) \nu_L$.  The standard convention is to assign $L(\nu_{L})=1$.

If no additional right handed neutrino field $N_R=({N_e} ~~{N_\mu} ~~{N_\tau})^T$ 
(satisfying $N_{R}=1/2 ~(1-\gamma_5)  N_{R}$ and with $L(N_R)=1$) is allowed, 
there isn't the possibility of including in the Lagrangian the Dirac mass term 
\beq {\cal L}_{Dir} \ni -\overline{\nu_L}^T m_D N_R + h.c.\eeq 
which, by defining $\nu=\nu_L+N_R$, can be written as 
$ {\cal L}_{Dir}\ni -{\overline\nu}^T m_D \nu$, 
$m_D$ being a matrix in flavour space.
The above term do not violate L, which is an additive quantum number.
 
For a massive charged particle of spin 1/2 one
needs four states, while only two are enough for an intrinsically neutral particle. 
If L is violated, there is no conserved quantum number that really makes 
neutrinos and antineutrinos different and a new type of mass term is possible. 
We can have a Majorana mass term
\bea {\cal L}_{Maj} &\ni & -\frac{1}{2} {\overline{(\nu^c)_R}}^T m_{LL} \nu_L+h.c.
=-\frac{1}{2} {\nu^t_L}^T C m_{LL} \nu_L+h.c. \nn\\
&\doteq & -\frac{1}{2} \nu_L^T m_{LL} \nu_L+h.c. \eea 
where $(\nu^c)_R=C \overline{\nu_L}^t$ and 
$C$ is the 4 by 4 matrix in Dirac space that implements charge conjugation.
We denote by $t$ the transposition in Dirac space, while maintaining $T$ for
the transposition in flavour space.
The Majorana mass term violates L by two units. 
Also, since in the SM $\nu_L$ is a weak isospin doublet, it transforms as a component
of an isospin triplet. In the following, as we are only interested in 
flavour indices and not in Dirac indices, we will simply
denote the Majorana term for left handed fields by $\nu_L^T m_{LL} \nu_L+h.c.$, omitting the 
Dirac matrix $C$ and the transposition $t$. Note that if L is violated and
$N_R$ also exists, then a second type of Majorana mass is also possible which is 
\beq {\cal L}_{Maj} \ni -\frac{1}{2} N_R^T m_{RR} N_R+h.c.~, \eeq where we again
omitted $C$ and $t$. Clearly it also violates L by two units, but, since $N_R$ is a gauge singlet, 
this term is invariant under the SM gauge group. 

In conclusion, if $N_R$ does not exist, we can only have a Majorana mass 
$m_{LL}$ if L is violated. 
If $N_R$  exists and L is violated, we can have both Dirac $m_{D}$ and Majorana 
masses $m_{LL}$ and $m_{RR}$.

If one wants to give masses to neutrinos but to avoid the conclusion that L is violated, 
then one must assume that $N_R$ exists and that neutrinos acquire Dirac 
masses through the usual Higgs mechanism as quark and leptons do. 
Technically this is possible. But there are two arguments against this possibility. \\
1) The first argument is that neutrino masses are extremely small so that the corresponding 
Yukawa couplings would be enormously smaller than those of any other fermion. 
Note that within each generation the spread of masses is by no more than a factor 
$10^{3\div5}$. But the spread between the $t$ quark and the heaviest neutrino would 
exceed a factor of $10^{11}$. \\
2) A second argument arises from the fact that once we introduce $N_R$ in the theory, 
then the L violating term $N^T_R m_{RR} N_R$ is allowed in the Lagrangian density 
by the gauge symmetry. 
In the minimal SM, i.e. without $N_R$, L and B conservation are 
accidental global symmetries that hold because there is no operator term of
dimension $\leq4$ that violates B and L but respects the gauge symmetry. 
On the other hand, in the presence of $N_R$, the dimension 3 operator 
$N_R^T m_{RR} N_R$ is gauge symmetric but violates L. 
We expect a mass factor in front of this operator in the Lagrangian
density, and in the absence of a protective symmetry, we expect it to be of the order of 
the cut off, i.e. of order $M_{GUT}$ or larger. 
Thus, L number violation is naturally induced by the presence of $N_R$, 
unless we enforce it by hand.
Once we accept L violation we gain an elegant explanation for the smallness of neutrino 
masses as they turn out to be inversely proportional to the large scale where lepton 
number is violated. 

\subsection{Smallness of Neutrino Masses}

Given that neutrino masses are extremely small, it is really difficult from the theory point 
of view to avoid the conclusion that L must be violated. In fact, it is only in terms of lepton
number violation that the smallness of neutrino masses can be explained as 
inversely proportional to the very large scale where
L is violated, of order $M_{GUT}$.

Assuming three neutrino flavours ${\nu'_L}= (\nu_e ~\nu_\mu ~\nu_\tau)^T$, 
consider the following two situations (in the neutrino interaction basis and after having 
diagonalized the charged lepton mass matrix).
\begin{itemize}
\item
If L is not conserved, even in the absence of $N_R$, Majorana masses can be generated 
for neutrinos by allowing in the Lagrangian the presence of dimension five 
operators of the form 
\beq 
O_5=-\frac{1}{2}\frac{(L^T \cdot \phi)~ \lambda~ (L \cdot \phi)}{M} \ni 
-\frac{1}{2}\frac{{\nu'}_L^T ~\lambda ~{\nu'}_L }{M}\langle \phi^0\rangle \langle \phi^0\rangle
\label{O5}
\eeq 
with $\phi=(\phi^+ ~ \phi^0)^T$ being the ordinary Higgs doublet, $L=({\nu'_L}~l_L)^T$ 
the weak lepton doublet, $\lambda$ a matrix in flavour space and
$M$ a large scale
($\cdot$ is put to remember the weak isospin). 
Neutrino masses generated by $O_5$ are then of the order 
$m_{\nu}\sim v^2/M$ for $\lambda_{ij}\sim {\rm O}(1)$, where 
$\langle \phi^0\rangle=v \sim {\rm O}(100~\rm{GeV})$ is the vacuum expectation 
value of the ordinary Higgs.
\item
The existence of a right handed neutrinos
is quite plausible because all GUT groups larger than $SU(5)$ 
require them. In particular, the fact that a field with the same quantum numbers of
a right handed neutrino completes the representation $16$ of 
$SO(10)$: $16=\bar 5+10+1$, so that all fermions of each family are contained in a single 
representation of the unifying group, is too impressive not to be significant. At least as a
classification group $SO(10)$ must be of some relevance. Assuming 
that there are both $N_{R}=(N_e ~N_\mu~ N_\tau)^T$ and L violation, 
the see-saw mechanism \cite{ss} is possible.
Consider the mass terms in the Lagrangian corresponding to Dirac and RR Majorana 
mass matrices (we consider LL Majorana mass terms as comparatively negligible) in 
the interaction basis in which charged leptons are diagonal:
\beq 
{\cal L}_{seesaw}=-\overline{{\nu'}_L}^T m_D N_R+
\frac{1}{2} N_R^T M_{RR} N_R+~ h.c.
\label{lag}
\eeq
The 3 by 3 matrices $m_D$ and $M_{RR}$ are the  Dirac and Majorana mass matrices in 
flavour space ($M_{RR}$ is symmetric, $M_{RR}=M_{RR}^T$, while $m_D$ is, in general, 
non hermitian and non symmetric).  One expects $m_D$ to come from the Higgs mechanism:
${\cal L}_{Yuk} 
%\ni - \phi  \cdot \overline{L}^T ~y_D~ N_R
\ni -\overline{{\nu'}_L }^T y_D N_R  \langle \phi^0 \rangle$. 
We expect the eigenvalues of $M_{RR}$ to be of order $M_{GUT}$ or more because 
RR Majorana masses are $SU(3)\times SU(2)\times U(1)$ invariant,  hence
unprotected and naturally of the order of the cut off of the low-energy theory.  
Since all $N_R$ are very heavy we can integrate them away and the resulting 
effective neutrino mass matrix is:
\beq
{\cal L}_{seesaw}=-\frac{1}{2} {\nu'}_L^T m_{\nu}^{eff} {\nu'}_L  +~ h.c.
\eeq
with \beq m_{\nu}^{eff}=m_D M_{RR}^{-1} m_D^T~~.
\eeq
This is the well known see-saw mechanism \cite{ss} result: the light neutrino masses 
are quadratic in the Dirac masses and inversely proportional to the large Majorana mass.  
If some $N_R$ are massless or light they would not be integrated away but
simply added to the light neutrinos. 
\end{itemize} 

Here we assumed that the additional non renormalizable terms from $O_5$ are 
comparatively negligible, otherwise they should
simply be added. After elimination of the heavy right-handed fields, at the level 
of the effective low energy theory, the two
types of terms are equivalent. In particular they have identical 
transformation properties under a chiral change of basis in
flavour space. The difference is, however, that in the see-saw mechanism, 
the Dirac matrix $m_D$ is presumably related to
ordinary fermion masses because they are both generated by the Higgs 
mechanism ($m_D=y_D v$) and both must obey GUT-induced constraints. 
Thus, if we assume the see-saw mechanism more constraints are implied. 
In particular we are led to the natural hypothesis that
$m_D$ has a largely dominant third family eigenvalue in analogy 
to $m_t$, $m_b$ and $m_{\tau}$ which are by far the largest
masses among $u$ quarks, $d$ quarks and charged leptons. 
Once we accept that $m_D$ is hierarchical it is very difficult to
imagine that the effective light neutrino matrix, generated by the see-saw 
mechanism, could have eigenvalues very close in absolute value.

Then, assuming that one neutrino mass, presumably the third generation neutrino mass 
$m_3$, is much larger than any other, we have $\Delta m^2_{atm}\approx m^2_3$, 
which means $|m_3| \approx 0.05$ eV. 
If we apply the see-saw formula assuming that the Majorana RR
masses lie around $M$, then, with $|m_3| \approx 0.05$ eV, ${m_D}_{33} \sim 200$ 
GeV (like the top quark mass or the Higgs vev), we find $M\sim 0.8 \cdot 10^{15}$ GeV, a value
amazingly close to $M_{GUT}$. This supports the initial assumption that indeed 
neutrino masses are related to the scale of L violation.

\subsection{Approximate Mixing Matrix in the Neutrino Sector}

Given the definition of the mixing matrix $U$ in eq. (\ref{U}) and the transformation 
properties of the effective light neutrino mass matrix $m_{\nu}^{eff}={m_{\nu}^{eff}}^T$:
\bea  
{\nu'}_L^T m_{\nu}^{eff} \nu'_L &= &\nu_L^T (U^T m_{\nu}^{eff} U) \nu_L \nn\\  
U^T m_{\nu}^{eff}U& = &{\rm Diag}[e^{i\phi_1}m_1,e^{i\phi_2}m_2,m_3]\equiv m_{diag}
\label{tr}
\eea  
where $m_{1,2,3}$ are real and positive,
we obtain the general form of $m_{\nu}^{eff}$:
\beq 
m_{\nu}^{eff}=U m_{diag} U^T~~.
\label{gen}
\eeq 
The MNS matrix $U$ can be parameterized in terms of three mixing angles and one phase, 
exactly as for the quark mixing matrix $V_{CKM}$. 
\beq  U= 
\left[\matrix{ c_{12} c_{13} & s_{12} c_{13}  & s_{13} \cr 
- s_{12} c_{23} e^{i \delta}-c_{12} s_{13} s_{23} & c_{12} c_{23} e^{i \delta}-s_{12} s_{13} s_{23}
 & c_{13} s_{23} \cr 
s_{12} s_{23} e^{i \delta}-c_{12} s_{13} c_{23}& -c_{12} s_{23} e^{i \delta}-s_{12}s_{13} c_{23}
& c_{13} c_{23}} 
\right] ~~~~~.
\eeq  
In addition we have the two phases 
$\phi_1$ and $\phi_2$ that are present because of the Majorana nature of neutrinos. 
Thus, in general, 9 parameters are added to the SM when non vanishing neutrino 
masses are included: 3 eigenvalues, 3 mixing angles and 3 CP violating phases 
\cite{cp1,cp2}.

Maximal atmospheric neutrino mixing and the requirement that the electron neutrino 
does not participate in the atmospheric oscillations, as indicated by the Super-Kamiokande
and CHOOZ data, lead directly to consider as good approximation:
$|U_{\mu3}|=|U_{\tau3}|$=$1/\sqrt{2}$ and $U_{e3}=0$.
Hence, $c_{23}=s_{23}=1/\sqrt{2}$ and $s_{13}=0, c_{13}=1$ (but different sign conventions 
are possible \cite{bar}) and the following structure of the
$U_{fi}$ ($f=e,\mu,\tau$, $i=1,2,3$) mixing matrix is obtained:
%(here we are not interested in CP violation effects: all matrices are taken real)
\beq  U_{fi}= 
\left[\matrix{ c_{12}& s_{12}&0 \cr 
-s_{12}/\sqrt{2}&c_{12}/\sqrt{2}&1/\sqrt{2}\cr 
s_{12}/\sqrt{2}&-c_{12}/\sqrt{2}&+1/\sqrt{2}     } 
\right] ~~~~~.
\label{ufi}
\eeq  
This result refers to the case of arbitrary solar mixing angle $s_{12}\equiv\sin{\theta_{sol}}$,
so that 
$c_{12}=s_{12}=1/\sqrt{2}$ holds for maximal solar mixing. 
In the limit $U_{e3}=0$, all CP violating effects vanish and we can neglect the 
additional phase parameter $\delta$ generally present in $U_{fi}$: the matrix $U$ is
real and orthogonal and equal to the product of a rotation by $\pi/4$ in the 23 plane times 
a rotation in the 12 plane.

While the simple parameterization of the matrix $U$ in eq. (\ref{ufi}) is quite useful
to guide the search for a realistic pattern of neutrino mass matrices, it should not be 
taken too literally. In particular, as stressed in section \ref{threeflan},
the data do not exclude a non-vanishing $U_{e3}$ element. 

\section{How to Reproduce the Large Atmospheric Mixing}

\subsection{Mass Hierarchies and Approximate Zeroth Order Textures}

Note that since we are assuming only two frequencies, given by 
\beq\Delta_{sol}= m^2_2-m^2_1,~~~~~~~
\Delta _{atm}\cong  m^2_3-m^2_{1,2}~~,\label{fre}
\eeq 
there are three possible hierarchies of mass eigenvalues
\cite{af,bar}:
\bea
        {\rm A}& : & |m_3| >> |m_{2,1}| \nonumber\\
        {\rm B}& : & |m_1|\sim |m_2| >> |m_3| \nonumber\\
        {\rm C}& : & |m_1|\sim |m_2| \sim |m_3|
\label{abc}
\eea 
(in case A there is no prejudice on the $m_1$, $m_2$ relation). 
For B and C different subcases are then generated
according to the relative sign assignments for $m_{1,2,3}$. 

Using eqs. (\ref{tr}) and (\ref{ufi}), for each case, by setting to zero the small masses, 
it is possible to find \cite{af,afrev} the effective light neutrino matrices which
automatically lead to maximal atmospheric mixing
\footnote{Note that here we are working in the basis where the charged lepton masses
are diagonal, and approximately given by $m_l={\rm Diag}[0,0,m_{\tau}]$. 
For model building one has to arrange both the charged lepton and the neutrino mass 
matrices so that the neutrino results coincide with those given here after diagonalization 
of charged leptons.}. Such zeroth order texture can be taken as guides in the
construction of specific models for neutrino masses and mixings.

For example, in case A, $m_{diag}={\rm Diag}[0,0,m_3]$ and we obtain
\beq A:~~~m_{diag}={\rm Diag}[0,0,1]~m_3~~~~~~\rightarrow~~~~~~ m_{\nu}^{eff}= 
\left[\matrix{ 0&0&0\cr
0&\frac{1}{2}&\frac{1}{2}\cr 
0&\frac{1}{2}&\frac{1}{2}    } \right] m_3~~~~~. 
\label{a}
\eeq 
In this particular case the results are the same for double and single maximal mixing. 
Note that the signs correspond to the phase convention adopted in eq. (\ref{ufi}). 

One example of case B follows by taking 
$m_{diag}={\rm Diag}[m_1,-m_1,0]$  with double maximal mixing, so that:  
\beq B:~~~m_{diag}={\rm Diag}[1,-1,0]~m_1~~~~\rightarrow~~~~~ m_{\nu}^{eff}= 
\left[\matrix{ 0&-\sq&\sq\cr
-\sq&0&0\cr
\sq&0&0}\right] m_1~~~. 
\label{caseB}
\eeq

A strong constraint arises in case C from the non observation of neutrino-less 
double beta decay which requires that the $ee$ entry of $m_{\nu}^{eff}$ must obey
$|(m_{\nu})_{ee}|\leq 0.2~{\rm eV}$ \cite{dbeta}. 
As observed in ref. \cite{GG}, if the average neutrino mass is around $1$ eV, it can only be  
satisfied if bimixing is realized (that is double maximal mixing, with solar neutrinos 
explained by the VO solution or maybe by the large angle MSW solution). 
The reason is that, as results from eqs. (\ref{tr}) and (\ref{ufi}),
$(m_{\nu})^{eff}_{ee}=m_1c_{12}^2+m_2 s_{12}^2$. 
For the required cancellation one needs opposite signs for $m_1$ and
$m_2$ and comparable values of $c_{12}^2$ and $s_{12}^2$, that is nearly maximal 
solar mixing. In this case, one example of case C is: 
\beq C:~~~m_{diag}={\rm Diag}[-1,1,1]~m_1~~~~\rightarrow~~~~~ m_{\nu}^{eff}= 
\left[\matrix{ 0&\sq&-\sq\cr
\sq&\frac{1}{2}&\frac{1}{2}\cr
-\sq&\frac{1}{2}&\frac{1}{2}}\right] m_1~~~. 
\eeq

It is possible to proceed similarly in all the other cases \cite{afrev}.
Once a solution to the solar neutrino problem is chosen and a set of 
suitable small perturbation terms is introduced, oscillation phenomena are unable to 
distinguish between the cases $A$, $B$ and $C$. However,
from the model building point of view, each texture represents an independent 
possibility in a zeroth order approximation.

\subsection{Degenerate Neutrinos}

Configurations B and C imply a very precise near degeneracy of squared masses.
For instance, case C with $m_i \sim 1$ eV would require a relative splitting 
$|\Delta m/m|\sim \Delta m^2_{atm}/2m^2\sim 10^{-(3\div 4)}$ 
and a much smaller one for solar neutrinos, especially if explained by the VO solution: 
$|\Delta m/m|\sim 10^{-(10 \div11)}$. 
Case C is the only one in which
neutrinos could represent a significant component of hot dark matter.
An intermediate possibility between the case of degenerate neutrino masses and the 
case of hierarchical ones is the situation $B$ in eq. (\ref{abc}). With two almost 
degenerate heavier states and a nearly massless neutrino we cannot reach a mass range
interesting for cosmological purposes. However, since the experimentally 
accessible quantities are the squared mass differences, case $B$ remains an open 
possibility.

As mentioned above, it appears to be unplausible that starting from hierarchical Dirac
matrices we end up via the see-saw mechanism into a nearly perfect degeneracy of 
squared masses. 
Thus models with degenerate neutrinos could only be natural if the dominant 
contributions directly arise from non renormalizable operators like $O_5$ in
eq. (\ref{O5}) because they are {\it a priori} unrelated to other fermion mass terms.  
The degeneracy of neutrinos should be guaranteed by some slightly broken symmetry,
which could also give the small terms required to go beyond the zeroth order 
approximation.  
Models based on discrete or continuous symmetries have been proposed \cite{fri}.

Notice however that, for the VO scenario, even if arranged at the GUT scale, 
it is doubtful that such a precise degeneracy could be stable against renormalization 
group corrections when running down at low energy unless it is protected by 
a suitable symmetry \cite{ello,brs}.  
For instance, working only with the light neutrinos, the $U(1)_Q$ charge 
$Q\equiv(L_e-L_\mu-L_\tau)$ allows at leading order \cite{hall} for the texture of 
the example in (\ref{caseB}). 
The non-vanishing entries $(m_\nu)_{12}$ and $(m_\nu)_{13}$ are  expected to be of 
the same order, although not necessarily equal. The vanishing entries can be 
filled by the ratio of the vev $\langle\varphi\rangle$ of a scalar field carrying two units 
of $Q$ and a mass scale $M$ providing the cut off to the low-energy theory:
\beq  m_\nu=m
\left[\matrix{
\epsilon&1&1\cr 1&\epsilon&\epsilon\cr 1&\epsilon&\epsilon} 
\right]~~~~~, 
\label{B1}
\eeq 
where $\epsilon=\langle\varphi\rangle/M$ and only the order-of-magnitudes are indicated. 
For values of $\epsilon$ smaller than 1 one obtains a small perturbation of the zeroth 
order texture. 
The mixing matrix has a large, not necessarily maximal, mixing angle in the 23 sector, 
a nearly maximal mixing angle in the 12 sector and a mixing angle of order $\epsilon$ 
in the 13 sector \cite{ghos}. The mass parameter $m$ should be  close to 
$10^{-(1\div2)}$ eV to provide the frequency required by the atmospheric oscillations. 
Finally, $m_2^2-m_1^2\sim m^2\epsilon$. In this model the VO solution to the solar neutrino 
deficit would require a very tiny breaking term, $\epsilon\sim {\rm O}(10^{-7})$.
The symmetry protects the degeneracy of $m_1$ and $m_2$ against 
renormalization group corrections \cite{brs,ibar}.

This mechanism is a typical example of such models which are very attractive at 
first glance, but which, after a deeper analysis, turns out to be incomplete.
In fact, the symmetry introduced is not able to account for the spectra of charged
leptons - which, in this model, would be roughly degenerate - and of quarks. 
Thus, in order to construct a complete model of fermion masses, one has to do some 
further assumption. For instance one can introduce an ad hoc flavour symmetry 
which fulfills this task without damaging the effective neutrino mass matrix.

\subsection{Hierarchical Neutrinos}

We now discuss models of type A with large effective light neutrino mass splittings 
and large mixings. 
Reconciling large splittings with large mixing would seem difficult. 
Indeed, one could guess that, in analogy to what is observed for 
quarks, large splittings correspond to small
mixings because only close-by states are strongly mixed. 
In particular it is of basilar importance to find a mechanism
which leads to the large atmospheric mixing.
Actually, there are mainly two ways in which this situation can be achieved.

\subsubsection{Large Mixing from Stretching}

The first one arises by stretching some parameters \cite{bapawi}. For instance,
one could start with not too small 23 mixings both in the charged lepton and 
in the effective neutrino mass matrices. For instance, 
take $\theta_l \sim \theta_\nu \sim \theta_C \sim 13^0$ 
where $\theta_C$ is the Cabibbo angle. If the relative phases are such
that the two angles just add, then from two not too small angles 
one can end up with a large one. Clearly the situation is not very
satisfactory because the largeness of the mixing is completely accidental rather than
descending as a consequence of some mechanism. 
In addition, one can expect at most the mixing to be large, but not at all to be 
exactly maximal.

\subsubsection{Large Mixing from the See-Saw}
\label{lmseesaw}  
Via the see-saw mechanism \cite{ss}, there are other two 
particularly simple subcases in which a large atmospheric mixing can be realized. 
They are based on the observation that, in a 2 by 2 matrix context, the requirement of 
large splitting and large mixings leads to a condition of vanishing determinant. 
For example the matrix
\beq m\propto 
\left[\matrix{ x^2&x\cr x&1    } 
\right]~~~~~. 
\label{md0}
\eeq 
has eigenvalues 0 and $1+x^2$ and for $x$ of O(1) the mixing is large. 
Thus, in the limit of neglecting small mass terms of order $m_{1,2}$, the demands 
of large atmospheric neutrino mixing and dominance of $m_3$ translate into the 
condition that the 2 by 2 subdeterminant 23 of the 3 by 3 mixing matrix vanishes. 
The problem is to show that this approximate vanishing can be
arranged in a natural way without unacceptable fine tuning.

Without loss of generality, leaving aside for the moment the possible presence
of flavour symmetries, one can go to the basis where both the charged 
lepton Dirac mass matrix $m_l$ and the Majorana matrix $M_{RR}$ of the 
right handed neutrinos are diagonal.
For simplicity, we start assuming that the role of the first generation 
is not crucial in the mechanism for the generation of neutrino masses, 
so than one can, with good approximation, work in the 2 by 2 case 
(in the next section this condition will be relaxed). 
Writing $m_D$ and $M_{RR}$, defined as in eq. (\ref{lag}), in the 
most general way
\beq
m_D=v \left[\matrix{a&b \cr c&1}\right],
~~~~~ M_{RR}=\left[\matrix{M_2&0\cr 0&M_3}\right]~~~~~,
\label{mdM}
\eeq
where $v$ is a vacuum expectation value, $a,b$ and $c$ are Yukawa 
couplings, then, via the see-saw, one obtains:
\beq
m_{\nu}^{eff}=\dd\frac{v^{2}}{M_3}
\left[\matrix{\dd\frac{a^{2}}{M_2}+\dd\frac{c^{2}}{M_3}~~&
\dd\frac{a~b}{M_2}+\dd\frac{c}{M_3} \cr &
\cr \dd\frac{a~b}{M_2}+\dd\frac{c} {M_3}~~&\dd\frac{b^{2}}{M_2}+\dd\frac{1}{M_3} }
\right]~~~~~.
\label{mnM}
\eeq
The request of large splittings among the light neutrino eigenvalues
is equivalent to demanding that the determinant of $m_{\nu}^{eff}$ is
much smaller than its trace. 
It is then possible to see at glance that two cases arise exhibiting large mixing, 
that is when the terms with respectively $M_3$ or $M_2$ at the denominator
are dominant.

\begin{itemize} 
\item {\bf Case I: Large mixing from $m_D$}\\
One simple example of the first case is realized if $M_2 \sim M_3$ and $a,b \ll 1$. 
In order to have a large splitting, one must have $c\sim 1$, that is the right handed 
neutrino of the third generation couples with the same strength \cite{King} to left handed 
$\nu_{\mu}$ and $\nu_{\tau}$. 
The heaviest mass for light neutrinos is then $m_{3}\sim v^{2}/M_3$. 
As already pointed out, in the hierarchical case the data from 
Super-Kamiokande give $m_{3}\sim 0.05$ eV so that, if one assumes that $v$ is
a typical weak scale, namely $200$ GeV, then $M_3 \sim 10^{15}$ GeV, just a bit
lower than the GUT scale.

It is worth to stress that this first mechanism makes use of {\it a strongly asymmetric Dirac 
matrix}, that is a matrix with a large left handed mixing already present. 
It has been observed \cite{af,bralav,berrossi,albrbarr,largem} that in $SU(5)$ left 
handed charged lepton mixings  
correspond to right handed mixings for down quarks (more on this later on). 
Large right handed quark mixings are not in contrast with experiment and 
viable GUT models that correctly reproduce the data on fermion masses and mixings 
can be constructed following this mechanism.

\item {\bf Case II: Large mixing from nothing}\\
An alternative possibility \cite{afm1} is to have dominance of the terms with
$M_2$ at the denominator. 
This is achieved for any $c<1$ if $a^{2},b^{2}> M_2/M_3$. 
The request for large splitting is then equivalent to the requirement that also $a\sim b$. 
Now it is the second generation right handed neutrino which is particularly 
light and which couples with the same strength \cite{King} to left handed $\nu_{\mu}$ 
and $\nu_{\tau}$. 
In order to be more specific, consider the following example with symmetric matrices. 
These matrices are interesting because, for instance,
one could want to preserve left right symmetry at the GUT scale. 
Then, the observed smallness of left handed quark mixings would
also demand small right handed mixings. 
Starting from 
\beq
 m_D=v
\left[\matrix{\ep&x\ep\cr x\ep&1    }
\right],~~~~~ M_{RR}^{-1}=\frac{1}{M_3}
\left[\matrix{r_2&0\cr 0&1    }
\right]~~~~~,
\label{mdMafm1}
\eeq
where $\ep$ is a small number, $x$ is of $O(1)$ and $r_2\equiv M_3/M_2$, 
then, via the see-saw, it is sufficient that $\ep^2 r_2\gg 1$ in order to have approximately:
\beq
m_{\nu}^{eff}=\frac{v^2}{M_3} 
\epsilon^2 r_2\left[\matrix{1&x\cr x&x^2}
\right]~~~~~.
\label{mnp1}
\eeq
The determinant is naturally vanishing so that the mass eigenvalues are 
widely split and for $x\sim 1$ the mixing is nearly maximal. It is exactly maximal if $x=1$.
The see-saw mechanism has created {\it large mixing from almost nothing}: all
relevant matrices entering the see-saw are nearly diagonal \cite{afm1,Smir}, 
that is they are diagonalized by transformations that go into the 
identity in the limit of vanishing $\ep$. 
Clearly, the crucial factorization of the small parameter $\epsilon^2$ only arises if 
the light Majorana eigenvalue is coupled to $\nu_\mu$ and
$\nu_\tau$ with comparable strength, that is $x\sim 1$.
An interesting feature of this second case, in connection with a possible
realization within a GUT scheme, is that it requires 
$M_3>v^2/m_3$, so that one can push $M_3$, the scale of L violation, beyond the GUT scale. 
This is a desirable feature because, for instance, it is expected in $SU(5)$ 
if right handed neutrinos are present and also in the breaking of 
$SO(10)$ to $SU(5)$.
Summarizing, the second case require a peculiar hierarchy in the 
Majorana eigenvalues in order to work, but it has however the good 
characteristic of being realized with nearly diagonal matrices. 
\end{itemize}

In the next chapter we present concrete examples which realize Case I and Case II 
in a 3 by 3 context, the desired textures being motivated by the introduction of 
a suitable flavour symmetry.

\clearemptydoublepage
\chapter[Models with Large Mixing]{Models For Hierarchical Neutrinos with Large Atmospheric Mixing}

\label{modneu}

\section{A Model with Large Mixing from $m_D$ }

By left handed mixing we mean non diagonal matrix elements that can only be eliminated 
by a large rotation of the left handed fields. Thus the question is how to reconcile large 
left handed mixings in the leptonic sector with the observed near diagonal form 
of $V_{CKM}$, the quark mixing matrix. 
Strictly speaking, since $V_{CKM}=U^{\dagger}_u U_d$, the individual matrices 
$U_u$ and $U_d$ need not be near diagonal, but $V_{CKM}$ does, while the analogue 
for leptons, $U$ of eq. (\ref{ufi}), cannot be near diagonal. 
However nothing forbids for quarks that, in the basis where $m_u$ is diagonal, 
the $d$ quark matrix has large non diagonal terms that can be rotated away by a pure
right handed rotation. 

Right handed mixings for quarks can correspond to left handed mixings 
for leptons.
In fact, in the context of SUSY $SU(5)$ \cite{ross} there is a very attractive hint of how 
this can be realized. 
In the $\bar 5$ of $SU(5)$ the $d^c$ singlet appears together with the lepton doublet 
$(\nu,e)$. The $(u,d)$ doublet and $e^c$ belong to the 10 and $\nu^c$ to the 1 and 
similarly for the other families. 
As a consequence, in the simplest model with mass terms arising from only Higgs 
pentaplets, the Dirac matrix of down quarks is the transpose of the charged lepton matrix:
$m^d_D=(m^l_D)^T$. 
Thus, indeed, a large mixing for right handed down quarks corresponds to a large 
left handed mixing for charged leptons because, at leading order we may have
\footnote{Here and in the following we write Dirac mass matrices with the 
convention $\bar{L} m_D R$.}:
\beq m^d_D=(m^l_D)^T=
\left[
\matrix{ 0&0&0\cr 0&0&0\cr 0&1&1}
\right]v_d
\eeq 
In the same simplest approximation with  5 or $\bar 5$ Higgs, the up quark mass 
matrix is symmetric, so that left and right mixing matrices are equal in this case. 
Then small mixings for up quarks and small left handed mixings for down quarks
are sufficient to guarantee small $V_{CKM}$ mixing angles even for large $d$ quark 
right handed mixings.  
If these small mixings are neglected, we expect:
\beq m^u_D=
\left[
\matrix{ 0&0&0\cr 0&0&0\cr 0&0&1}
\right]v_u~~~.
\eeq 
The crucial fact is that {\it when the charged lepton matrix is diagonalized the large 
left handed mixing of the charged leptons is transferred to the neutrinos}. 
To see this, note that in $SU(5)$ we can always diagonalize the $u$ mass matrix 
by a rotation of the fields in the 10, the Majorana matrix $M$ by a rotation of the 1 
and the effective neutrino matrix $m_\nu$ by a rotation of the $\bar 5$. 
In this basis the $d$ quark mass matrix fixes $V_{CKM}$ and the charged lepton mass
matrix fixes $U$, the neutrino mixing matrix. 

\subsection{The Choice of Horizontal Abelian Charges}

We give here an explicit example \cite{af} of the mechanism under discussion 
in the framework of a unified SUSY $SU(5)$ theory with an additional 
$U(1)_F$ flavour symmetry \cite{fro}. 
This model is to be taken as merely indicative, in that some important problems, like,
for example, the cancellation of chiral anomalies, are not tackled here. 
Most importantly, in the following minimal version, it has the unpleasant feature of
predicting the same spectrum for down quarks and charged leptons.
The problem of the differentiation of these spectra will be addressed in the following
chapters.
Despite this, we find it impressive that the general pattern of
all what we know on fermion masses and mixings is correctly reproduced 
at the level of orders of magnitude.  
We regard the present model as a low-energy effective theory valid at energies close 
to $M_{GUT}\ll M_{Pl}$. 
We can think to obtain  it by integrating out the heavy modes from an unknown 
underlying fundamental theory defined at an energy scale  close to $M_{Pl}$. From 
this point of view the gauge anomalies generated by the light supermultiplets listed 
below  can be compensated by another set of supermultiplets with masses above 
$M_{GUT}$, already eliminated from the  low-energy theory. In particular, we assume
that these  additional supermultiplets are vector-like with respect to $SU(5)$ and 
chiral with respect to $U(1)_F$. 
Their masses are then naturally expected to be of the order of the $U(1)_F$
breaking scale, which, in the following discussion, turns out to be near $M_{Pl}$. 
It is possible to check explicitly the possibility of canceling the gauge anomalies 
in this way but, due to our ignorance about the fundamental theory,  it is not
particularly instructive to illustrate the details here. 

In this model the known generations of quarks and leptons are contained in  triplets
$\Psi_{10}^a$ and $\Psi_{\bar 5}^a$, $(a=1,2,3)$ transforming as $10$ and 
${\bar 5}$ of $SU(5)$, respectively. Three more $SU(5)$ singlets $\Psi_1^a$ describe 
the right handed neutrinos. We start by discussing the Yukawa coupling allowed 
by $U(1)_F$-neutral  Higgs multiplets $\varphi_5$ and $\varphi_{\bar5}$ in the $5$ and 
${\bar 5}$ $SU(5)$ representations and by the flavon $X$, singlet of $SU(5)$ with 
$F=-1$.
 
We assign to these fields the following $F$-charges:
\begin{table}[h]
\begin{center}
\begin{tabular}{|c|c|c|c|c|}
\hline  & & && \\ field & $\Psi_{10}$ & $\Psi_{\bar 5}$&$\Psi_1$ & $X$\\ 
& & & &\\ \hline  & && &\\
$F $ & $(3,2,0) $ & $(3,0,0)$ & $(1,-1,0)$ & $-1$ \\ & & & &  \\ \hline
\end{tabular} \end{center} 
\caption{$F$ charge assignments.}
\label{afch}
\end{table}
   
\subsection{Charged Fermion Mass Matrices}

In the quark sector we obtain 
\footnote{In eq. (\ref{mdirquark}) the entries denoted by 1 in $m_D^u$ and $m_D^d$  are 
not necessarily equal. As usual, such a
notation allows  for $O(1)$ deviations. 
Also remember that we write Dirac mass matrices with the 
convention $\bar{L} m_D R$.}:
\beq m_D^u=(m_D^u)^T=
\left[
\matrix{
\lambda^6&\lambda^5&\lambda^3\cr
\lambda^5&\lambda^4&\lambda^2\cr
\lambda^3&\lambda^2&1}
\right]v_u~~,~~~~~~~ m_D^d=
\left[
\matrix{
\lambda^6&\lambda^3&\lambda^3\cr
\lambda^5&\lambda^2&\lambda^2\cr
\lambda^3& 1 &1}
\right]v_d~~,
\label{mdirquark}
\eeq  from which we get the order-of-magnitude relations:
\bea m_u:m_c:m_t & = &\lambda^6:\lambda^4:1 \nonumber\\ 
m_d:m_s:m_b & = &\lambda^6:\lambda^2:1
\eea and 
\beq V_{us}\sim \lambda~,~~~~~ V_{ub}\sim \lambda^3~,~~~~~ V_{cb}\sim \lambda^2~.
\eeq 

Here $v_u\equiv\langle \varphi_5 \rangle$, 
$v_d\equiv\langle \varphi_{\bar 5} \rangle$  and $\lambda$ denotes the ratio between 
the vev of $X$ and an ultraviolet cut off identified with the Planck mass 
$M_{Pl}$: $\lambda\equiv\langle X \rangle/M_{Pl}$.
To correctly reproduce the observed quark mixing angles, we  take $\lambda$ of the 
order of the Cabibbo angle. The fermion spectrum depends on $\langle X \rangle$ and 
$M_{Pl}$ only through the ratio $\lambda$. In chapter \ref{real5} we will see how 
$\lambda$ can be dynamically generated by minimizing the $X$-dependent
effective potential of the theory.
For non-negative $F$-charges, the elements of  the 
quark mixing matrix $V_{CKM}$ depend only on the charge
differences of the left handed quark doublet \cite{fro}. 
Up to a constant shift, this defines the choice in Table \ref{afch}.
Equal $F$-charges for $\Psi_{\bar 5}^{2,3}$ are then required to fit
$m_b$ and $m_s$. We will comment on the lightest quark masses later on.

At this level, the mass matrix for the charged leptons  is the transpose of $m_D^d$:
\beq m_D^l=(m_D^d)^T
%=\left[
%\matrix{
%\lambda^6&\lambda^3&\lambda^3\cr
%\lambda^5&\lambda^2&\lambda^2\cr
%\lambda^3&1&1
%\right]
\eeq and we find:
\beq m_e:m_\mu:m_\tau  = \lambda^6:\lambda^2:1 
\eeq 
The $O(1)$ off-diagonal entry of $m_D^l$ gives rise to a large left handed mixing 
in the 23 block which corresponds to a large right handed mixing in the $d$ mass matrix. 
Thus, this situation corresponds to Case I of section \ref{lmseesaw}. 

\subsection{Neutrino Sector}

Two different situations arise allowing or not for the presence of 
the additional anti-flavon field $\bar X$, with $F=1$. 
Let us first consider what happens allowing for the pair $X$ and $\bar X$.

\subsubsection{Two Flavons}

In the neutrino sector, the Dirac and Majorana mass matrices are given by:
\beq m_D=
\left[
\matrix{
\lambda^4&\lambda^2&\lambda^3\cr
\lambda&\lambda'&1 \cr
\lambda&\lambda'&1}
\right]v_u~~,~~~~~~~~ M=
\left[
\matrix{
\lambda^2&1&\lambda\cr 1&\lambda'^2&\lambda'\cr
\lambda&\lambda'&1}
\right]{\bar M}~~,
\eeq where $\lambda'\equiv\langle\bar X\rangle/M_{Pl}$ and ${\bar M}$  denotes the large 
mass scale associated to the right handed neutrinos: ${\bar M}\gg v_{u,d}$.

After diagonalization of the charged lepton sector and after integrating out the heavy 
right handed neutrinos we obtain the following neutrino mass matrix in the low-energy 
effective theory:
\beq m_\nu=
\left[
\matrix{
\lambda^6&\lambda^3&\lambda^3\cr
\lambda^3&1&1\cr
\lambda^3&1&1}\right]{v_u^2\over {\bar M}}
\label{mnuaf}
\eeq 
where we have taken $\lambda\sim\lambda'$. The $O(1)$ elements in the 
23 block are produced by combining the large left handed mixing induced by the charged 
lepton sector and the large left handed mixing in $m_D$. 
A crucial property of $m_\nu$ is that, as a result of the see-saw mechanism and of the 
specific $U(1)_F$ charge assignment, the determinant of the 23 block is 
{\it automatically} of $O(\lambda^2)$.

It is easy to verify that the eigenvalues of $m_\nu$ satisfy the relations:
\beq m_1:m_2:m_3  = \lambda^4:\lambda^2:1~~.
\eeq 
The atmospheric neutrino oscillations require 
$m_3^2\sim 10^{-3}~{\rm eV}^2$. From eq. (\ref{mnuaf}), taking $v_u\sim 250~{\rm GeV}$, 
the mass scale ${\bar M}$ of the heavy Majorana neutrinos turns out to be close 
to the unification scale, ${\bar M}\sim 10^{15}~{\rm GeV}$. The squared mass 
difference between the lightest states is  of $O(\lambda^4)~m_3^2$,
appropriate to the MSW solution  to the solar neutrino problem. 
Finally, beyond the large mixing in the 23 sector,
$m_\nu$  provides a mixing angle $s_{12} \sim (\lambda/2)$ in the 12 sector, close to the 
range preferred by the SMA-MSW solution. 
In general $U_{e3}$ is non-vanishing, of $O(\lambda^3)$.

In general, the charge assignment under 
$U(1)_F$ allows for non-canonical kinetic terms that represent an additional source 
of mixing. Such terms are allowed by the underlying flavour symmetry and it would 
be unnatural to tune them to the canonical form. The results quoted up to now  remain
unchanged after including the effects related to the most general kinetic terms, 
via appropriate  rotations and rescaling in the flavour space (see also ref. \cite{lns}).

\subsubsection{Only One Flavon Field}
\label{ooff}
A general problem common to all models dealing with flavour is that of recovering the 
correct  vacuum structure by minimizing the effective potential of the theory. 
It may be noticed that the presence of two multiplets $X$ and
${\bar X}$ with opposite $F$ charges could hardly be reconciled, without adding extra 
structure to the model, with a large common vev for these fields, due to possible 
analytic terms of the kind $(X {\bar X})^n$ in the superpotential
\cite{irges}.
%\footnote{We thank N. Irges for bringing our attention on this point.}.  
It is then interesting to explore the consequences of allowing only the negatively 
charged $X$ field in the theory.

It can be immediately recognized that, while the quark mass matrices of eq. (\ref{mdirquark}) 
are unchanged, in the neutrino sector the Dirac and Majorana matrices get modified into:
\beq m_D=\left[\matrix{
\lambda^4&\lambda^2&\lambda^3\cr
\lambda&0&1\cr
\lambda&0&1}
\right]v_u~~,~~~~~~~~ 
M=\left[\matrix{
\lambda^2&1&\lambda\cr 1&0&0\cr
\lambda&0&1}\right]{\bar M}~~.
\eeq 
The zeros are due to the analytic property of the superpotential that makes impossible 
to form the corresponding $F$ invariant by using $X$ alone. 
These zeros should not be taken literally, as they will be eventually filled by
small terms coming, for instance, from the diagonalization of the charged 
lepton mass matrix and from the transformation that put the kinetic terms into 
canonical form. It is however interesting to work out, in first approximation, the case  
of exactly zero entries in $m_D$ and $M$, when forbidden by $F$.

The neutrino mass matrix obtained via see-saw from $m_D$ and $M$ has the same 
pattern as the one displayed in eq. (\ref{mnuaf}).
A closer inspection reveals that the determinant of the 23 block is identically zero, 
independently from $\lambda$. This leads to the following pattern of masses:
\beq m_1:m_2:m_3  = \lambda^3:\lambda^3:1~~,~~~~~m_1^2-m_2^2 = {\rm O}(\lambda^9)~m_3^2~.
\eeq 
Moreover the mixing in the 12 sector is almost maximal:
\beq {s\over c}={\pi\over 4}+{\rm O}(\lambda^3)~~.
\eeq 
For $\lambda\sim 0.2$, both the squared mass difference $(m_1^2-m_2^2)/m_3^2$  
and $\sin^2 2\theta_{sol}$ are remarkably close to the values  required by the VO 
solution to the solar neutrino problem. This property  remains reasonably stable 
against the perturbations induced by small terms (of order $\lambda^5$) replacing the 
zeros, coming from the diagonalization of the charged lepton sector  
and by the transformations that render the kinetic terms canonical. 

We find quite interesting that also the VO solution, requiring  an intriguingly 
small mass difference and a bimaximal mixing, can be reproduced, at least at the 
level of order of magnitudes, in the context of a "minimal" model of flavour compatible 
with SUSY $SU(5)$. In this case the role played by supersymmetry  is essential, 
a non-supersymmetric model with $X$ alone not being distinguishable 
from the version with both $\bar X$ and $X$, as far as low-energy flavour 
properties are concerned.

In chapter \ref{real5} we will again exploit this mechanism based on the presence of
only one flavon $X$, but with slightly different charge
assignments for the matter field $\Psi_{\bar 5}$, so that the LOW solution to the solar
neutrino problem will naturally emerge.

\subsection{Outlook}
\label{outlook}

We have constructed two examples which realize Case I of section \ref{lmseesaw}.
Obviously, the order of magnitude description offered by this model is not intended 
to account for all the details of fermion masses. Even neglecting the parameters 
associated with the $CP$ violating observables, some of the relevant observables are
somewhat marginally reproduced.   
For instance we obtain $m_u/m_t\sim \lambda^6$  which is perhaps too large. 
However we find it remarkable that in such a simple scheme most of the 12 independent  
fermion masses and the 6 mixing angles turn out to have the correct order of magnitude. 
Notice also that our model prefers large values of
$\tan\beta\equiv v_u/v_d$. This is a consequence of the equality 
$F(\Psi_{10}^3)=F(\Psi_{\bar 5}^3)$. 
In this case the Yukawa couplings of top and bottom quarks  are expected to 
be of the same order of magnitude, while the large
$m_t/m_b$ ratio is attributed to $v_u \gg  v_d$ (there may be factors $O(1)$ modifying 
these considerations, of course). 
%We recall here that in supersymmetric grand unified models large values 
%of $\tan\beta$  are one possible solution to the problem of reconciling the boundary  
%condition $m_b=m_\tau$ at the  GUT scale with the low-energy data \cite{largetb}.  
To keep $\tan\beta$ small, one could  suppress $m_b/m_t$ by adopting 
different $F$-charges for the  
$\Psi_{\bar 5}^3$ and $\Psi_{10}^3$. Alternatively,
as will be explicitly shown in chapter \ref{real5}, one can 
assign F-charges also to the Higgs pentaplets.

Additional contributions to flavour changing processes (both in the quark and in 
the lepton \cite{vis2} sectors) and to $CP$ violating observables are generally expected 
in a SUSY GUT. However, a reliable estimate of the corresponding effects would 
require a much more detailed definition of the theory than attempted here. 
Crucial ingredients such as the mechanism of supersymmetry breaking and its 
transmission to the observable sector have been ignored.
We are implicitly assuming that the omission of this aspect of the flavour problem 
does not substantially alter our discussion. 
For instance, it has been pointed out \cite{buch} that in 
$SU(5)\otimes U(1)_F$ based models with large $\tan \beta$ and 
radiatively induced flavour mixing, one generally 
expects the branching ratio of the flavour changing process $\mu \rightarrow
e \gamma$ to be comparable to the present experimental upper bound
(the exact values depending on the particular model considered). 
On the other hand, for $\tan \beta =O(1)$, the branching ratio turns out to be smaller
by about four orders of magnitude. 

A common problem of all $SU(5)$ unified theories based on a minimal Higgs structure 
is represented by the relation $m_D^l=(m_D^d)^T$ that, while leading to the successful 
$m_b=m_\tau$ boundary condition at the GUT scale, provides the wrong prediction
$m_d/m_s=m_e/m_\mu$ (which, however, is an acceptable order of magnitude equality). 
In section \ref{secfmm}, we will describe in some detail this problematic feature
of minimal GUT models and briefly review the proposed solutions. 
An interesting possibility to overcome this problem and improve the picture of fermion 
masses will be adopted in the model of chapter \ref{real5}, where we will
exploit the presence of an additional supermultiplet transforming in the $75$ representation
of $SU(5)$. 
%We can easily overcome this problem and improve the picture \cite{eg} by introducing 
%an additional supermultiplet ${\bar\theta}_{24}$ transforming in the adjoint representation 
%of $SU(5)$ and  possessing a negative $U(1)_F$ charge,
%$-n~~(n>0)$.  Under these conditions, a positive
%$F$-charge $f$ carried by the matrix elements 
%$\Psi_{10}^a \Psi_{\bar 5}^b$ can be compensated  in several different ways by monomials of the kind
%$({\bar\theta})^p({\bar\theta}_{24})^q$, with
%$p+n q=f$. Each of these possibilities represents an independent contribution to 
%the down quark and charged lepton mass
%matrices, occurring with an unknown coefficient of O(1). Moreover the product 
%$({\bar\theta}_{24})^q \varphi_{\bar 5}$ contains both the ${\bar 5}$ and the 
%$\overline{45}$ $SU(5)$ representations,  allowing for a differentiation between the down 
%quarks and the charged leptons. The only, welcome, exceptions are  given by the O(1) 
%entries that do not require any compensation and, at the leading order, remain the same 
%for charged leptons and  down quarks. This preserves the good $m_b=m_\tau$ prediction. 
%Since a perturbation of O(1) in the subleading matrix elements  is sufficient to cure the bad
%$m_d/m_s=m_e/m_\mu$ relation, we can safely assume that  
%$\langle{\bar\theta}_{24}\rangle/M_{Pl}\sim\lambda^n$, to preserve the correct 
%order-of-magnitude predictions in the remaining sectors. 

\section{A Model with Large Mixing from Nothing}

Without loss of generality we can go to a basis where both the charged lepton 
Dirac mass matrix $m_D^l$ and the Majorana matrix $M$ for the right handed 
neutrinos are diagonal. 
In fact, after diagonalization of the charged lepton Dirac mass
matrix, we still have the freedom of a change of basis for the right handed neutrino fields, 
in that the right handed charged lepton and neutrino fields, as opposed to left handed 
fields, are uncorrelated by the $SU(2) \times U(1)$ gauge
symmetry. We can use this freedom to make the Majorana matrix diagonal: 
$M^{-1}=V^Td_MV$ with $d_M={\rm Diag}[1/M_1,1/M_2,1/M_3]$. 

In order to understand how to generalize to the 3 by 3 context the mechanism 
already described in Case II of section \ref{lmseesaw} (see eq. (\ref{mdMafm1})),
it is useful to start by considering symmetric matrices.  
These matrices are very interesting because, for example, one could want to
preserve left-right symmetry at the GUT scale. 
Then, the observed smallness of left handed mixings for quarks would
also demand small right handed mixings.
 
One simple class of examples is the following one \cite{afm1}. We start from
\beq
 m_D=v 
\left[\matrix{\epsilon''&\epsilon'& y~\epsilon'\cr
\epsilon'&\epsilon &x~\epsilon\cr y~\epsilon' &x~\epsilon &1      } 
\right],~~~~~ M^{-1}=\frac{1}{\Lambda} 
\left[\matrix{ r_1&0&0\cr 0& r_2&0\cr 0&0& r_3    } 
\right]~~~~~,
\label{mdM3}
\eeq
where, unless otherwise stated, $x$ and  $y$ are $O(1)$; 
$\epsilon$, $\epsilon'$ and $\epsilon''$ are independent small numbers and 
$r_i\equiv M_3/M_i$. Note that the 23 blocks of these matrices 
closely match eq. (\ref{mdMafm1}).  
We expect $\epsilon''\ll\epsilon'\ll\epsilon\ll 1$ and, perhaps, also $r_1\gg r_2\gg r_3=1$, 
if the hierarchy for right handed neutrinos follows the same pattern as for 
known fermions. 
Depending on the relative size of the ratios $r_i/r_j$, $\epsilon/\epsilon'$ and 
$\epsilon'/\epsilon''$, we can have models with dominance of any of the $r_{1,2}$. 

For example, we set $x=1$ (keeping $y$ of O(1)) and assume 
$r_2\epsilon^2\gg r_1\epsilon'^2,r_3$, together with  $r_2 \epsilon'^2\gg r_1 \epsilon''^2$ and 
$r_2\epsilon\gg r_1 \epsilon''$. In this limit, with good accuracy we obtain:
\beq 
m_{\nu}=
\frac{v^2}{\Lambda}r_2\epsilon^2\left[\matrix{\dd{\epsilon'^2\over \epsilon^2}&\dd{\epsilon'\over\epsilon}&
\dd{\epsilon'\over\epsilon}\cr
\dd{\epsilon'\over\epsilon}&1+\dd\frac{r_1\epsilon'^2}{r_2\epsilon^2}&1 \cr
\dd{\epsilon'\over\epsilon}&1&1+\dd\frac{r_3}{r_2\epsilon^2}    } 
\right]~~~~~. 
\label{mnFI}
\eeq 
Note that, for $\epsilon\epsilon''\le\epsilon'^2$:
\beq
{\rm Det}[m_{\nu}]=\left(\frac{v^2}{\Lambda}\right)^3 r_1 r_2 r_3 \epsilon'^4~~~~~.
\label{det}
\eeq
Approximate eigenvalues, in units of $v^2/\Lambda$ and with $m_3>m_2>m_1$, are given by
\beq
m_3\sim 2 r_2\epsilon^2,~~~~~~m_2\sim \frac{1}{2}r_1\epsilon'^2,~~~~~~m_1\sim 
\frac{r_3\epsilon'^2}{\epsilon^2}\label{la1}
\eeq
for $r_1\epsilon'^2>r_3$ or, alternatively,
\beq
m_3\sim 2 r_2\epsilon^2,~~~~~~m_2\sim \frac{1}{2}r_3,~~~~~~m_1\sim 
\frac{r_1\epsilon'^4}{\epsilon^2}\label{la2}
\eeq
for $r_1\epsilon'^2<r_3$. Having set $x=1$ the atmospheric neutrino mixing is nearly maximal. 
The solar neutrino mixing is instead generically small in these models, being proportional to 
$\epsilon'/\epsilon$. Thus the SMA-MSW solution is obtained. It is easy to find
set of parameter values that lead to an acceptable phenomenology within these solutions. 
As an illustrative example we take:
\beq
\epsilon\sim \lambda^4~~,~~~~~\epsilon'\sim \lambda^6~~,~~~~~\epsilon''\sim\lambda^{12}~~,
\eeq
\beq
r_1\sim\lambda^{-12}~~,~~~~~r_2\sim\lambda^{-9}~~,
\eeq
where $\lambda\sim\sin\theta_C$, $\theta_C$ being the Cabibbo angle. 
The neutrino mass matrices become
\beq
m_D=v
\left[\matrix{\lambda^{12}&\lambda^6&\lambda^6\cr
\lambda^6&\lambda^4&\lambda^4\cr \lambda^6&\lambda^4&1      } 
\right]~~,~~~~~~M=\Lambda
\left[\matrix{\lambda^{12}&0&0\cr
0&\lambda^9&0\cr 0&0&1      } 
\right]
\label{ex1}
\eeq
and, in units of $v^2/\Lambda$, we obtain
\beq
m_3\sim 1/\lambda~~,~~~~~m_2\sim 1~~,~~~~~m_1\sim\lambda^4~~.
\eeq
The solar mixing angle $\theta_{12}$ is
of order $\lambda^2$, suitable to the SMA-MSW solution. Also $\theta_{13}\sim\lambda^2$.

It is also possible to arrange the
parameters in eq. (\ref{mdM3}) in such a way that $r_1$ is dominant. 
Phenomenologically viable models can also be constructed in this case, 
always of the SMA-MSW type. 
For instance, we can take $y=1$ (keeping $x$ of $O(1)$) in order to obtain a 
large atmospheric mixing angle, and further assume the dominance of the $r_1$ terms, 
namely:
$\epsilon'^2 r_1\gg \epsilon^2 r_2,~r_3$, with $\epsilon''^2 r_1\gg \epsilon'^2 r_2$ and 
$\epsilon'' r_1\gg \epsilon r_2,~r_3$. 
In this case we can approximate the light neutrino mass matrix as follows:
\beq 
m_{\nu}=
\frac{v^2}{\Lambda}r_1\epsilon'^2
\left[\matrix{\dd{\epsilon''^2\over\epsilon'^2}
&\dd{\epsilon''\over\epsilon'}&\dd{\epsilon''\over\epsilon'}\cr
\dd{\epsilon''\over\epsilon'}&1+\dd\frac{r_2\epsilon^2}{r_1\epsilon'^2}&1 \cr
\dd{\epsilon''\over\epsilon'}&1&1+\dd\frac{r_3}{r_1\epsilon'^2}    } 
\right]~~~~~. 
\label{mnFII}
\eeq 
As an example we consider the following choice of parameters:
\beq
\epsilon\sim \lambda^4~~,~~~~~\epsilon'\sim \lambda^6~~,~~~~~\epsilon''\sim\lambda^8~~,
\eeq
\beq
r_1\sim\lambda^{-14}~~,~~~~~r_2\sim\lambda^{-6}~~.
\eeq 
In this case we have
\beq
 m_D=v
\left[\matrix{\lambda^8&\lambda^6&\lambda^6\cr
\lambda^6&\lambda^4&\lambda^4\cr \lambda^6&\lambda^4&1      } 
\right]~~,~~~~~~M=\Lambda
\left[\matrix{\lambda^{14}&0&0\cr
0&\lambda^6&0\cr 0&0&1      } 
\right]~~~~~.
\label{ex2}
\eeq
In units of $v^2/\Lambda$, we find
\beq
m_3\sim 1/\lambda^2~~,~~~~~m_2\sim 1\gg m_1
\eeq
and $\theta_{12}\sim\lambda^2$.

As anticipated in section \ref{lmseesaw}, an interesting feature of these textures, 
in connection with a possible realization 
within a GUT scheme, is that $m_3$ is much larger than $v^2/\Lambda$ by construction. 
This means that the lepton number breaking scale $\Lambda$ can be pushed up to the 
order of the GUT scale or even beyond. 
A large value of $\Lambda$, $\Lambda\geq M_{GUT}$, is naturally expected in $SU(5)$ 
if right handed neutrinos are present and in $SO(10)$ if the grand unified group is broken 
down to $SU(5)$ close to or slightly below the Planck mass. 
We recall that if $m_3$ is approximately given by $v^2/\Lambda$ and $v\sim 250~{\rm GeV}$,
then $\Lambda\sim 10^{15}~{\rm GeV}$.

Models based on symmetric matrices are directly compatible with left-right 
symmetry and therefore are naturally linked with $SO(10)$. This is to be confronted with
models - like those described in the previous section -
that have large right handed mixings for quarks, which, in $SU(5)$, 
can be naturally translated into large left handed mixings for leptons. 
In this connection it is interesting to observe that the proposed textures for the neutrino 
Dirac matrix can also work for up and down quarks. For example, the matrices
\beq
 m_D^u\propto
\left[\matrix{0&\lambda^6&\lambda^6\cr
\lambda^6&\lambda^4&\lambda^4\cr \lambda^6&\lambda^4&1      } 
\right]~~,~~~~~~ m_D^d\propto
\left[\matrix{0&\lambda^3&\lambda^3\cr
\lambda^3&\lambda^2&\lambda^2\cr \lambda^3&\lambda^2&1      } 
\right]~~,
\label{mud}
\eeq
where for each entry the order of magnitude is specified in terms of 
$\lambda\sim \sin{\theta_C}$, lead to acceptable mass matrices and mixings. 
In fact $m_u:m_c:m_t=\lambda^8:\lambda^4:1$ and $m_d:m_s:m_b=\lambda^4:\lambda^2:1$. 
The $V_{CKM}$ matrix receives a dominant contribution from the down sector in that 
the up sector angles are much smaller than the down sector ones. 
%For this reason we also take $(m_d)_{13}\sim \lambda^3$ which in the limit of neglecting 
%the up sector angles becomes $V_{ub}$.
The same kind of texture can also be adopted in the charged lepton sector
\footnote{Notice that the possibility of treating on equal footing all fermion
mass matrices is also present in the framework of an extended flavour democracy.
See for instance Ref. \cite{abjs} and references therein.}.

It is simple to realize that models of this sort cannot be derived 
from a broken $U(1)_{F}$ horizontal symmetry, with a single field that breaks 
spontaneously $U(1)_{F}$. 
For symmetric matrices, left and right $U(1)$ charges must be equal: then, for 
example, if $m_{33}\sim l^{2q_3} \sim 1$ or $q_3=0$, then $m_{22}\sim l^{2q_2}$ is 
different from $m_{23}\sim l^{q_2}$ etc. 

However, the symmetry requirement is not really necessary to produce a large 
atmospheric mixing angle from neutrino mass matrices characterized by small mixings. 
Indeed, by considering for simplicity the 2 by 2 case, instead of eq. (\ref{mdMafm1}),
we could equally well have 
started from a Dirac mass matrix of the kind:
\beq
 m_D\propto 
\left[\matrix{\epsilon& x'\cr x\epsilon &1    } 
\right]~~,~~~~~ M^{-1}\propto 
\left[\matrix{r_2&0\cr 0&1    } 
\right]~~~~~,
\label{mdMM}
\eeq
with $x'$ any number smaller than or equal to one. 
Once the condition $\epsilon^2 r_2\gg1$ is satisfied, the neutrino mass 
matrix of eq. (\ref{mnp1}) is obtained again, independently from $x'$. 
The generalization of this mechanism to the full 3 by 3 case is straightforward, 
and analogous to the one discussed above for symmetric matrices.
This leaves much more freedom for model building and does not necessarily 
bound the realization of these schemes to left-right-symmetric scenarios.

The solar mixing angle is generically small in the class of models explicitly 
discussed above. However, small mixing angles in the Dirac and Majorana 
neutrino mass matrices do not exclude a large solar mixing angle.
For instance, this is generated from the asymmetric, but nearly
diagonal mass matrices:
\beq
m_D=v
\left[\matrix{\lambda^6&0&0\cr
\lambda^6&\lambda^4&0\cr 0&\lambda^4&1      } 
\right]~~,~~~~~~M=\Lambda
\left[\matrix{\lambda^{12}&0&0\cr
0&\lambda^{10}&0\cr 0&0&1      } 
\right]~~~~~.
\label{ex22}
\eeq
They give rise to a light neutrino mass matrix of the kind:
\beq 
m_\nu=
\left[
\matrix{
\lambda^2&\lambda^2&0\cr
\lambda^2&1&1\cr
0&1&1}\right]{v^2\over \lambda^2\Lambda}~~~~~,
\label{mnu3}
\eeq 
which is diagonalized by large $\theta_{12}$ and $\theta_{23}$ and small $\theta_{13}$.
The mass hierarchy is suitable to the LMA-MSW solution. 

\subsection{Renormalization Effects}

These results are stable under renormalization from the high-energy 
scale $\Lambda$ where the mass matrices are produced down to
the electroweak scale. Indeed, suppose that we start with a 
supersymmetric theory whose superpotential $w$ at the scale 
$\Lambda\ge M_{GUT}$ is given by:
\beq
w=Q y^u  U^c H_u+  Q y^d  D^c H_d+ L  y^\nu N^c  H_u+ L y^e E^c H_d
  +{1\over 2} N^c M N^c~~~,
\eeq
where 
\beq
y^e(\Lambda)=    
\left[
\matrix{0& 0\cr 0& y_\tau}
\right]~~,~~~~
y^\nu(\Lambda)=    
\left[
\matrix{\epsilon& x~\epsilon\cr x~\epsilon& 1}
\right]~~,~~~~
M(\Lambda)=    
\left[
\matrix{M_2& 0\cr 0& M_3}
\right]~~.
\eeq
Here we focus on the two heaviest generations. The matrices 
$y^{u,d,e,\nu}$ describe the Yukawa couplings of the theory,
related to the mass matrices through $m_{u,\nu}=y^{u,\nu} v_u$ 
and $m_{d,e}=y^{d,e} v_d$, where $v_{u,d}$ are the
vacuum expectation values of the neutral scalar components
in the superdoublets $H_{u,d}$. The mass scales
are ordered according to $\Lambda\ge M_3\gg M_2\gg M_W$,
$M_W$ denoting the electroweak scale. 
At the scale $M_2$, after integrating out the superfields
$N^c$ associated to the heavy right handed neutrinos, the 
mass matrix for the light neutrinos, including the 
renormalization effects, is given by:
\beq
m_\nu(M_2)=
{v_u^2 \over M_2}
\left[
\matrix{({y^\nu}_{22})^2& y^\nu_{22}~ y^\nu_{32}\cr
y^\nu_{22}~ y^\nu_{32}& ({y^\nu}_{32})^2} 
\right]
+ 
O( \frac{1}{ M_3})~~~.
\eeq
The couplings ${y^\nu}_{22}$ and ${y^\nu}_{32}$ are evaluated
at the scale $M_2$. If the running of $M$ is computed by
by taking $\epsilon=0$, then $M_2$ stays approximately constant when 
going from $\Lambda$ down to $M_2$. Notice that, if we neglect the terms
of $O(1/M_3)$, the determinant of $m_\nu(M_2)$ vanishes,
even including the renormalization effects. These effects 
modify the mixing angle at the scale $M_2$ and they depend on how
the Yukawa coupling ${y^\nu}_{22}$ and ${y^\nu}_{32}$ change
from $\Lambda$ to $M_2$. The running of the neutrino Yukawa couplings
is governed by the equation:
\beq
{d y^\nu\over dt}={1\over 16\pi^2} 
\left[
{\rm tr}(3 y^u {y^u}^\dagger + y^\nu {y^\nu}^\dagger )
-4\pi(3 \alpha_2+{3\over 5}\alpha_1)
+3 y^\nu {y^\nu}^\dagger + y^e
 {y^e}^\dagger \right]~{y^\nu} ~~,
\eeq
from $\mu=\Lambda$ to $\mu=M_3$, with $t={\rm log}\mu$.
At $M_3$, the superfield $N^c_3$ is integrated out.
This produces terms of order $1/M_3$, which are neglected
in the present discussion. From $\mu=M_3$ down to
$\mu=M_2$ the running of ${y^\nu}_{22}$ and ${y^\nu}_{32}$
is described by an equation formally identical to
the previous one, with 
\beq
y^\nu\equiv
\left[
\matrix{y^\nu_{22}& y^\nu_{23}\cr y^\nu_{32}& y^\nu_{33}}
\right]
\to
y^\nu\equiv
\left[\matrix{y^\nu_{22} \cr y^\nu_{32}} \right]~~.
\eeq
To assess the size of the renormalization effects on 
${y^\nu}_{22}$ and ${y^\nu}_{32}$, we have integrated numerically
the above equations in the approximation of constant $y^{u,e}$
and $\alpha_{1,2}$. This should provide a rough estimate of the effects.
As an example we take $\Lambda=10^{18}$ GeV,
$M_3=10^{16}$ GeV and $M_2=10^8$ GeV, $\alpha_2=(5/3) \alpha_1=1/24$; 
we choose $y_\tau=0.6$, corresponding
to the large $\tan\beta\equiv v_u/v_d$ regime, which enhances
the renormalization effects of $y^\nu$.
Assuming as initial conditions 
${y^\nu}_{22}(\Lambda)={y^\nu}_{23}(\Lambda)={y^\nu}_{32}(\Lambda)=0.05$ and 
${y^\nu}_{33}(\Lambda)=1$,
we find that ${y^\nu}_{22}$ and ${y^\nu}_{32}$
are modified by about 10\% at the
scale $M_2$. 
The mixing angle, nearly maximal at the scale $\Lambda$, 
remains close to maximal:
$\sin^2 2\theta_{23}\sim 0.98$ in the example at hand.

The running from $M_2$ down to the electroweak scale does
not appreciably affect neither the vanishing determinant
condition, nor a nearly maximal neutrino mixing angle.
This can be seen by analyzing the renormalization
group equation for $m_\nu$, which, in the one-loop approximation,
is given by:
\beq
{d m_\nu\over d t}={1\over 8\pi^2}
\left\{
\left[
{\rm tr}(3 {y^u}^\dagger y^u)
-4\pi(3 \alpha_2+{3\over 5}\alpha_1)
\right]m_\nu
+{1\over 2}
\left[
m_\nu {y^e}^\dagger y^e +({y^e}^\dagger y^e)^T m_\nu
\right]
\right\}~~~.
\eeq
The solution is:
\beq
m_\nu(M_W)
=
C
\left[
\matrix{{m_\nu}_{22}(M_2)&  {m_\nu}_{32}(M_2)~B\cr
{m_\nu}_{32}(M_2)~B& {m_\nu}_{33}(M_2)~B^2}
\right]~~,
\label{sol1}
\eeq
where
\beq
C={\rm exp}\left[{\dd{\int_{t_0}^t c(t') dt'}}\right]~~,~~~~
B={\rm exp}\left[{\dd{\int_{t_0}^t b(t') dt'}}\right]~~,~~~~~~~
t_0={\rm log}(M_W)~~,~~~~t={\rm log}(M_2)~~
\eeq
and
\beq
c(t)={1\over 8\pi^2}
\left[
{\rm tr}(3 {y^u}^\dagger y^u)-4\pi(3 \alpha_2+{3\over 5}\alpha_1)
\right](t)~~,
\eeq
\beq
b(t)={y_\tau^2(t)\over 16\pi^2}~~.
\eeq
We see from eq. (\ref{sol1}) that if the determinant of $m_\nu$
vanishes at the scale $M_2$, then it vanishes also at the weak
scale. The mixing angle at the weak scale is given by:
\beq
\sin^2(2 \theta)(M_W)=\frac{4 {m_\nu}_{32}^2(M_2) B^2}
{4 {m_\nu}_{32}^2(M_2) B^2+
({m_\nu}_{33}(M_2) B^2-{m_\nu}_{22}(M_2))^2}~~~.
\eeq
The condition of maximal mixing at the scale $M_2$ is
${m_\nu}_{33}(M_2)={m_\nu}_{22}(M_2)$. It is easy to see
that, if this condition is met, then the first correction to
$\sin^2(2 \theta)(M_W)$ is of second order in the parameter
$(B-1)$. Numerically, we find that $\sin^2(2 \theta)(M_W)$ is 
reduced only by a of few percent.

Summarizing, the renormalization effects do not appreciably
modify the results obtained at the tree level for eigenvalues
and mixing angle of the neutrino mass matrix. They
manifest themselves as $O(1)$ coefficients that, for all
practical purposes, can be
absorbed in the definition of the classical parameters.
An analogous analysis can be carried out in the 3 by 3 case,
with similar results. This is in agreement with the analysis
of ref. \cite{ello}, as far as the case of hierarchical neutrino
masses is concerned.

\subsection{An Explicit GUT Example with Broken Flavour Symmetry}

From our general discussion it is clear that some amount of correlation between the 
Dirac and the Majorana sectors is required in order to reproduce, at the 
order-of-magnitude level, the observed pattern of oscillations 
(assuming from now on the SMA-MSW solution for solar neutrinos). 
In order to show that this correlation could be induced by an underlying 
flavour symmetry and, at the same time, to present an explicit grand unified example 
of the class of textures considered here, we \cite{afm1} sketch the features of an 
$SU(5)$ model
with a spontaneously broken flavour symmetry based on the gauge group 
$U(1)_{A}\times U(1)_{B}$.
 
The three generations are described by ${ \Psi_{10}}^a$ and ${ \Psi_{\bar 5}}^a$, 
$(a=1,2,3)$ transforming as $10$ and ${\bar 5}$ of $SU(5)$, respectively. 
Three more $SU(5)$ singlets ${ \Psi_1}^a$ describe the right handed neutrinos. 
Here we focus on the Yukawa coupling allowed by Higgs multiplets 
$\varphi_5$ and $\varphi_{\bar5}$ in the $5$ and ${\bar 5}$ $SU(5)$ representations 
and by three flavons, $\varphi_A^-$, $\varphi_B^-$ and $\varphi_B^+$, taken, 
to begin with, as $SU(5)$ singlets. The charge assignment of the various fields is 
summarized in Table \ref{afm1ch}. 
We assume that the flavon fields develop the following set of vevs:
\beq
\langle\varphi_A^-\rangle\sim\lambda~\Lambda~~,~~~~~
\langle\varphi_B^-\rangle\sim\lambda~\Lambda~~,~~~~~
\langle\varphi_B^+\rangle\sim\Lambda~~,
\eeq
where $\Lambda$ denotes the cutoff of the theory and provides the mass scale 
suppressing the higher-dimensional operators invariant under the gauge and the 
flavour groups.
\begin{table}[h]
\begin{center}
\begin{tabular}{|c|c|c|c|c|c|c|c|c|}   
\hline        & & & & & & & &\\                
 & ${\Psi_{10}}$ & ${\Psi_{\bar 5}}$ & ${\Psi_1}$ & $\varphi_5$ & $\varphi_{\bar 5}$ 
& $\varphi_A^-$ & 
$\varphi_B^-$ & $\varphi_B^+$\\& & & & &&&&\\
\hline
& & & & & & & & \\
$SU(5)$ & 10 & ${\bar 5}$ & 1 & 5 & ${\bar 5}$ & 1 & 1 & 1\\ 
& & & & & & & & \\
\hline
& & & & & & & & \\
${\rm U(1)}_{\rm A}$ & (3,2,0) & (3,1,0) & (6,3,0) & 0 & 0 & $-$1 & 0 & 0\\
& & & & & & & & \\
\hline                        
& & & & & & & & \\
${\rm U(1)}_{\rm B}$ & (1,0,0) & ($-$2,$-$2,$-$1) & (0,2,0) & 0 & 0 & 0 & $-$1 & +1\\
& & & & & & & & \\
\hline
\end{tabular}                             
\end{center}
\caption{Quantum numbers of matter and flavon multiplets.}
\label{afm1ch}
\end{table} 

In the quark sector we obtain 
\footnote{In eq. (\ref{mquark}) the entries denoted by 1 in $m_D^u$ and $m_D^d$ 
are not necessarily equal. As usual, such a notation allows 
for $O(1)$ deviations.}
\beq m_D^u=(m_D^u)^T=
\left[
\matrix{
\lambda^8&\lambda^6&\lambda^4\cr
\lambda^6&\lambda^4&\lambda^2\cr
\lambda^4&\lambda^2&1}
\right]v_u~~,~~~~~~~ m_D^d=
\left[
\matrix{
\lambda^6&\lambda^4&\lambda^3\cr
\lambda^5&\lambda^3&\lambda^2\cr
\lambda^3&\lambda&1}
\right]v_d~~,
\label{mquark}
\eeq 
from which we get the order-of-magnitude relations:
\bea m_u:m_c:m_t & = &\lambda^8:\lambda^4:1 \nonumber\\ m_d:m_s:m_b & = &\lambda^6:\lambda^3:1
\eea and 
\beq V_{us}\sim \lambda~,~~~~~ V_{ub}\sim \lambda^3~,~~~~~ V_{cb}\sim \lambda^2~.
\eeq 
Here $v_u\equiv\langle \varphi_5 \rangle$, $v_d\equiv\langle \varphi_{\bar 5} \rangle$.
Notice that $V_{ub}$ and $V_{us}$ are dominated by the down-quark contribution.
The mass matrix for the charged leptons  is the transpose of $m_D^d$:
\beq m_D^l=(m_D^d)^T
%=\left[
%\matrix{
%\lambda^6&\lambda^3&\lambda^3\cr
%\lambda^5&\lambda^2&\lambda^2\cr
%\lambda^3&1&1
%\right]
\eeq and we find
\beq m_e:m_\mu:m_\tau  = \lambda^6:\lambda^3:1~~~~. 
\eeq 
At this level we obtain the well-known prediction $m_b=m_\tau$, together with the 
unsatisfactory relation $m_d/m_s=m_e/m_\mu$ 
(which, however, is an acceptable order of magnitude equality).

In the neutrino sector, the Dirac and Majorana mass matrices are
given by:
\beq m_D=
\left[
\matrix{
\lambda^9&\lambda^6&\lambda^3\cr
\lambda^7&\lambda^4&\lambda\cr
\lambda^6&\lambda^4&1}
\right]v_u~~,~~~~~~~~ M=
\left[
\matrix{
\lambda^{12}&\lambda^{11}&\lambda^6\cr \lambda^{11}&\lambda^{10}&\lambda^5\cr
\lambda^6&\lambda^5&1}
\right]\Lambda~~.
\eeq
Notice that while in the model under consideration $m_D$ is asymmetric, 
it is diagonalized by small unitarity transformations, i.e. transformations that go 
to the identity in the limit of vanishing $\lambda$, in both the left and the right sectors.
After diagonalization of the charged-lepton sector and after integrating out the heavy 
right handed neutrinos, we obtain the following neutrino mass 
matrix in the low-energy effective theory:
\beq 
m_\nu=
\left[
\matrix{
\lambda^4&\lambda^2&\lambda^2\cr
\lambda^2&1&1\cr
\lambda^2&1&1}\right]{v_u^2\over \lambda^2\Lambda}~~~~.
\label{mnu}
\eeq

The $O(1)$ elements in the 23 block are produced only by the interplay of the 
left handed mixing in $m_D$ and the hierarchy in the Majorana sector. 
The contribution from the charged lepton sector is subleading.
An important property of $m_\nu$ is that, as a result of the see-saw mechanism 
and of the specific U(1)$_{\rm F}$ charge assignment, the determinant of the 23 block 
is of $O(\lambda^2)$ (in units of $(v_u^2/\lambda^2\Lambda)^2$).
This gives rise to the desired hierarchy between atmospheric and solar frequencies.
It is easy to verify that the eigenvalues of $m_\nu$ satisfy the relations:
\beq 
m_1:m_2:m_3  = \lambda^8:\lambda^2:1~~.
\eeq 
The atmospheric neutrino oscillations require 
$m_3^2\sim 3.5\times 10^{-3}~{\rm eV}^2$. 
From eq. (\ref{mnu}), taking $v_u\sim 250~{\rm GeV}$, the mass scale ${\Lambda}$ of the
heavy Majorana neutrinos turns out to be close to $2\cdot 10^{16}~{\rm GeV}$.
The squared mass difference between the lightest states is  of $O(\lambda^4)~m_3^2$,
appropriate to the MSW solution of the solar neutrino problem. 
Finally, beyond the large mixing in the 23 sector, $m_\nu$  provides a mixing 
angle $\theta_{12} \sim \lambda^2$ in the 12 sector, close to the range preferred 
by the SMA-MSW solution. 
In general $U_{e3}$ is non-vanishing, of $O(\lambda^2)$. 
We find it encouraging that the general
pattern of all that we know on fermion masses and mixings is correctly 
reproduced at the level of orders of magnitude. 

This model is to be taken merely as an illustration, among many other possibilities, 
of the scenario outlined: small mixing angles for the fundamental fermions 
giving rise to a large mixing angle for the light neutrinos. 
For this reason we do not attempt to address some important problems,
such as the cancellation of chiral anomalies, which we implicitly postpone 
to an energy scale higher than the unification scale. 
We have not dealt here with the problem of recovering the correct 
vacuum structure by minimizing the effective potential of the theory.
Also, the order of magnitude description offered by this model is not intended 
to account for all the details of fermion masses. Even neglecting the parameters 
associated with the CP violating observables, some of the relevant
observables are somewhat marginally reproduced. For instance,
as already emphasized,
a common problem of all $SU(5)$ unified theories, based on a minimal Higgs structure, 
is represented by the relation $m_D^l=(m_D^d)^T$, which, while leading to the 
successful $m_b=m_\tau$ boundary condition at the GUT scale, provides the
wrong prediction $m_d/m_s=m_e/m_\mu$. We might overcome this problem
and improve the picture \cite{elg} by assuming that the flavon fields 
$\varphi_A^-$, $\varphi_B^{\pm}$ or a suitable subset of them transform in 
the adjoint representation of $SU(5)$.
%introducing additional supermultiplets 
%transforming in the adjoint representation of SU(5), possessing non trivial $U(1)_A\times U(1)_B$ charges
%and developing VEVs smaller than $\Lambda$.
%This would offer more freedom in compensating the non-vanishing $A$ and $B$ charges of the fermion bilinears.
%At the same time the product 
The product $(24)^q {\bar 5}$ (where $q$ is a positive integer)
contains both the ${\bar 5}$ and the $\overline{45}$ $SU(5)$ representations and the 
couplings would allow for a differentiation between the down quarks and the 
charged leptons. Again, since the purpose of the present model was only 
illustrative, we do not insist here on recovering fully realistic mass matrices. 
Finally, additional contributions to flavour-changing processes and 
to CP-violating observables are generally expected in a supersymmetric GUT. 
However, a reliable estimate of the corresponding effects
would require a much more detailed definition of the theory than attempted here. 
Crucial ingredients such as the mechanism of supersymmetry breaking and its 
transmission to the observable sector have been ignored in the present note. 

In this example we have given up a possible symmetric structure of the textures in favour 
of a relatively simple and economic realization of the underlying flavour symmetry.
It is possible, within the $SU(5)$ or $SO(10)$ gauge theories, to provide slightly more 
elaborate examples where this symmetric structure is manifest.
Here we have insisted on requiring both the Dirac matrices and the Majorana matrix
$M$ to be quasi-diagonal (that is with all off-diagonal entries {\it ij} suppressed with 
respect to the largest of the diagonal {\it ii} and {\it jj} ones). 
In some alternative models, all Dirac matrices are indeed quasi-diagonal, 
but large mixings are present in the Majorana matrix $M$. For instance,
the idea of a large atmospheric mixing angle from intrinsically
small mixings in the Dirac mass matrices has recently been advocated
in the context of a $U(2)$ flavour theory \cite{bcr}. 
In this model the Dirac neutrino mass matrix is  
the one displayed in eq. (\ref{mdM3}), with $y$ of O$(\epsilon)$ 
(i.e. $(m_D)_{13}\ll (m_D)_{12}$) and $\epsilon''\sim\epsilon'^2$. 
The $U(2)$ symmetry favours an $r_1$ dominance, by adequately
suppressing the $M_{11}$ element. However, to have both $\nu_\mu$ and $\nu_\tau$
with approximately the same coupling to the light Majorana mass eigenstate,
this should possess a component of order $\epsilon'$ along the $\nu^c_3$
direction. In turn, this property is only exhibited if also the $M_{33}$
entry is sufficiently small
\footnote{Technically, this suppression can be realized in the framework of 
an $SO(10)$ gauge group, by assuming that there are no flavour singlets
that couple to $\nu^c_3\nu^c_3$.}. As a consequence, there are two heavy Majorana
states with similar masses and there is a large mixing in the 23 Majorana
sector. In this case, large mixings, which are absent from the Dirac mass matrices, arise from the Majorana sector, 
a physically similar possibility but technically different from that examined here.

\subsection{Outlook}

Summarizing, in most models that describe neutrino oscillations with nearly maximal 
mixings, there appear large mixings in at least one of the matrices $m_D$, $m_D^l$, $M$ 
(i.e. the neutrino and charged-lepton
Dirac matrices and the right-handed Majorana matrix). 
Here we have discussed the peculiar possibility 
that large neutrino mixing is only produced by the see-saw 
mechanism starting from all nearly diagonal matrices.
Although this possibility is certainly rather special, 
we have shown that models of this sort can
be constructed without an unrealistic amount of 
fine tuning and are well compatible with grand unification ideas
and the related phenomenology for quark and lepton masses.

\clearemptydoublepage
\chapter{Problems of SUSY GUTs}

\label{pfsg} 

The idea that all particle interactions merge into a unified theory 
at very high energies \cite{uth} is so attractive, for both aesthetical and phenomenological 
reasons, that this concept has become widely accepted by now. 
Indeed, the quantitative success of coupling unification in SUSY Grand Unified Theories 
(GUTs) \cite{sgut} can be hardly interpreted as just a coincidence.
In addition, support has also come from the developments on neutrino oscillations 
\cite{SKatm},
which point to lepton number violation at a scale close to the coupling unification one.

Despite the fact that they constitute promising ideas beyond the standard model,
{\it SUSY GUTs are in a not so safe ground as at first sight could appear}. 
This chapter is devoted to a critical reanalysis of those problematic
aspects found in constructing realistic models.   

Assuming gauge unification, minimal models of GUTs based on $SU(5)$, 
$SO(10)$, ... have been considered in detail. 
Also, many articles have addressed particular aspects of GUTs models 
like proton decay, fermion masses and, recently, neutrino masses and mixings.
In the previous chapter, indeed, the attention has focused on
the phenomenology of neutrino masses and mixings.    
But on the one hand, {\it minimal models are not plausible as they need a large amount of 
fine tuning} and are therefore highly unnatural, for example with respect to the 
doublet-triplet splitting problem or proton decay. 
On the other hand, {\it analyses of particular aspects of GUTs often leave aside the 
problem of embedding the sector under discussion into a consistent whole}. 

So the problem arises of going beyond 
minimal toy models by formulating sufficiently realistic, not
unnecessarily complicated, relatively complete models that can 
serve as benchmarks to be compared with experiment. 
More appropriately, instead of "realistic" one should say 
"not grossly unrealistic" because it is clear that many important
details cannot be sufficiently controlled and assumptions must be made.
Finally, the success or failure of such program can decide whether 
a stage of gauge unification is a likely possibility or not. 

In this chapter we will review SUSY GUT's basic weak-points and analyze critically 
the current status of solutions. As will be seen, most of these solutions are partial,
have some problematic feature and are in general quite complicated.
The crucial problem is to find a consistent picture of the Higgs sector which leads
to light doublets but triplets heavy enough to 
escape current bounds on proton decay \cite{pbounds}. In addition, the field content must 
not destroy coupling unification.
During this survey, we will devote special attention to those solutions
that will adopted in the following chapter, 
where an example of semi-realistic model will be proposed. 

%%%%%%%%%%%%%%%%%%%%%%%%%%%%%%%%%%%%%%%%%%%%%%%

\section{The Doublet-Triplet Splitting Problem}

The phenomenological motivation for weak scale SUSY is the solution of the
hierarchy problem. While the mass of the Higgs doublet should be around $100$ GeV,
both coupling unification and non observation of higgsino mediated proton decay 
require the Higgs triplet mass to lie around $10^{16}$ GeV, the precise value 
depending on the details of the model under discussion. 
Achieving this doublet-triplet splitting \cite{dtspl} is the most compelling problem one has to face in 
GUTs, because it requires to fine-tune a parameter at the level of $14$ orders of magnitude. 
It is worth to remind that, with respect to this problem, SUSY GUTs stay in a much better 
situation than non SUSY GUTs. Thanks to non-renormalization theorems, 
in the former the fine-tuning has to be done only once, that is at tree level,
while in the latter it has to be done at all orders of the perturbative expansion.   

After a short presentation of the transparent way in which 
the doublet-triplet splitting problem appears in the case of minimal $SU(5)$, 
this section has the purpose of summarizing the status of the most interesting 
mechanisms which avoids this fine-tuning also insisting on their weak-points. 
In many models, in fact, even if one succeeds in achieving the desired splitting at tree level,
this could be upset by a lifting of the doublet mass, due to either radiative 
corrections or non renormalizable operators. In addition, one has to be careful
because some of the possible mechanisms can destroy coupling unification. 
 
\subsection{The Doublet-Triplet Splitting Problem in Minimal $SU(5)$ }
\label{23splprob}

In minimal SUSY $SU(5)$ the Higgs sector consists of one chiral superfields in the 
adjoint representation, $\Sigma$, and one pair of fundamental and antifundamental, 
$H, \overline H$.  The breaking of $SU(5)$ is provided by    
the following superpotential terms: 
\begin{equation}
w_{Higgs}=M_{24} ~{\Sigma}^A_B {\Sigma}^B_A 
+ \lambda  ~{\Sigma}^A_B {\Sigma}^B_C {\Sigma}^C_A
+c~ {\overline H}_A  {\Sigma}^A_B H^B + M_5 ~{\overline H}_A H^A ~~~~~~,
\label{wmin}
\end{equation}
where $A, B, ...=1, ..., 5$ and summation over repeated indices has to be understood.
In the limit of unbroken SUSY, minimization of this part of superpotential gives
the following $SU(3)\otimes SU(2) \otimes U(1)$ symmetric vevs
\footnote{Actually there are three supersymmetric degenerate minima:
in one of them $SU(5)$ is unbroken, while in the other two it is broken respectively
to $SU(4)\otimes U(1)$ and $SU(3)\otimes SU(2) \otimes U(1)$. The degeneracy
could eventually be lifted and the MSSM vacuum chosen due to supergravity corrections.}: 
\bea
\langle \Sigma \rangle &=&  \dd\frac{2}{3} \dd\frac{M_{24}}{\lambda}~diag(2, 2, 2, -3, -3) ~~~~~~,\\
\langle H \rangle &=& 0~~~,\\
\langle \overline H \rangle &=& 0~~~.\nn
\eea
This superpotential also ensure heavy masses to those $\Sigma$ multiplets 
which are not eaten by the Higgs mechanism (see Table \ref{gutthmin}).
Due to the structure of the $\Sigma$ vacuum, doublets and triplets acquire different 
masses:
\beq
M_D=M_5 - \frac{2 c}{\lambda} M_{24}~~,~~~M_T= M_5 +\frac{4 c}{3 \lambda}  M_{24}~~~.
\eeq  
To have light doublets one has to impose that the parameters $M_5$ and 
$M_{24}$, which should assume values around $10^{16}$ GeV, satisfy
\beq
M_5 \cong  \frac{2 c}{\lambda} M_{24} ~~~~
\eeq
with an accuracy of at least the $14^{th}$ decimal, so that $M_D=O(100 {\rm GeV})$.
This hierarchy problem is inherent to all GUTs. 
SUSY doesn't explain the origin of the two scales, but renders this constraint stable 
against radiative corrections, so that, at least, the fine-tuning has to be done only once. 
Non SUSY GUTs, in contrast, need the tuning to be repeated at all perturbative orders.

Similar considerations also apply to the case of minimal $SO(10)$, where $H, \overline H$
lie in the $10$ representations of $SO(10)$.

\section{Solutions to the Doublet-Triplet Splitting Problem}

The actual problem is to achieve the doublet-triplet splitting in a natural way, without
fine-tuning. 
Some solutions that overcome this problem have been proposed, based on the introduction
of Higgs representations with useful group-theoretical properties, which do the job
of allowing a mass only for triplets but not for doublets.

Several attempts have been done and someone of the most important 
will be summarized in the following.
Despite the fact that basic ideas are indeed very elegant, emphasis will be given to
some problems which remain unsolved.  
In addition, it turns out that embedding these solutions in a realistic 
and not too cumbersome framework is not an easy task.

\subsection{The Missing Doublet Mechanism}

One of the most economical solution the doublet-triplet splitting problem is 
represented by the Missing Doublet Mechanism (MDM) \cite{mdm}. 
It works in $SU(5)$, which is the lower rank group containing 
the standard model, and is realized just replacing the $24$ of Higgses by a $75$, $Y$, and 
by further adding a pair of $50, \overline{50}$ Higgses. 
Unfortunately, this proposal suffers of some weakness which, however, 
can be overcome. 

\subsubsection{Description of the Mechanism}

The part of the superpotential responsible for the breaking of $SU(5)$ reads 
\beq  
w_{Higgs}=c_1~Y^{A B}_{C D}~ Y^{C D}_{E F}~ Y^{E F}_{A B}
        +M_Y~Y^{A B}_{C D}~ Y^{C D}_{A B}~~~~~,
\eeq
where, as usual, $A,B,...=1,2,...,5$.
These terms provide $Y$ with a vev 
$\langle Y \rangle \equiv M_Y/(2 c_1) \sim M_{GUT}$ and give a mass to all
physical components of $Y$, i.e. those that are not absorbed by the Higgs mechanism
(see Table \ref{htrmdm}). 
The SUSY vacuum of the $75$ breaks $SU(5)$ uniquely to $SU(3)\otimes SU(2)\otimes U(1)$: 
\beq
\langle Y_{C D}^{A B} \rangle=- \dd \langle Y \rangle   
          \left[ \dd\left( {\delta}_{\gamma }^{\alpha } {\delta}_{\delta }^{\beta  }
                      +2 {\delta}_{c}^{a } {\delta}_{d}^{b}
                  - \dd\frac{1}{2} {\delta}_{C}^{A} {\delta}_{D}^{B} \dd\right) 
                  - \left(A \leftrightarrow B\right) \dd\right]~~~~~~,
\eeq
where Greek indices assume values from 1 to 3 and Latin indices from 4 to 5. 

The doublet-triplet splitting is realized by the following superpotential terms,
\begin{eqnarray}
w_{MDM}=c_2~H^{A} Y^{D E}_{B C} {H_{50}}^{B C F G} {\epsilon}_{A D E F G}~
+~c_3 {\overline H}_{A} Y^{B C}_{D E} {H_{\overline{50}}}_{B C F G} 
                {\epsilon}^{A D E F G}~\nonumber\\
+~M_{50}~{H_{50}}^{A B C D} {H_{\overline{50}}}_{A B C D} ~~~~~~~~~~~~~~~.
\label{wwwmdm}
\end{eqnarray} 
Note that the direct term $M_5 {\overline H}_A H^A$ must not to appear.
The splitting occurs because the $50$ of chiral superfields contains a 
$(\bar 3,1,1/3)$, i.e. a coloured antitriplet, $SU(2)$ singlet (of electric charge 1/3) 
but no colourless doublet (1,2,-1/2).
Denoting the colour triplets contained in $H$, ${\overline H}$, $H_{50}$, 
${H_{\overline{50}}}$ by $H_{3u}$, $H_{3d}$, $H'_{3d}$ and $H'_{3u}$ respectively, 
after $SU(5)$ breaking: 
\beq
w_{MDM} \ni -4 \sqrt{3} c_2 \langle Y\rangle~ H'_{3d} H_{3u} 
- 4 \sqrt{3} c_3 \langle Y\rangle~ H_{3d} H'_{3u} +  M_{50}~ H'_{3d} H'_{3u} +...
\label{wmdm}
\eeq
where dots stand for mass terms for the other multiplets contained in $50, \overline{50}$. 
Thus, $w_{MDM}$ doesn't produce any mass term for the doublets but, 
on the other hand, the Higgs colour triplets mix with the analogous states in the 
$50$. The resulting triplet mass matrix is of the see-saw form:
\beq
\hat{m}_T= \left[ \begin{array}{cc}
 0& - 4 \sqrt{3} c_2 \langle Y\rangle \\
 - 4 \sqrt{3} c_3\langle Y\rangle& M_{50}   \end{array}\right]~~~~~. 
\label{hatmt}
\eeq
The eigenvalues of the matrix $\hat{m}_T \hat{m}_T^\dagger$ are the squares of:
\begin{equation}
m_{T1,2}=\frac{1}{2}\left[
\sqrt{M_{50}^2+48~(c_2+c_3)^2\langle Y\rangle^2}\pm \sqrt{M_{50}^2+
48~(c_2-c_3)^2\langle Y\rangle^2}\right]~~~.
\label{autmdm}
\end{equation}
Note that $m_{T1}m_{T2}=~48~ c_2~ c_3~ \langle Y\rangle^2$, so that
triplets are naturally heavy.
The effective mass that enters in the dimension 5 operators with $|\Delta B=1|$ is
the following combination of physical masses: 
\begin{equation}
m_T=\frac{{{m_T}_1} {{m_T}_2}}{M_{50}}=\frac{48~ c_2~ c_3~ \langle Y\rangle^2}
{M_{50}}
\label{mTmin}~~~.
\end{equation}

In the literature there is confusion \cite{yam,his,mnty,ber} on the numerical coefficients
of the previous relations and also on the values for the heavy masses
of the $75$. In particular, many authors seem to forget the coefficient $4 \sqrt{3}$
in eqs. (\ref{wmdm}), (\ref{hatmt}) and following. But this factor is important, because it
gives in eq. (\ref{mTmin}) the large coefficient $48$ that cannot be neglected when 
evaluating the effective triplet mass which suppress proton decay amplitudes. 
The formulae above have been rederived,
by carefully paying attention to the group-theoretical factors.
Appendix \ref{appgroup} contains a brief summary of this procedure.

\subsubsection{Possible Criticisms}
\label{mdmcrit}
Criticism against the MDM is often moved by the common fear that its large 
rank representations are dangerous, mainly for two arguments.  
The first one is that, from the point of view of string theory, it is really hard to find 
examples containing such huge representations. 
From a more conservative point of view, however, embedding GUTs into 
a string theory is for sure interesting but not at all compelling.
The second argument is that the large content of matter fields in the MDM causes 
the gauge coupling to blow up before $M_{Pl}$. In principle, a theory which is 
non perturbative before $M_{Pl}$ is acceptable. 
%However, it hides an exacerbation 
%of potential problems with non renormalizable operators, as emphasized in ref. \cite{ran}.
  
Apart from the two previous considerations, the MDM in its simplest version has mainly 
a weak point.
No matter what symmetry is used to forbid the direct mass term for ${\overline H} H$ in the
superpotential, the doublet Higgs could take mass from non renormalizable 
operators of the form \cite{ran}:
\beq
\frac{ {\overline H} H Y Y}{M_{Pl}}~~~~.
\eeq 
If one believes Planck suppressed operators consistent with the theory are present, 
the doublet has a too big mass. The problem is exacerbated in MDM because
the coupling blows up at a low scale and it is probably the associated strong scale 
which would suppress such operators.
A promising way to overcome this obstacle was introduced in ref. \cite{ber} and is based
on the introduction of an additional anomalous $U(1)$ flavour symmetry.

As will be clarified, the solution of the doublet-triplet splitting problem in
terms of the $50$, ${\overline{50}}$ and $75$ representations is very suggestive because
it automatically leads to improve the prediction of $\alpha_3 (m_Z)$ and
at the same time relaxes the constraints from proton decay. 

\subsection{The Sliding Singlet Mechanism}

New possibilities emerge when introducing an additional singlet field $S$.
The simpler one \cite{slidings} is the Sliding Singlet Mechanism. It is realized by just 
adding to eq. (\ref{wmin}) the term
\beq
f S {\overline H}_A H^A~~~~.
\eeq
where $f$ is an adimensional coefficient. At tree level and considering also soft SUSY 
breaking terms, the vev of $S$ can be arranged
to cancel the large contributions to the doublet mass coming from the other terms.
However, it is well known that this mechanism is unstable against radiative corrections
\cite{slidko}, because tadpole diagrams induce the doublet to take a too big mass 
$\sim \sqrt{m_{3/2} M_{GUT}}$.
 
\subsection{The Missing VEV Mechanism}

The Missing VEV Mechanism \cite{mvm} (MVM), 
which apply to the case of $SO(10)$, is based on the presence of a
$45$ Higgs representation. With respect to $SU(5)$, $45=1+24+10+\overline{10}$,
so that, in practice, having a singlet and a $24$, it is based on the same idea of the 
Sliding Singlet. This can occur because the $45$ representation is not constrained 
to be traceless. In fact, the $45$ can have
a vev of the form $\langle 45 \rangle\sim diag(1, 1, 1, 0, 0) \otimes i \sigma$.
The term $10_1 ~45 ~10_2$,  gives mass to triplets but not to doublets. Note that 
two $10$ are required due to the antisymmetry of $45$, so that there are four doublets
and four triplets. 

This mechanism seems very nice at first glance, but worrisome at second.
In fact the problem arises of giving mass to just two doublets (or two
linear combinations of them). This could be done, for instance, by introducing 
a term $M_2 10_2 10_2$ and forbidding any direct coupling with $10_1 10_1$. 
Heavy doublets have then a mass $\sim M_2$ and are in no way connected with
the light ones. 
But there is also the need for triplets heavy enough to agree with
experimental bounds on proton decay, and this is not very easy. For instance,
by adding $M_2 10_2 10_2$, the triplets mass matrix $\hat m_{T}$ turns out to be 
of a see-saw form (exactly as in the case of the MDM but with the $45$ instead
of the $75$ of $SU(5)$, see eq. (\ref{mTmin})). 
The effective mass entering proton decay amplitudes
is then proportional to $ \langle 45 \rangle^2 / M_2 $. But in order to achieve the 
needed suppression of proton decay amplitudes, we need 
$M_2 \le 10^{-(2 \div 3)} \langle 45 \rangle \sim M_{GUT}$.
Apart from the fact that the not-so-heavy doublets would give dangerously large threshold 
corrections, this requirement is clearly a fine-tuning.

The simplest versions of the MVM are then not realistic, but
at the price of introducing a good deal of new fields,
various models have been realized in the past in which the parameter playing the role 
of $M_2$ is suppressed because of symmetry reasons. Also the new fields produce 
large threshold corrections, so that one cannot exclude the accidental
possibility of a general compensation of all their contributions.
 
However, dealing with a large number of GUT fields is not very attractive,
because the predictivity of the model is highly reduced and, most importantly, 
the success of coupling unification within the MSSM would be merely a lucky coincidence.   
In addition, due to the increased
number of Higgs representations that generally appears in many models,
there is the problem of actually implementing a potential to get the desired minimum,
paying attention to possible flat directions and without destabilizing the minimum 
with higher order terms.     
In fact, in order to lower the rank of $SO(10)$, one needs also one (or more) $16$.
But the couplings between $16$ and $45$ can destabilize the the desired vev of the
$45$. 
The literature is rich of interesting models \cite{babu, ber&tav,al&barr} 
where this sort of compromise between the internal elegance and the necessity of 
agreeing with phenomenology  appears clearly. 

\subsection{Higgses as pseudo-Goldstone Bosons}

An interesting approach to the doublet-triplet splitting problem is obtained if one
consider the Higgs doublets to be pseudo-Goldstone bosons related to a spontaneously 
broken accidental global symmetry of the Higgs sector. This distinguishes
the doublets from the triplets in a very non trivial way. In this scenario the 
Higgs sector of theory
does possess a spontaneously broken accidental global symmetry and the Higgs
doublets are the associated Goldstone bosons.  
The accidental global symmetry is explicitly broken in the Yukawa sector. The
Higgs doublets become pseudo-Goldstone bosons with a mass determined by the
scale of soft SUSY breaking terms.
Because of non-renormalization theorems, this mass is protected against radiative
corrections. 

Despite that the idea is very elegant, 
the difficulty one finds in its actual realization is
to motivate the accidental symmetry in a natural way. 
For instance, models involving a gauged $SU(5)$ symmetry and a global $SU(6)$ \cite{hgb1} 
unfortunately require an unpleasant fine-tuning of some parameters of the superpotential.
In order to justify the accidental symmetry, an alternative possibility \cite{hgb2} 
is to assume that it arises because two sectors of the chiral superfields responsible 
for gauge symmetry breaking do not mix. But such a separation has to be imposed ad hoc.

%%%%%%%%%%%%%%%%%%%%%%%%%%%%%%%%%%%%%%%%%%%%%%%

\section{Proton Decay}

Proton decay is maybe the most interesting consequence of GUTs. 
Since a long time it has been representing a very hard challenge 
from the experimental point of view. The Kamiokande detector was construct
for this purpose and, after Super-Kamiokande, maybe
a third-generation detector will be constructed.
From the theoretical perspective this subject is not less engaging: the actual value 
of proton lifetime (if any) is determined by such an enormous 
amount of physics, from the GUT scale down to $1$ GeV, that it offers the possibility 
of exploring various aspects of particle physics at the same time.
The theoretical prediction is thus not at all accurate and strongly model dependent, 
but the common opinion is that it should lies just in the range already at hand 
of experimental searches or not very far.

A lot of weighty informations have already come from these studies:
\begin{itemize}
\item non SUSY $SU(5)$ is excluded because, due to the lightness of heavy gauge 
bosons, it predicts a lifetime induced by dimension 6 operators which is much lower 
than present bounds. 
Same conclusions hold for greater groups which break in one step to $SU(5)$, like $SO(10)$.
\item minimal SUSY $SU(5)$ is in very serious troubles: with SUSY particles
lighter than $1$ TeV, dimension 5 operators would induce proton to decay mainly into
$K^+ \bar{\nu}$ at an exceedingly fast rate, in conflict with experiment.
The case of $SO(10)$ is less definite, because the prediction
of proton lifetime is linked to the particular Higgs content chosen to realize
the breaking of $SO(10)$. However, it turns out that also in those models where
as few as possible Higgs representations are introduced \cite{bapawi}, the
prediction of proton lifetime should not be far from the range explored by 
present searches.    
\end{itemize}  

Prompted by the recent Super-Kamiokande improvements on proton partial lifetimes 
bounds \cite{lastSK}, a renewed interest on this subject has arisen. Since GUTs without SUSY 
turned out to be disappealing, the attention has focused mainly on the question 
whether it is possible to construct examples of realistic SUSY GUT models. 
In such models, achieving a sufficient suppression of proton decay amplitudes from 
dimension 5 operators is of crucial importance. In fact, since the actual prediction of 
proton lifetime depends significantly of the details of the model, one is interested in
arranging things in such a way to escape conflict with bounds. 

To understand how this can be done reasonably, that is without introducing fine-tuning,
it is useful to have a close look at the calculation of proton decay lifetime in the simplest
possible case, namely minimal SUSY $SU(5)$.  
In this framework we will summarize the principal steps of the calculation
of proton lifetime induced by dimension 5 operators \cite{five}. We will treat
independently Yukawa couplings of doublets and triplets. Also down quarks and 
charged leptons mass matrices will not be correlated, so that the calculation can be easily 
adapted also to non minimal models. 
We will show in some detail that, generically, the amplitude is dominated by only 
two terms of the Lagrangian and we will discuss, also quantitatively, how large can be 
the effect of the two CP violating phases (in addition to the CKM one) which enter 
this amplitude. We will comment on the source of the various uncertainties and give
numerical examples.
These have to be compared with the bounds collected in Table \ref{datip} 
from ref. \cite{lastSK} .  

\begin{table}[t]
\begin{center}
\begin{tabular}{|c|c|}
\hline 
&\\Channel & Lower Limits (90 \% CL)\\
&$\tau / B (10^{33} yr)$ \\ 
&\\
\hline & \\ $K^+ \bar \nu$& $1.9$\\
$\pi^+ \bar \nu$& $0.025$\\ 
$e^+ \pi^0$& $4.4$\\
$\mu^+ \pi^0$& $3.3$\\
& \\
\hline
\end{tabular}
\end{center}
\caption{Lower Limits on Proton Decay Channels \cite{lastSK}.}
\label{datip}
\end{table}

\subsection{Dimension Five Operators}

The starting point is the most general renormalizable superpotential which could come
from $SU(5)$. 
Yukawa couplings of doublets and triplets will be considered uncorrelated as well as 
down quarks and charged lepton mass matrices. The discussion can thus be adapted
for any $SU(5)$ based model.
The possible presence of any GUT superfield with
vanishing vev will not alter the following discussion.   

It is useful to write down the superpotential in terms of the MSSM fields. 
The triplets $H_{3u}\sim (3,1,-1/3),
H_{3d}\sim (\bar 3,1,1/3)$,  
together with the two light Higgs doublets $H_{2u},H_{2d}$ stay in a pair of $SU(5)$ 
pentaplets $H, \overline H$. 
Omitting colour and weak isospin indices, in the interaction basis the terms of the 
superpotential which are relevant to proton decay read:
\bea
w &\ni & Q^T y_u U^c H_{2u} +Q^T y_d D^c H_{2d} + L^T y_e E^c H_{2d} \nn\\
&+& Q^T \hat{A} Q H_{3u}+ {U^c}^T \hat B E^c H_{3u} +Q^T \hat C L H_{3d} +
{U^c}^T \hat D D^c H_{3d}\\
&+& m_T H_{3u} H_{3d}\nn~~~~~,
\eea
where $Q$, $L$, $U^c$, $D^c$ and $E^c$ denote as usual the chiral multiplets associated to 
the three fermion generations, $y_{u,d,e}, \hat A, \hat B, ...$ are the $GUT$ scale 
Yukawa matrices of doublets and triplets and $m_T$ is the triplets mass. 

By integrating out the colour triplets \cite{enr, susygut, hmy} one obtains the following 
effective superpotential
\beq
w_{eff}=\frac{1}{m_T}\left[Q \hat{A} Q~
Q \hat{C} L + U^c \hat{B} E^c~
U^c \hat{D} D^c\right]+...~~~~~,
\label{weffp}
\eeq
dots standing for terms that do not violate baryon or lepton number.
The former (latter) term, which involves four fields with only left (right) chirality, is referred 
to as LLLL (RRRR) operator. 
$w_{eff}$ generates dimension 5 operators with two fermions and two sfermions
in the Lagrangian.

At lower scales these dimension five operators give rise to the four-fermion operators
relevant to proton decay, via a "dressing'' process, that is through 1-loop diagrams with
exchange of charginos and neutralinos. 
It turns out that the relevant contributions arise from 
chargino exchange \cite{enr,five,dress}. 
When considering the operator $QQQL$, the most important dressing comes
from wino exchange. There are four different terms in the Lagrangian, 
here called $O^L_i, (i=1, ...,4)$, coming from the exchange of a wino  
\footnote{Gluino dressing contributions cancel among each other in case of
degeneracy between first two generations of squarks \cite{ehnt,gluino}.}. 
Charged higgsino exchange provides instead the most important dressing of 
$U^c E^c U^c D^c$ \cite{goto}.
Also in this case four Lagrangian terms arise, here called $O^R_i, (i=1, ...,4)$.
The conventions adopted, the definition of the loop function $f$  
and all these operators are explicitly given in appendix \ref{appprot}
and in Table \ref{operatgen}.

\subsection{Hunting Dominant Operators}

The situation would then appear quite complicated. 
However, in the case of minimal $SU(5)$, where 
\beq
- 2 \hat A= \hat B =y_u=y_u^T~~~,~~~~~- \hat C = \hat D = y_d =y_e^T~~~~,
\label{masssu5}
\eeq
it turns out that, in each category, 
only one term gives the relevant contribution 
to proton decay amplitudes and that the contributions from the other three terms are
suppressed by at least two powers of the Cabibbo angle, $\lambda^2$. 
This is expected to be approximately true also in those models where the matrices
$\hat A, \hat B, ...$ do not differ significantly from those of the minimal case.

\begin{table}[t] 
\begin{center}
\begin{tabular}{|c|c c c c|}
\hline
& & &  & \\
$O^L_1$ &$ {[ L_d^T (\hat A+{\hat A}^T) L_d] }_{i j} {[ L_u^T \hat C L_e] }_{k l}$ &$\rightarrow$
& ${[V^T y_u^{(d)} K^+ V]}_{i j}{[V^* y_d^{(d)}]}_{kl}$ & $\equiv {c^L_1}_{i j k l}$ \\ %3
& & &  &\\

$O^L_2$ & $ {[ L_u^T (\hat A+{\hat A}^T) L_d] }_{i j} {[ L_d^T \hat C L_e] }_{k l}$  & $\rightarrow $ 
& ${[y_u^{(d)} K^+ V]}_{i j} {[y_d^{(d)}]}_{kl}$& $\equiv {c^L_2}_{i j k l}$ \\  %1
& & &  &\\

$O^L_3$& $ {[ L_u^T (\hat A+{\hat A}^T) L_d] }_{i j} {[ L_u^T \hat C L_e] }_{k l}$ & $\rightarrow$ 
& ${[y_u^{(d)} K^+ V]}_{i j} {[V^* y_d^{(d)}]}_{kl}$ & $\equiv {c^L_3}_{i j k l}$\\  %2
& & &  &\\

$O^L_4 $ & ${[ L_u^T (\hat A+{\hat A}^T) L_u] }_{i j} {[ L_d^T \hat C L_e] }_{k l}$  & $\rightarrow$  
& ${[y_u^{(d)} K^+]}_{i j} {[y_d^{(d)}]}_{kl}$& $\equiv {c^L_4}_{i j k l}$\\  %4
& &  & &\\
\hline
& & & &\\
$O^R_1$ &$ {[ L_d^+ y_u^* \hat B y_e^+ L_e^*] }_{k l}  {[ R_u^* \hat D R_d^+] }_{i j} $ & $\rightarrow$ 
& ${[V^+ ( y_u^{(d)} )^2 V y_d^{(d)}  ] }_{k l} {[K V^* y_d^{(d)} ]}_{i j}$ & $\equiv {c^R_1}_{k l i j}$ 
\\ % 3
& & & &\\
$O^R_2$ & ${[ R_u^* \hat B y_e^+ L_e^*] }_{k l} {[ L_d^+ y_u^* \hat D R_d^+] }_{i j}$ & $\rightarrow$ 
& ${[y_u^{(d)} V y_d^{(d)}]}_{k l} {[V^+ y_u^{(d)} K V^* y_d^{(d)}]}_{i j}$& $\equiv {c^R_2}_{k l i j}$
\\ %4
& & & &\\
$O^R_3 $ & $ {[ R_u^*  \hat B  R_e^+] }_{k l}  {[ L_d^+ y_u^*  \hat D y_d^+ L_u^*] }_{i j} $& $\rightarrow$ 
& ${[y_u^{(d)} V]}_{k l} {[V^+ y_u^{(d)} K V^* (y_d^{(d)})^2 V^T]}_{i j}$ & $\equiv {c^R_3}_{k l i j}$
\\ % 1
& & & &\\
$O^R_4$ & $ {[ L_d^+  y_u^* \hat B  R_e^+] }_{k l}  {[ R_u^* \hat D y_d^+ L_u^*] }_{i j}$& $\rightarrow$ 
& ${[V^+ ( y_u^{(d)} )^2 V]}_{k l} {[K V^* (y_d^{(d)})^2 V^T]}_{i j}$ & $\equiv {c^R_4}_{k l i j}$
\\ % 2
& & & &\\
\hline
\end{tabular}
\end{center}
\caption{In the case of minimal $SU(5)$, the combinations in square brackets of 
Table \ref{operatgen} can be expressed in terms of the diagonal Yukawa couplings 
of doublets and in terms of the CKM matrix $V \equiv V_{CKM}$. See appendix 
\ref{appprot} for the conventions adopted here.}
\label{opmin}
\end{table}
From eqs. (\ref{masssu5}), 
also the following relations follow
\beq
R_u^\dagger  = L_u K~~~, L_e = R_d^\dagger~~~~,R_e= L_d^\dagger~~~~~~,
\label{trasf}
\eeq
where $R_{u,d,e}, L_{u,d,e}$ are defined in Appendix \ref{appprot} 
and $K$ is a diagonal matrix of phases \cite{ncafasi,hmy}
with determinant one. The
two degree of freedom it contains are CP violating phases, in addition to the 
CKM one.   
Since proton decay amplitudes are insensitive to the overall phase, it is not 
restrictive to choose $K$ to be of the following form
\beq
K=diag(e^{i {\phi }_1} , e^{i {\phi }_2},1)~~~~.
\label{kappa}
\eeq

In the mass eigenstates basis and making use of the previous definitions, 
the four-fermion operators of Table \ref{operatgen} get greatly simplified.
In fact, it turns out that the combinations in square brackets can be expressed
just in terms of the diagonal Yukawa couplings of doublets 
and in terms of the CKM matrix, as shown in Table \ref{opmin}.
This means that no trace is left of right handed mixings of fermions.

The relative importance of the various operators can be approximately understood 
expressing them in terms of powers of the Cabibbo angle $\theta_C = \lambda$. 
If the masses of the sparticles in the loop
are such that they allow $f$ (defined in appendix \ref{appprot}) 
to vary in a small range, so that it can be quite safely 
factored out, and remembering the approximate GUT scale relations 
\bea
m_u:m_c:m_t={\lambda}^8 : \lambda ^4 : 1\nn\\
m_d:m_s:m_b={\lambda}^4 : \lambda ^2 : 1
\eea
\beq
V_{CKM}\sim \left[\begin{array}{ccc}
   1& \lambda & \lambda^3 \\
   \lambda & 1   & \lambda^2 \\
   \lambda^3 & \lambda^2 & 1  
\end{array} \right]
\eeq
one immediately obtains Table \ref{pow}.

\begin{table}[h] 
\begin{center}
\begin{tabular}{|c||c| c| c|  c|}
\hline
& & & &  \\
& $c^L_1$ &$c^L_2$ & $c^R_1$ & $c^R_2$ \\
& & & & \\
\hline 
\hline
& & & & \\
$ p \rightarrow K^+ {\bar \nu}$ 
&$\lambda ^{10}\tan \beta$ & $\lambda ^{12}\tan \beta$
& $\lambda ^{10}\tan^2 \beta$  &  $\lambda ^{22}\tan^2 \beta$ \\
&$(\bar \nu_{\mu,\tau})$ & $(\bar \nu_{\mu})$  
& $(\bar \nu_{\tau})$ & $(\bar \nu_{\mu,\tau})$\\
& & & &\\
\hline
& & & &\\
$p \rightarrow \pi^+ \bar \nu$ 
&$\lambda ^{11}\tan \beta$ & $\lambda ^{14}\tan \beta$
&  $\lambda ^{11}\tan^2 \beta$  &  $\lambda ^{25}\tan^2 \beta$\\
& $(\bar \nu_{\mu,\tau})$ & $(\bar \nu_{\mu})$ 
& $(\bar \nu_{\tau})$ & $(\bar \nu_{\mu,\tau})$\\
& & & & \\
\hline
\hline
& & & &\\
& $c^L_3$ &$c^L_4$ & $c^R_3$ & $c^R_4$ \\
& & & & \\
\hline 
& & & &\\
$ p \rightarrow K^0 l^+ $ 
&$\lambda ^{14}\tan \beta$ & $\lambda ^{12}\tan \beta$
&  $\lambda ^{17}\tan^2 \beta$  &  $\lambda ^{14}\tan^2 \beta$ \\
& $(\mu^+)$ & $(\mu^+)$  & $(e^+)$ & $(\mu^+)$ \\
& & & &\\
\hline
& & & &\\
$p \rightarrow \pi^0 l^+ $ 
&$\lambda ^{13}\tan \beta$ & $\lambda ^{14}\tan \beta$
&  $\lambda ^{18}\tan^2 \beta$  &  $\lambda ^{15}\tan^2 \beta$\\
& $(\mu^+)$ & $(e^+)$ & $(e^+)$ & $(e^+)$\\
& & & & \\
\hline
\end{tabular}
\end{center}
\caption{Relative importance of the combinations in square brackets  
appearing in Table \ref{opmin}, expressed in terms of powers of the 
Cabibbo angle $\theta_C = \lambda$. Note that we summed over neutrino
flavours, which are experimentally indistinguishable.}
\label{pow}
\end{table}

One can see at glance that $O^L_1, O^R_1$ are, within their groups, the dominant operators.
The contributions to proton decay amplitudes from other operators are 
suppressed by at least $\lambda^2$ with respect to them, and can be safely neglected 
in the present estimate. The calculation then proceeds as summarized in Appendix 
\ref{appprot}.

Generally, the dominant channel is $p \rightarrow K^+ \bar{\nu}$. However,
also $p \rightarrow \pi^+ \bar{\nu}$ is important.
In fact, the difference of a factor $\lambda\sim 0.25$ doesn't produce a factor
$1/16$ in rates. The $\pi^+$ channel is favoured by phase space for a factor of $2$,
as appears from Table \ref{chir}. Another factor $4$ 
comes from Wick theorem. It results that $\pi^+ $ is suppressed by only a factor of $1/2$ in
rate, with respect to $K^+$.    
As will be explained later, under certain circumstances linked to the two CP violating
phases, the channel with $\pi^+$ could even be the dominant one. 
% per goto non e' cosi' perche prende RRR altissimo

As pointed out in \cite{goto}, $O^L_1, O^R_1$  can be comparable.
For large $\tan \beta$ and/or $m_{\tilde h} \gg m_{\tilde w}$, $O^R_1$ dominates.
But large values of  $\tan \beta$ are discouraging because they lead to faster
proton decay, exacerbating a possible contrast with experiment.
For small  $\tan \beta$ and $m_{\tilde h} \leq  m_{\tilde w}$, 
$O^L_1$ gives the dominant contribution.

\subsection{Proton Decay from Non-Renormalizable Operators}

It has been pointed out \cite{nroforp} that also non-renormalizable operators 
can lead to proton decay. If one views GUTs as effective rather than fundamental theories,
then one has to consider also the role of non-renormalizable operators.
The following operators, where it is understood that we are taking light generations, 
are completely allowed by gauge symmetry,
\beq
w_{eff} \ni  c_L \frac{Q Q Q L}{\Lambda}  + c_R \frac{U^c  E^c U^c  D^c}{\Lambda}~~~~~
\label{nonren}
\eeq
and lead, after the dressing process already described, to proton decay.  
Consider for instance the first term. It will be dressed as the LLLL operator
arising from triplet exchange, so that, evaluating the ratio between its proton 
decay amplitude $\mathcal{A}_{nr}$ and the amplitude from heavy triplet 
exchange $\mathcal{A}_3$, loop functions cancel each other. The difference
between the first terms of (\ref{weffp}) and (\ref{nonren}) resides only in their 
multiplicative coefficients and in the mass at denominator.
Approximately, for the dominant channel $p \rightarrow K^+ \bar \nu$,
with $\Lambda \sim M_{Pl}$, one finds 
\beq
\frac{\mathcal{A}_{nr}}{\mathcal{A}_3} \sim \frac{c_L }{M_{Pl}} \frac{m_T}
{\lambda^{10} \tan \beta} 
\eeq
For small $\tan \beta$ and $m_T \sim 2 \cdot 10^{16}$ GeV
we have something like $\mathcal{A}_{nr} / \mathcal{A}_3 \sim c_L \lambda^{-7} $,
which means a reduction of a factor $10^{8\div 9}$ in lifetime for $c_L=O(1)$.
With the former values of $\tan \beta$ and $m_T$, as we will see, the expectation
for proton lifetime is roughly $10^{31} $ yr, which is lower then actual bounds.
For non renormalizable operators one should then expect a lifetime of around
$10^{22}$ yr. To restore agreement with the bounds, one has to impose
$c_L$ to be at least smaller than $\lambda^8 \sim 10^{-(5 \div 6)}$. 
This is indeed a very large fine-tuning. 
The most natural solution is to impose flavour symmetries, which could 
give a valid reason to justify the required smallness of $c_L$ thus adequately 
suppressing these disturbing non renormalizable operators.
Flavour symmetries are usually largely employed in model building, because
they provide a clue to understand the observed picture of fermion masses and mixings. 
So, it would be interesting to study if, with the tool represented by flavour
symmetries, one can solve both problems.

\section{How to Prolong Proton Lifetime}

There are several parameters entering the formula for proton lifetime in minimal $SU(5)$,
so that one can should be careful not to underestimate theoretical
errors. 

For instance, by applying the chiral Lagrangian technique,
one obtains the formulas for the calculation of proton decay rates which are given
in Table \ref{chir}. One can see that the hadronic matrix elements $\alpha=- \beta$ are 
very important ingredients, because proton rate scale as their second power.
Unfortunately, these quantities are not very well known. 
Commonly, their quoted range \cite{albe} is $0.03 \div 0.003$ GeV$^3$ 
and the smallest value used in the calculations. However, it is worth to stress
that with the largest value, the rate would be increased by roughly two orders 
of magnitude.

However, one could also construct elaborated models
where one or more of the important parameters change with respect to the minimal
model. So, it is of crucial importance to analyze how much the prediction 
for proton lifetime is affected if one moves a certain parameter in its possible range.
In what follows we will briefly give a survey of the most interesting circumstances
in which proton lifetime can be raised. Finally some numerical example will be provided.
These are to be compared with those given in the next chapter, where a concrete
model will be examined.

\subsection{Altering Triplets Yukawa Couplings}

Clearly, the previous discussion about dominant operators, apply only to those models 
where the matrices $\hat A, \hat B, ...$ don't differ significantly from those of the minimal 
case. If, for some reason, such matrices are much more suppressed than 
in the minimal case, then proton decay rates would be reduced.
This possibility has been explored \cite{inserz} in the literature and,
in general, the suppression of certain matrix elements follows from
the group-theoretical construction of the model. One can also investigate
if the same strategy employed to suppress the Yukawa couplings of
triplets could also be effective in differentiating charged leptons and squarks
spectra.
Even if this is certainly an attractive possibility, it seems not very effective
in increasing proton lifetime. Indeed, when implemented in a realistic framework,
it is difficult, by only this mean, to adequately suppress all channels at the same time.

\subsection{Raising the Triplet Mass }

A really crucial parameter is the mass $m_T$. 
The dependence is strong, $\tau_p \sim m_T^{-4}$, so that 
a large value of $m_T$ would drastically decrease the rates for all channels.
As will be shown in the following, when considering the effect of heavy thresholds
on the running of the gauge couplings, it follows that in minimal $SU(5)$ and with SUSY
particles lighter then $1$ TeV, $m_T$ should not exceed $10^{16}$ GeV. This value
causes proton decaying at a too fast rate, in conflict with experiment. Minimal SUSY $SU(5)$
is then in serious troubles.  
Interestingly, a large value for $m_T$ is instead preferred in the MDM. 
Notice that in this case $m_T$ is not
a physical quantity but an effective one, given by eq. (\ref{mTmin}), and could in principle
even exceed $M_{Pl}$. In the next chapter we will discuss a realistic version 
of the MDM which maintains this attractive feature.
On the contrary, in those realizations of the MDM and MVM 
where a large number 
of heavy representations lead to huge threshold corrections (often of both signs), 
the correlation between the value of $m_T$ and the prediction for $\alpha_3 (m_Z)$ is lost.

\subsection{Moving the Two CP Violating Phases}

There are phases entering the proton decay amplitudes \cite{ncafasi} which don't affect 
any of the low energy observables, like for instance fermion masses and mixings 
\footnote{In the diagonalization of the mass matrices, these phases can always be 
absorbed through suitable fields redefinitions, so that only one phase is left, the 
$V_{CKM}$ one.}, but which
play a really interesting role, because they can produce sizeable effects on proton rates.
In fact, proton lifetime can easily span over one order of magnitude, just allowing
the phases to vary in their range.

As usual, it is instructive to see this firstly in minimal $SU(5)$, and then comment
on its extensions.

\subsubsection{The phases of minimal $SU(5)$}
% nath prd 32 85 2348 sbaglia perche' non conosceva la vckm
% dice che per pi non ci puo essere cancellazione come per K
% crede che per pi lo stau domini decisamente su scharm
% invece anche per pi c'e' interferenza
% da vedersi e' se si trova nella stessa regione di k o no
% dipende da Vckm (vub/vcb )e dalla sua delta

% goto prende RRRR enorme quindi non riesce mai a sopprimerlo
% come non riesce per K.

In minimal $SU(5)$ the two phases can be clearly identified. 
Following the definition of eqs. (\ref{trasf}) and (\ref{kappa}), 
after a bit of work,
one can see how they enter in the decay amplitude. 
In particular, considering the dominant operators $O_1^L, O_1^R$, one realizes that the
dominant amplitudes are proportional to the following quantities
\bea
\mathcal{A}(K^+ \bar \nu_{\mu})& \propto &(e^{i \phi_2} a^L_{\tilde c} 
+ a^L _{\tilde t} )\nn\\
\mathcal{A}(K^+ \bar \nu_{\tau})& \propto & (e^{i \phi_2} a^L_{\tilde c} 
+ a^L _{\tilde t}+e^{i \phi_1} a^R_{\tilde t^c} )~~~~~~~,
\label{phmin}
\eea
where $a^L_{\tilde c}, a^L _{\tilde t}$ and $a^R_{\tilde t^c}$ are complex numbers
which contain the piece of amplitude coming respectively from $\tilde c$
and $\tilde t$ for wino dressing and from $\tilde t^c$ for higgsino dressing.
$a^L_{\tilde c}, a^L _{\tilde t}$ are comparable, so that, depending on $\phi_2$, 
they can interfere constructively or destructively in the amplitude. 
One can have three possibilities.
Firstly, if the contribution from higgsino exchange which enters
in $\mathcal{A}(K^+ \bar \nu_{\tau})$ is smaller than the contribution from wino,
a global suppression of the channel $p \rightarrow K^+ \bar \nu$ can be achieved
\cite{dress} for certain values of $\phi_2$. 
Note that, in this situation, the role of $\phi_1$ is irrelevant.
Secondly, if also $a^R_{\tilde t^c}$ is of the same order of 
$a^L_{\tilde c}, a^L _{\tilde t}$, also $\phi_1$, one can never achieve at the same time 
a suppression of both $\mathcal{A}(K^+ \bar \nu_{\mu})$ and $\mathcal{A}(K^+ \bar \nu_{\tau})$.
Summing their corresponding rates, there isn't a sizeable variation 
of the total rate for $p \rightarrow K^+ \bar \nu$ over the whole region of the space
spanned by $(\phi_1,\phi_2)$.
The third and last possibility is that $a^R_{\tilde t^c}$ dominates over the other contributions,
in which case any effect of interference is totally lost. However this situation would
require very large values of $\tan \beta$, leading to unacceptably fast proton decay.

In the first case we obtained a suppression of the main decay channel
$p \rightarrow K^+ \bar \nu$. What about the channel $p \rightarrow \pi^+ \bar \nu$, for which
the rate is in general expected to be just a factor of 2 smaller? In the first case, for
those $\phi_2$ values which produce destructive interference in the $K^+ \bar \nu$ rate,
it becomes the dominant channel.
In the second case, instead, it is always subdominant \cite{goto}.

\subsubsection{The phases in generic models} 

When constructing realistic $SU(5)$ models which break the relation $y_d=y_e^T$, 
then, in general, $\hat A, \hat B, ...$ are no more linked to fermion masses. 
However, it is reasonable to expect their entries to be of the same order
of magnitude, that is:
\beq
- 2 \hat A\sim \hat B \sim y_u~~~,~~~~~- \hat C \sim \hat D \sim y_d \sim y_e^T~~~~.
\eeq
In this situation one again finds the same results of the minimal
case. In fact, the dominant amplitudes can be still written as
\bea
\mathcal{A}(K^+ \bar \nu_{\mu})& \propto &( b^L_{\tilde c} 
+ b^L _{\tilde t} )\nn\\
\mathcal{A}(K^+ \bar \nu_{\tau})& \propto & ( b^L_{\tilde c} 
+ b^L _{\tilde t}+b^R_{\tilde t^c} )~~~~~~~,
\eea
where $b^L_{\tilde c}, b^L _{\tilde t}$ and $b^R_{\tilde t^c}$ are complex numbers
which contain the piece of the amplitude coming respectively from $\tilde c$
and $\tilde t$ for wino dressing and from $\tilde t^c$ for higgsino dressing.
Thus, all the comments made when discussing eq. (\ref{phmin}) still apply.
For instance, when $b^R_{\tilde t^c}$ is not important, interference occurs
depending on the relative phase between $b^L_{\tilde c}, b^L _{\tilde t}$.

\subsection{Playing with SUSY Particles}

In the proton decay amplitude the loop function $f$, defined in eq. (\ref{floop}), 
plays an important role. It is obtained by dressing the dimension 5 
operators in order to convert them into four-fermions operators. 
This function depends on the masses of SUSY 
particles which run in the loop.
To have an idea of the range of values in which $f$ can lie,
it is not a too bad approximation to assume that all squarks and sleptons
are degenerate, with common mass $m$. In this case, the loop function
for the exchange of the chargino $\tilde{c}$ of mass $m_{\tilde{c}}$ greatly simplifies
\beq
f_{\tilde{c}}=\frac{1}{16 \pi^2} \frac{m_{\tilde{c}}}{(m^2-m^2_{\tilde{c}})}\left[
m^2 - m^2_{\tilde{c}} -m^2_{\tilde{c}}  \log
\frac{m^2}{m^2_{\tilde{c}}}\right]~~~.
\eeq 
The behaviour of $f$ is the following:
by fixing $m$, it grows until $2 m_{\tilde{c} }$ and then decreases;
by fixing $m_{\tilde{c} }$, it grows when $m$ decreases. 
In the limit $m_{\tilde{c} } \ll m$, then 
\beq
 f_{\tilde{c}} \propto \frac{m_{\tilde{c}} }{ m^2}~~~~.
\eeq
This means that if sfermions are much more heavy than charginos,
proton decay rates get greatly suppressed. However, it is not
possible to push $m$ towards values greater then $1$ TeV, if one 
wants SUSY to be the solution of the hierarchy problem.

The hypothesis of degeneracy (or nearly degeneracy) of sfermions is useful 
because the various $f$ can be factorized, that is, summing over $\bar \nu$ flavours, 
one can symbolically write:
\beq
\Gamma (p \rightarrow K^+ \bar \nu) \propto |f_{\tilde{w}} ~a_L +f_{\tilde{h}} ~ a_R|^2
\eeq
where $a_{L,R}$ represent the contribution to amplitude 
coming respectively from wino and higgsino dressing. 
If $m_{\tilde{w}}=m_{\tilde{h}}$ the $a_L$ contribution is the dominant one.
Dominance of $a_R$ occurs only if $m_{\tilde{w}}\ll m_{\tilde{h}}$ and/or $\tan \beta$
is large.  
To quantify the uncertainty on $\tau_p $ induced by the SUSY spectrum,
one has to study the behaviour of $f$ for reasonable values of $m_{\tilde w, \tilde h}$ 
and sfermion masses. Considering typical SUGRA derived spectra with values of the soft
breaking parameters in the range from $100$ GeV to $1$ TeV, one finds a variation 
of about one order of magnitude in rates.  

%To evaluate the loop function $f$
%we also need the spectrum of the supersymmetric particles. As an example
%we take here the same spectrum considered in table 1, with $m_{SUSY}=250$ GeV.
%This leads to a squark mass of about $800$ GeV, a slepton mass of approximately $350$ GeV
% a wino mass of 250 GeV and a charged higgsino mass of 125 GeV.

\subsection{Numerical Examples}
\label{numexam}
The previous observations are clarified by the following example, where
we compare two different set of textures for fermion masses which
lead to exactly equal phenomenology, that is they both give the same 
masses and the same $V_{CKM}$.

The first set 
\footnote{In our convention Dirac mass terms are given by ${\bar L} m R$.}
 is the well known Georgi-Jarlskog proposal \cite{g&j}
\beq
\begin{array}{lr}
y_u^{GJ}=\left[\matrix{ 0& a \lambda^6 & 0 \cr
  a  \lambda^6 & 0 & b \lambda^2 \cr
   0 &  b \lambda^2  & 1 } \right]~~~~,&
y_d^{GJ}=\left[\matrix{
0 & g \lambda^3 &0 \cr
g' \lambda^3 & j \lambda^2 & 0 \cr
0 & 0 & 1    } \right]~.\\
& \\ \end{array}\eeq
For $\tan \beta =3$ and with $\lambda =0.23$, 
the values $a=0.799$, $b=0.813$ and $g=0.005+0.372 i$, $g'=0.004-0.285 i$, $j=0.387$
give: $(m_u,m_c,m_t)=(0.0012,0.28,154)$ GeV, $(m_d,m_s,m_b)=(0.0009,0.24,1.17)$ GeV,
$|V_{u d}|=0.22$, $|V_{u b}|=0.027$, $|V_{c b}|=0.043$ and the Jarlskog invariant 
$I_G=2.2 \cdot 10^{-5}$.
These GUT scale values, when run to low energies, are compatible with the
measured fermion masses and mixings.
The second set is obtained by replacing $y_d^{GJ}$ with
the following matrix possessing a large right-handed mixing
in the 23 plane:
\beq
\begin{array}{lr}
y_d^{HO}=\left[\matrix{ 0& A \lambda^3 & 0 \cr
  A^*  \lambda^3 & 0 & B \lambda^2 \cr
   0 &  1 & 1 } \right]~~~~\end{array}~~~.\eeq
This particular form for the down quark mass matrix has been studied in detail by 
Hagiwara and Okamura \cite{hagi}.
For  $\tan \beta =3$ and $A=0.274+0.475 i$, $B=- 0.217-0.711 i$, one obtains
exactly the previously quoted values of masses and mixings.
These two set, due to the different values of CP violating phases,
give different predictions for proton lifetime.
Keeping $m_T=2 \cdot 10^{16}$ GeV, $\alpha=- \beta=  0.003$ GeV$^3$,
$m_{\tilde w}=250$ GeV, $ m_{\tilde h}=125$
GeV and assuming sfermions to be degenerate with $m_{\tilde {sf}}=837$ GeV,
then one predicts the values shown in Table \ref{comp}.

\begin{table}[h] \begin{center}
\begin{tabular} {|c|c|c|}
\hline & &\\ & $\tau_p/B(p \rightarrow K^+ \bar \nu)$ & $\tau_p/B(p \rightarrow \pi^+ \bar \nu)$\\
 &$10^{31} yr$ & $10^{31} yr$ \\ & &\\ \hline & &\\
 1$^{st}$ set (GJ) & $5.6$ & $2.8$ \\
& & \\
 2$^{nd}$ set (HO)& $0.7$ & $1.2$ \\
& &\\
\hline
\end{tabular}
\end{center}
\caption{Comparison of the predictions for proton lifetime between
the two set of mass matrices.}
\label{comp}
\end{table}

The values of the various parameters adopted here are suitable for the case of 
minimal $SU(5)$. As pointed out, they suffer from large theoretical uncertainties.
However one can recognize that minimal $SU(5)$ is in serious troubles, predicting 
a mean value for proton lifetime which is smaller then the present bound by
two or three orders of magnitude.

In addition, one can also conclude that, even in the best constrained case of minimal 
$SU(5)$, results depend on the textures utilized, even if they give completely
equivalent low energy phenomenology. The predictions differ for about
one order of magnitude and also the main channel is different. 
This discrepancy has to be attributed entirely to the two CP violating phases.  
In fact, it is possible to recognize that the first case is an example of destructive 
interference for the amplitude of $K^+ \bar \nu$ (see eq. (\ref{phmin})). 
The dominant mode is then $\pi^+ \bar \nu $. 
In the second case, where no destructive interference occurs,
proton lifetime is smaller by about one order of magnitude and 
the mode $\pi^+ \bar \nu $ is slightly subdominant.
The result is that an uncertainty of about one order of magnitude arises
just from the ignorance of the input textures.

One can wonder about the effect on rates induced by taking appropriate
textures for $y_e$, and not just $y_d^T$, as required by minimal $SU(5)$.
As will be discussed in the next section, one simple and well known possibility
is to assume that certain elements of these matrices arise from a $\bar{45}$ of Higgses
and not from a $\bar 5$. In the former case, the element of $y_e$ will
be multiplied by a factor $- 3$ with respect to the corresponding one of $y_d^T$. 
The matrices $\hat C, \hat D$, arising from the interaction of the $\bar 5$ with
matter, will have zero in the corresponding entry. 
For the first set, Georgi and Jarlskog pointed out that realistic Yukawa couplings 
for charged leptons can be obtained if the (22) element comes entirely from a $\bar{45}$,
so that $m_e=1/3 ~m_d, m_{\mu}=3 m_s, m_{\tau}=m_b$.
For the second set, these relations can be achieved if ${y_d}_{2,3}$ comes from
a $\bar{45}$:
\beq
\begin{array}{lr}
y_e^{GJ}=\left[\matrix{
0 & g' \lambda^3 &0 \cr
g \lambda^3 & -3 j \lambda^2 & 0 \cr
0 & 0 & 1    } \right]~~~~~~~,&
y_e^{HO}=\left[\matrix{ 0& A^* \lambda^3 & 0 \cr
  A  \lambda^3 & 0 & 1 \cr
   0 &-3 B \lambda^2  & 1 } \right]~~~~. \\ \end{array}\eeq
Correspondingly, ${\hat C}_{2 2}^{GJ}={\hat D}_{2 2}^{GJ}=0$ and
${\hat C}_{2 3}^{HO}={\hat D}_{2 3}^{HO}=0$.
It easy to adapt the calculation of proton decay lifetime to this particular case.
Under the same assumptions made to obtain Table \ref{comp} one now obtains Table
\ref{compbis}.

\begin{table}[h] \begin{center}
\begin{tabular} {|c|c|c|}
\hline & &\\ & $\tau_p/B(p \rightarrow K^+ \bar \nu)$ & $\tau_p/B(p \rightarrow \pi^+ \bar \nu)$\\
 &$10^{31} yr$ & $10^{31} yr$ \\ & &\\ \hline & &\\
 1$^{st}$ set (GJ) & $5.8$ & $2.9$ \\
& & \\
 2$^{nd}$ set (HO)& $0.8$ & $1.5$ \\
& &\\
\hline
\end{tabular}
\end{center}
\caption{Comparison of the predictions for proton lifetime between
the two set of mass matrices in the realistic case.}
\label{compbis}
\end{table}
The results are very similar to those of the minimal case.
Indeed this is a simple and very particular example, 
however, it is true also in general that the use of simple realistic 
textures also for charged leptons doesn't have very sizeable effects
on proton lifetime. Achieving a certain degree of suppression through a differentiation 
between triplet and doublet Yukawa couplings is a technically realizable possibility
only for much more complicated models, where a number of ad hoc 
assumptions must be made \cite{inserz}.
% btv e nath
It is clear that one is interested in looking for such mechanisms that lead
to construct non minimal models where
proton lifetime can be pushed comfortably
above experimental bounds. As pointed out, a resolutive way to do this,
would be to have triplets much more heavy that in the minimal case.
In the next chapter we will discuss
one of these models. 

%%%%%%%%%%%%%%%%%%%%%%%%%%%%%%%%%%%%%%

\section{Fermion Masses and Mixings}
\label{secfmm}

The aesthetically attractive fact of putting in the same representation conceptually 
different objects like quarks and leptons has mainly two consequences.
The most apparent one is the fact that quarks and leptons mass matrices turn out
to be strictly correlated. Despite the difficulties found, this can give 
interesting starting points to face the flavour problem. 
The second consequence is the possibility of violating baryon and
lepton numbers. As shown, this leads to the striking prediction of proton lifetime,
with values in the region already partially excluded by present bounds. 
Minimal models fail in correctly predicting both mass relations and proton lifetime.

In the following we will review the usual methods which can be adopted to
share some light on the resolution of the first problem. It is clear that
finally one is interested in understanding if these mechanisms are compatible
or even help in the solution of the other problems like proton decay and the
other typical GUT problems. 
  
\subsection{Mass Relations}
\label{massrel}
The evident problem of minimal models is the prediction of wrong mass relation 
for down quarks and charged leptons at $M_{GUT}$. 
Assuming an exact $R$-parity discrete symmetry under which matter fields are
odd and $H$, $\overline{H}$ are even,  
if the superpotential contains the following Yukawa interaction terms
\begin{equation}
w_{Yuk}=\frac{1}{4}  \Psi_{10}^{A B}~ G_u~ \Psi_{10}^{C D}~H^E \epsilon_{A B C D E}
+\sqrt{2}~ \Psi_{10}^{A B}~ G_d~ {\Psi_{\bar 5}}_A~ {\overline H}_B
\end{equation}
where $A,B,...=1,...,5$, $H$, $\overline H$ are the pentaplets which contain the pair 
of light doublets and $G_u, G_d$ are $3\times 3$ matrices in flavour space.
Writing $w_{Yuk}$ in terms of MSSM multiplets, the interactions with 
doublets read:
\beq 
w_{Yuk} \ni  Q^T G_u U^c H_{2u} +Q^T G_d D^c H_{2d} + L^T G_e E^c H_{2d} ~~~.
\eeq
Then we get the following Dirac mass matrices at $M_{GUT}$:
\begin{equation}
{\hat m}_{u}=G_{u}~ \frac{v_u}{\sqrt{2}}~~~,~~~~~{\hat m}_{e}^T={\hat m}_{d}=G_{d}~ \frac{v_d}{\sqrt{2}}~~~~~,
\end{equation}
where $v_u$, $v_d$ parameterize the vevs of $H$, ${\overline H}$.
The sharp transposition relation ${\hat m}_{e}^T={\hat m}_{d}$ is in conflict with what is expected
on phenomenological grounds, by evolving the values of down quarks and charged lepton
masses from their values measured at the electroweak scale to the GU scale, i.e. 
the Georgi-Jarlskog relations:
\begin{equation}
m_e \cong \frac{1}{3} m_d~~~,~~~~~m_{\mu} \cong 3 m_s~~~,~~~~~m_{\tau} \cong m_b~~~.
\end{equation}
This problem is even exacerbated in minimal $SO(10)$,
where ${\hat m}_u={\hat m}_u^T={\hat m}_{\nu D} \propto 
{\hat m}_d={\hat m}_d^T={\hat m}_e$.
However, at least in $SO(10)$ there is naturally the place for right-handed
neutrinos. 

On the contrary, in $SU(5)$ the inclusion of neutrino masses
represents an additional problem. 
A very simple way out is the addition of a gauge singlet field $\Psi_1$, which contains
right-handed neutrinos. 
If one views $SU(5)$ as descending from $SO(10)$, this is of course not only natural
but also necessary.
Then, in the superpotential these terms are allowed
\begin{equation}
\Psi_1~  G_\nu~ {\Psi_{\bar 5}}_A~ H^A
- \frac{1}{2}~ M~ \Psi_1~ G_M~ \Psi_1
\end{equation}
which realize the see-saw mechanism. If $M$ is just a bit smaller than $M_{GUT}$, then
the atmospheric neutrino mass scale is naturally obtained. 
However one should emphasize that, because the term which gives mass to right-handed
neutrinos doesn't break $SU(5)$, then $M$ should in principle be expected of order
of the cut-off of the theory, that could even be as large as $M_{Pl}$.  
The advantage of groups like $SO(10), SU(4) \otimes SU(2) \otimes SU(2)$ is that they 
directly possess a particle with the same quantum numbers of right-handed neutrinos, 
whose mass scale thus results for dynamical reasons of order of the GUT scale. 

\subsection{How to Achieve Realistic Spectra}

The flavour problem consists in explaining the origin of the 
difference between the three fermionic families, difference which
causes their masses and mixings to be 
organized in strongly and bizarre hierarchical structures.
Probably this problem is linked to the more fundamental one: why just three
families?
It is clear that the idea of GU cannot alone answer these deep questions.
The only thing it can offer is a much more constrained framework with respect to the
MSSM, where one can look for the criterion according to which the different mass 
matrices are linked one to each other. 
One can thus learn a lot about what would be allowed to be present at scales higher
then the electroweak one and what could not at all be there.

To understand the situation it is convenient to take $SU(5)$ as reference.
This is not at all restrictive because $SO(10)$ can be always thought in terms
of its $SU(5)$ content. The problems one has to face are two:
firstly a mechanism to produce a progressive suppression in the masses of the three
generation, secondly a way to reproduce the correct values of masses and mixings.
The usual road taken to overcome the first problem is by the use of flavour
symmetries \cite{fro}, for instance a continuous $U(1)$. This strategy
is very successful, even if one is left with the discomfort of having done
ad hoc assumptions for the charges.  
Apart from this, it turns out that it is very difficult to realize
the correct spectra without tuning some parameter, even in the more complicated
extensions of minimal GUT models.

Possible ways for accounting for neutrino masses and mixings have been already
discussed. Here we will review the main strategies one can use to break the
bad transposition relation between down quarks and charged leptons.
Since $10 \times \bar 5 = 5 + 45$, a mass term for quarks and leptons can be
realized only with a $\bar 5$ and a $\bar{45}$. The $\bar 5$, which is of course nothing
that the pentaplet containing one of the light Higgs doublets, predicts the sharp
transposition relation. So, to have a realistic spectrum there is the absolute need for
a $\bar{45}$, which succeeds in breaking the unwanted transposition.
The $\bar{45}$ can be a "true" or an effective one.
The first scenario, with a true $\bar{45}$, was proposed a long time ago.
Now it seems not very attractive because the true $\bar{45}$ contains
a true doublet, which needs to be massive. 
One can then introduce another $45$ to construct a direct mass term (and also
to canceling anomalies). 
But it is well known that the two additional doublets cannot be light. If this were
the case then the prediction for $\alpha_3 (m_Z)$ assuming coupling unification
would be given by:
\beq
\alpha_3 (m_Z)=\alpha_{em} (m_Z) \frac{56-2 \delta n_H}{(120+6 \delta n_H)sin^2 \theta (m_Z)
-(24+3\delta n_H)}
\eeq
where $\delta n_H$ is the number of light extra doublets, in addition to the two of the MSSM.
Taking $\alpha_{em} (m_Z)=1/128$ and $sin^2 \theta (m_Z)=0.23$, one has:
\bea
\alpha_3 (m_Z)&=&0.12~~~~~~~~~~~~~~~~~~~~~~~~~\delta n_H=0~~~~\rm{(MSSM)}\\
\alpha_3 (m_Z)&=&1.13~~~~~~~~~~~~~~~~~~~~~~~~~\delta n_H=2~~~~~~~~.
\eea
It is then absolutely unavoidable: to give a GUT scale mass to the four triplets;
to maintain at the electroweak scale two linear combinations
of the doublets in $\bar 5,\bar{45}$ and $5, 45$ but raising at the GUT scale 
their two orthogonal combinations. 
Here come the difficulties, because this asymmetric splitting between doublets
and triplets is a much more complicated version of the usual doublet triplet splitting
problem.

To escape from these difficulties one can make complete use of the possibility
of working with non renormalizable operators and explore the field offered by 
the effective $\bar{45}$. The simplest possible mechanism is the well known
proposal made by Ellis and Gaillard \cite{elg} of considering the $\bar{45}$ of the
product $\bar 5 \times 24=\bar 5+\bar{45}+\bar{70} $. This is indeed a very economical 
solution, because the $\bar 5$ is already there and the $24$, which must have non vanishing
vev, can be naturally the one responsible for the breaking of $SU(5)$.
Thus, this mechanism uses just the field content of minimal $SU(5)$.
The following operator
\beq
\sqrt{2}~ {\Psi_{10}^{A B}}_{i}~ {G_d}_{ij}~ {{\Psi_{\bar 5}}_C}_{j}~ {\overline H}_B
\frac{ \langle {\Sigma}^C_A  \rangle}{\Lambda}
\eeq
where $\Lambda$ is the cut-off, gives ${G_d}_{i j}={y_d}_{i j}=- 2/3 {y_e}_{j i}$.
Note that both channels $\bar 5$ and $\bar{45}$ of the product $\bar 5 \times 24$
contribute to the mass.  
Also operators with more insertions of $\Sigma$ can be useful in realistic examples.
The number of the invariants obtained by contracting the indices in all possible
ways grows with the number of insertions made, so that one can use the freedom
in choosing their relative coefficient to fit the masses. This is however
a kind of proof that realistic mass spectra are realizable rather than a genuine
prediction of the model.

There are other possibilities, in addition to the use of a $24$.
In the MDM, for instance, the $75$ responsible for the breaking of $SU(5)$
can be successfully well employed, without the introduction of additional fields.
In fact, since $\bar 5 \times 75 =\bar {45}+\bar{50}+\bar{280}$, in this case only the 
pure $\bar{45}$ channel contributes. For instance
\beq
\sqrt{2}~ {\Psi_{10}^{A B}}_i~ {G_d}_{ij}~ {{\Psi_{\bar 5}}_C}_j~ {\overline H}_D
\frac{ \langle Y^{C D}_{A B}  \rangle}{\Lambda}
\eeq
gives ${G_d}_{i j}={y_d}_{i j}= - 1/3 {y_e}_{j i}$.
In the next chapter the strategy of inserting the $75$ to obtain a phenomenologically
acceptable spectrum of fermions will be employed in the context of a realistic
version of the MDM.

%%%%%%%%%%%%%%%%%%%%%%%%%%%%%%%%%%%%%%%%%%%%%%%%
\section{Coupling Unification}

The most outstanding support to the idea of Grand Unification comes from the quantitative 
success of coupling unification in SUSY GUTs \cite{sgut}. 
Assuming just the field content of the MSSM (with two light Higgses), in the leading 
order approximation (LO), that is at one loop, the RGE for
the running coupling measured at $m_Z$ read:

\beq
\frac{1}{  {\alpha}_i  ( Q ) }=\frac{1}{\alpha_i (m_Z)}+\frac{b_i}{2 \pi} \ln{ \left( \frac{Q}{m_Z}\right)}
\label{1loop}
\eeq

with
\beq
b_1 =\frac{33}{5}~~~~,~~~~~b_2=1~~~~,~~~~~b_3=-3~~~.
\eeq

\begin{figure}[p]
\begin{center}
{\bf \Large SUSY Standard Model}
\end{center}
\vspace*{1.cm}
\centerline{\psfig{file=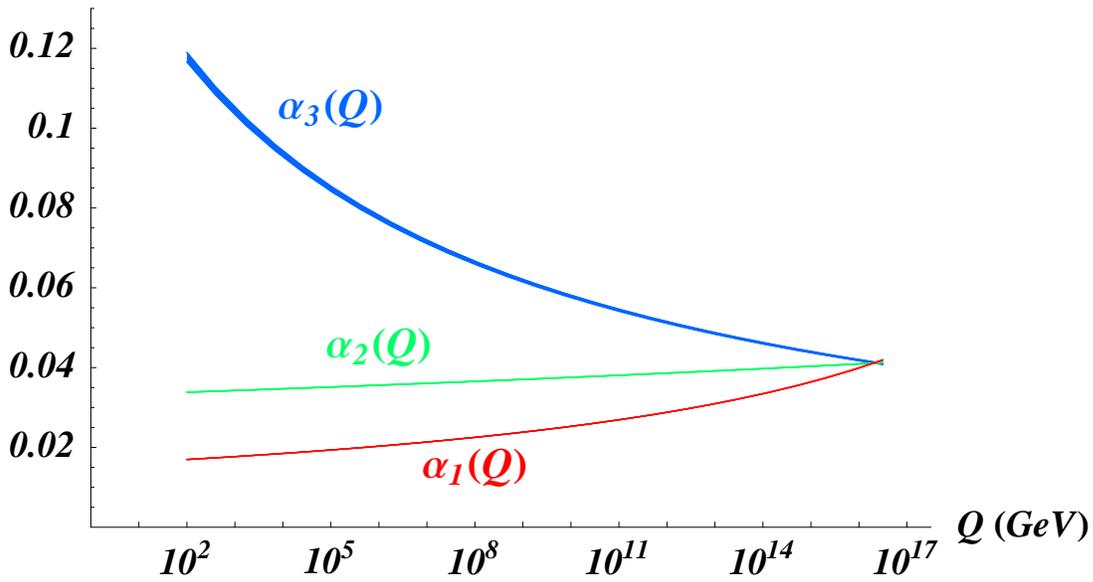,width=1.1\textwidth}}
\caption{Evolution of the running gauge couplings within the MSSM. 
The curves are obtained in the 1 loop approximation by evolving the 
gauge couplings from their measured values at $m_Z$ \cite{PDG}: 
${\alpha}_1(m_Z)=0.016945 \pm 0.000007$, 
${\alpha}_2(m_Z)=0.033813 \pm 0.00003$ 
(from $\alpha_{em} (m_Z)=127.934 \pm 0.027$, $\sin^2{\theta}(m_Z)=0.23117 \pm 0.00016$,
$m_Z=91.1872 \pm 0.0021$ GeV) and  
${\alpha}_3(m_Z)=0.1184 \pm 0.0012$.}
\label{runmssm}
\end{figure}

Nowadays it is possible, thanks to the quite precise 
measurements of the three running gauge coupling at $m_Z$, to single out a not 
too broad range for the scale of SUSY unification $M_{GUT}$. 
Fig. \ref{runmssm} shows the flow of the gauge coupling in the one loop approximation
starting at $m_Z$ with the PDG data of last summer \cite{PDG}
, that is, in the $\overline{MS}$ scheme:  
$\alpha_{em} (m_Z)=127.934 \pm 0.027$, $\sin^2{\theta}(m_Z)=0.23117 \pm 0.00016$,
$m_Z=91.1872 \pm 0.0021$ GeV and the world average 
$\alpha_3 (m_Z)=0.1184 \pm 0.0012$.
As can be clearly seen, unification is very successful and the associated scale  
$M_{GUT}$ turns out to be in the small range 
$(1.9 \div 2.1) \cdot 10^{16}$ GeV, with the previously quoted errors. 

\begin{figure}[p]
\begin{center}
{\bf \Large Standard Model}
\end{center}
\vspace*{1.cm}
\centerline{\psfig{file=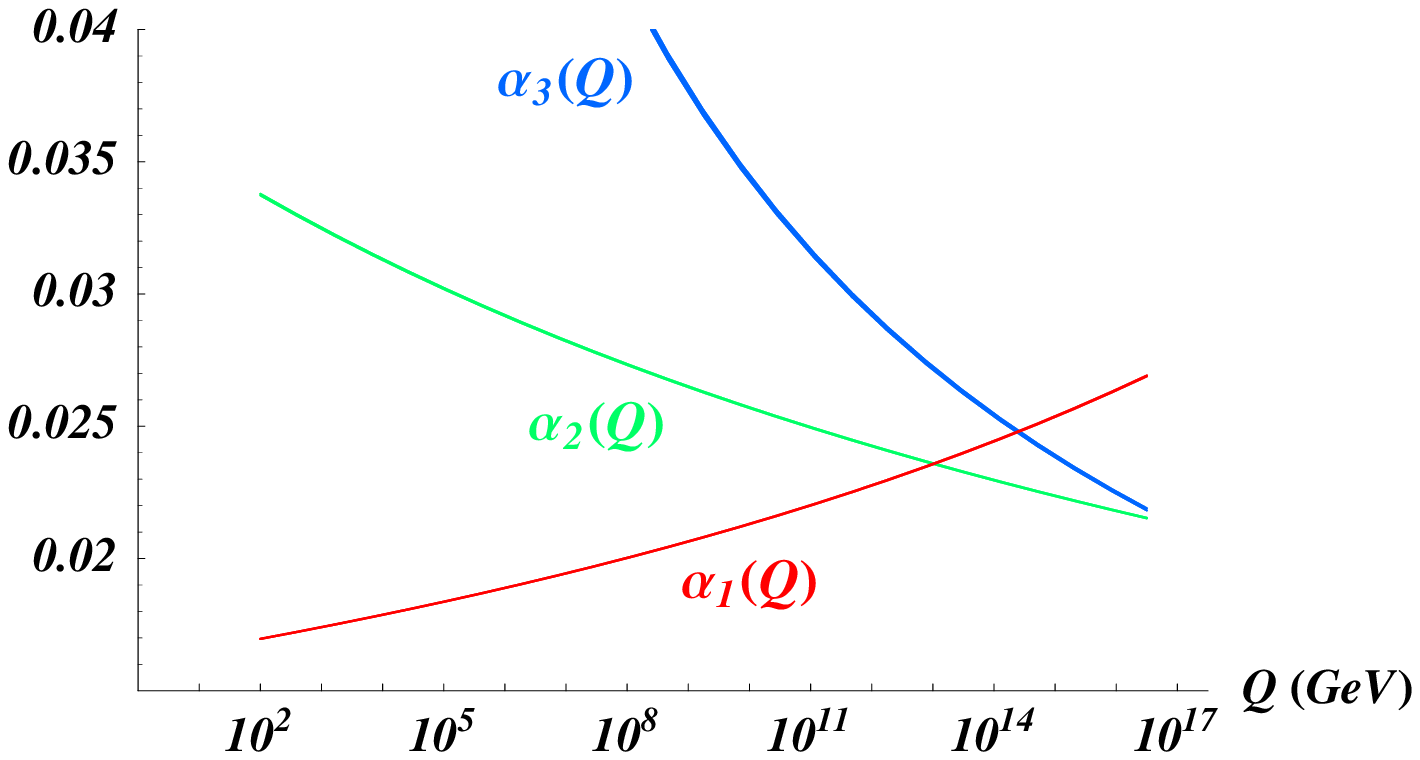,width=1.1\textwidth}}
\caption{The same as fig. \ref{runmssm} but within the SM.}
\label{runms}
\end{figure}

Now it is also well established that GUT without SUSY is neatly disfavoured, 
as appears clearly from the comparison of figs. {\ref{runmssm}} and {\ref{runms}}.
For the SM equations analogous to (\ref{1loop}) hold, but with 
$b_1 =41/10$, $b_2=-19/6$, $b_3=-7$.

Going beyond LO turns out to be necessary because the corrections from two
loop running on the prediction for $\alpha_3 (m_Z)$ are of the order of 10\%. 
It is reasonable however to disregard three
loop running, which would give correction of only about 1\% on $\alpha_3 (m_Z)$.
In the following we will describe the calculations relative to the two loop prediction 
of $\alpha_3 (m_Z)$ starting from $\alpha_1 (m_Z), \alpha_2 (m_Z)$ and assuming 
coupling unification. We will give numerical results for the case of minimal $SU(5)$.
They will be useful for comparison with the analogous predictions obtained in the MDM
case, whose analysis will be present in detail in the next chapter.

Different GUT models give different predictions for $\alpha_3 (m_Z)$ because
the contribution to its running coming from threshold effects is sizeable.
If one is convinced that the success of coupling unification in the LO approximation
is not just a coincidence, then, in model building, one is constrained to 
introduce as few as possible GUT fields, which in addition have to lie in not too large 
representations.
It is clear that working with a lot of field is equivalent to have a lot of degrees
of freedom at disposal, which can be arranged ad hoc to reproduce the data.
The elegance of the model is completely lost, but at least one possess a sort
of existence proof of the realizability of low energy data starting from a GUT 
scenario. Working with few fields, instead, it is much more hard to give
a completely satisfactory picture of particle phenomenology. However,
this latter point of view is the only one from which a true deeper understanding
of fundamental constituents could come.
%, since physics is much more simple thanwe can suspect.

\subsection{Coupling Unification Beyond the Leading Order Approximation}

It is well known that, going beyond the LO approximation, in the minimal version 
of SUSY $SU(5)$ the central value of $\alpha_3 (m_Z)$ required by the
constraint of coupling unification is somewhat large: 
$\alpha_3 (m_Z)\approx 0.13$ \cite{yam,bag,cla,lp,cpp}. 

%1loop
From the one-loop renormalization group
evolution of the couplings in the MSSM with only two light Higgs doublets, eqs. (\ref{1loop}), 
by evolving two couplings from their values measured at $m_Z$ and imposing
the constraint of unification, one can run back to low energies with the third coupling 
thus giving a prediction for its value at $m_Z$.
In this procedure it is preferable to take as input the two best measured couplings,
that is $\alpha_1 (m_Z), \alpha_2 (m_Z)$. 
In fact,  defining in the ${\overline{MS}}$ scheme 
$\alpha \equiv \alpha_{QED}(m_Z)$, 
$\sin^2{\theta}\equiv \sin^2{\theta}(m_Z)$,
\beq
\alpha_1 (m_Z) =\frac{ \alpha}{1-\sin^2{\theta}} ~~~,~~~\alpha_2 (m_Z) =\frac{ \alpha}{\sin^2{\theta}}
\eeq
and the predictions are:
\bea
\ln{(M^{(0)}_{GUT}/m_Z)}&=&\frac{\pi (3-8\sin^2{\theta})}{14 \alpha}\nonumber\\
a_{5}^{(0)}&=&\frac{28\alpha}{36\sin^2{\theta}-3}\nonumber\\
a_{3}^{(0)}&=&\frac{28\alpha}{60\sin^2{\theta}-12}
\label{loeq}
\eea
where $a_3\equiv \alpha_3(m_Z)$, $a_5\equiv \alpha_5 (M_{GUT})$ and the superscript 
$(0)$ refers to a quantity evaluated in the LO approximation. 
It is worth to stress that the LO results are the same in
minimal SUSY $SU(5)$ and in all its extensions where the MSSM $\beta$ functions 
apply until $M_{GUT}$. 

By using
%sostituire!
$\alpha^{-1}=127.934$ and $\sin^2{\theta}=0.231$ one 
obtains $a_{3}^{(0)}=0.118$ and 
$M^{(0)}_{GUT}=2.1~10^{16}$ GeV.

%beyond LO
To go beyond the LO one must include two loop effects in the running of gauge couplings, 
threshold effects at the scale $m_{SUSY}$, close to the electroweak scale, 
and threshold effects at the large scale $M_{GUT}$. 
The unification scale $M_{GUT}$ is not univocally defined beyond LO and here 
we choose to identify it with the mass of the $SU(5)$ superheavy gauge bosons $m_V$. 
%The results derived in ref. \cite{yam} can be written in the following form:
Following ref. \cite{yam} it is possible to include these effects by writing:
\beq
\frac{1}{  {\alpha}_i  ( m_Z ) }=\frac{1}{a_5}+\frac{b_i}{2 \pi} 
\ln{ \left( \frac{M_{GUT}}{m_Z}\right)}
+\frac{\delta_i}{\pi}
\eeq
where $a_5=\alpha_5 (M_{GUT})$ and the next to leading corrections are contained in 
$\delta_i$, $i=1, 2, 3$ and can be separated according to their origin. 
The logarithmic contribution coming from two loop running, $\delta_i (2)$, and the 
contributions of SUSY and GUT thresholds,  respectively $\delta_i (l)$ and $\delta_i (h) $:
\beq
\delta_i=\delta_i (2)+\delta_i (l)+\delta_i (h)
\label{deltai}
\eeq

The LO eqs. (\ref{loeq}) are thus improved by:
\bea
\ln{\left( \frac{M_{GUT}}{m_Z}\right) }&=&\ln{ \left(    \frac{M_{GUT}^{(0)} }{m_Z} \right) }+\delta_X
\nn\\
\frac{1}{a_5}&=&\frac{1}{a_5^{(0)}}+\delta_5\nn\\
a_3&=&\frac{a_{3}^{(0)}}{(1+a_{3}^{(0)}\delta_3)}
\label{impr}
\eea
where the various $\delta_{X,5,3}$ are linear combinations of $\delta_i$:
\bea
\delta_X &=& \frac{10}{28} (\delta_2-\delta_1)~~~\nn\\
\delta_5 &=& -\frac{1}{28} (33 \delta_2- 5 \delta_1)~~~\nn\\
\delta_3 &=& \frac{1}{7 \pi} (5 \delta_1- 12 \delta_2 +7 \delta_3)~~~
\eea

In order to obtain the prediction for $a_3$ it is necessary to specify
the various contributions which enters $\delta_3 = \delta_3 (2) +\delta_3 (l)+ \delta_3 (h)$.

\subsubsection{Two-loop running}

The one from two-loop running is finite and given by:
\beq
\delta_i (2)=\sum_{j=1}^{3} \frac{b_{i j}}{2 b_j} 
\ln{ \left(   1+\frac{b_j}{2 \pi} a_5^{(0)} \ln{\left(  \frac{M_{GUT}^{(0)}}{m_Z}\right)} \right)}
\eeq 
where
\beq
b_{i j}=\left[\begin{array}{ccc}
 199/50& 27/10 & 44/5 \\
  9/10 & 25/2   &  12 \\
   11/10 & 9/2 & 7  
\end{array} \right]
\eeq

In particular 
\beq \delta_3 (2)=-0.823~~~~. \label{delta32} \eeq

\subsubsection{SUSY Thresholds}

The one loop thresholds from SUSY particles read: 
\beq
\delta_i (l) = -\sum_{j} b_i^{(l)} (j)  \ln{ \left( \frac{m_{j}}{m_Z} \right)} 
\eeq
where $j$ runs over all light particles of mass $m_j$ and $b_i^{(l)} (j)$ can be read from
Table \ref{sthres}.
\begin{table}[h]
{\begin{center}
\begin{tabular}{|c|c|c|c|}   
\hline
& & & \\                         
sparticle & $ b_1^{(l)} (j) $ & $ b_2^{(l)} (j)$  & $b_3^{(l)} (j) $ \\ 
& & &\\
\hline \hline
& & &\\
gluinos & $0$ & $0$ & $1$ \\
& & &\\
winos & $0$ &$2/3$ &$0$ \\ 
& & &\\
higgsinos & $1/5$ & $1/3$ & $0$\\
& & &\\
extra Higgses & $1/20$ & $1/12$ & $0$\\
& & &\\
$\widetilde{q}_L$ & $1/60$ & $1/4$ & $1/6$\\
& & &\\
$\widetilde{u}_R$ & $2/15$ & $0$ & $1/12$\\
& & & \\
$\widetilde{d}_R$ & $1/30$ & $0$ & $1/12$\\
& & & \\
$\widetilde{l}_L$ & $1/20$ & $1/12$ & $0$\\
& & & \\
$\widetilde{e}_R$ & $1/10$ & $0$ & $0$\\
& & & \\
\hline 
\end{tabular} 
\end{center}}
\caption{SUSY thresholds.}
\label{sthres}
\end{table}

To give a concrete example, one has to specify a certain SUSY spectrum. 
We will consider as representative the one given in Table {\ref{tabsusysp}},
where the additional freedom related to the parameters $m_0$, $m_{1/2}$, $\mu$, $m_H$ 
can be fixed by choosing $0.8 m_0=0.8 m_{1/2}=2 \mu=m_H$ and taking as a definition
$m_{SUSY}\equiv m_H$, so that all particle masses can be expressed in term of 
$m_{SUSY}$.

The result is then
\beq
\delta_3 (l)=-0.503+\frac{19}{28 \pi} \ln{\left(  \frac{m_{SUSY}}{m_Z} \right)}~~~.
\label{delta3l}
\eeq

\subsubsection{GUT Thresholds}

Threshold contributions from heavy particles can be expressed as:
\beq
\delta_i (h) =c_i -\sum_{j} b_i^{(j)} (h)  \ln{ \left( \frac{m_{j}}{m_{GUT}} \right)} 
\label{deltaih}
\eeq
with $c_1=0, c_2=1/6, c_3=1/4$ and the sum is over all heavy particles.

To specialize to the case of minimal $SU(5)$ one has to consider its heavy spectrum.
After the Higgs mechanism, 12 complex fields of the $24$ representation  
(the 24 would-be Goldstone bosons) are eaten by the $24_V$ of gauge vector bosons, which
thus become massive. The other 12 complex fields in the $24$ acquire instead a heavy mass.
In addition to them, there are the colour triplets of the $5$ and $\overline 5$ representation. 
Thresholds can be evaluated according to:
\begin{table}[hp] 
{\begin{center}
\begin{tabular}{|c|c|c|c|c|}   
\hline
& & & & \\                         
state $(SU(3),SU(2),U(1))$ & $ b_1^{(h)} (j) $ & $ b_2^{(h)} (j)$  & $b_3^{(h)} (j) $ & mass  \\ 
& & & &\\
\hline \hline
& & & &\\
$(8,1,0)$ & $0$ & $0$ & $3/2$ & $m_{\Sigma}$ \\
& & & &\\
$(1,3,0)$ & $0$ &$1$ &$0$ & $m_{\Sigma}$ \\ 
& & & &\\
$(1,1,0)$ & $0$ & $0$ & $0$ & $0.2 m_{\Sigma}$\\
& & & &\\
\hline 
& & & &\\
$(3,1,-1/3), (\overline 3,1,1/3) $& $1/5$ & $0$ & $1/2$ & $m_T$\\
& & & &\\
\hline 
\end{tabular} 
\end{center}}
\caption{GUT thresholds. $m_\Sigma \equiv 5 M_{24}$, with $M_{24}$ defined 
as in section \ref{23splprob}.}
\label{gutthmin}
\end{table}

For this case one obtains
\beq
\delta_3^{min} (h) =-0.0114 -\frac{9}{14 \pi} \ln{ \left(  \frac{m_T} {M_{GUT}} \right)}
+\frac{3}{14 \pi} \ln{ \left(  \frac{m_{\Sigma}} {M_{GUT}} \right)}~~~.
\label{delta3h}
\eeq

\subsubsection{Results}
\label{coupres}
The first of eqs. (\ref{impr}) turns out to be very useful.
The contributions to its corrective term $\delta_X$ coming from two-loop running and
SUSY thresholds induced by the spectrum of Table {\ref{tabsusysp}} are:
\bea
\delta_X (2)&=&0.469\\
\delta_X (l)&=&-0.036-\frac{25}{84} \ln{\left(  \frac{m_{SUSY}}{m_Z} \right)}~~.
\eea 
Thresholds from GUT particles within minimal $SU(5)$ give:
\beq
\delta_X (h)=0.060-\frac{5}{14} \ln{\left(  \frac{m_\Sigma}{M_{GUT}} \right)}
+\frac{1}{14} \ln{\left(  \frac{m_T}{M_{GUT}} \right)}~~.
\eeq
Here in minimal $SU(5)$ it turns out clearly, but it is true in general that, 
keeping fixed $m_{SUSY}$, the first of eqs. (\ref{impr}) can be considered as a relation 
between the GUT masses. In this case it relates $m_V=M_{GUT}$ to $ m_\Sigma, m_T$.

Substituting in $\delta_3 (h)$ the expression thus obtained for $M_{GUT}$,
one can always write  $\delta_3$ according to the following expression: 
\beq
\delta_3 = k+\dd\frac{1}{2 \pi}  \ln{\left(  \frac{m_{SUSY}}{m_Z} \right)} 
-\dd\frac{3}{5 \pi} \ln{\left(  \frac{m_T}{M_{GUT}^{(0)}} \right)}
\eeq
This expression for $\delta_3$ is general because we have subdivided the various 
contributions in a numerical coefficient plus logarithmic contributions. 
The logarithmic term with $m_{SUSY}$ comes from particles with masses near $m_{SUSY}$.
Similarly, the logarithmic term with $m_T$ arises from particles with masses of 
order $m_T\approx M_{GUT}$. The definition of $m_T$ is the mass of Higgs colour triplets 
in the minimal model or the effective mass that, 
in realistic models, plays the same role for proton decay \cite{yam,bag}. 
Notice that that there isn't any dependence on $m_\Sigma$.
The term indicated with $k=k(2)+k(SUSY)+k(M_{GUT})$ contains the contribution of 
two loop diagrams to the running couplings, $k(2)$, the threshold contribution from 
states near $m_{SUSY}$, $k(SUSY)$, and the threshold contribution from states near 
$M_{GUT}$, $k(M_{GUT})$. This way of subdividing the contribution is useful 
because the threshold contributions would vanish if all states had the mass $m_{SUSY}$ or 
$M_{GUT}$ so that only mass splittings contribute to $k(SUSY)$ and $k(M_{GUT})$.

\begin{figure}[p]
\centerline{\psfig{file=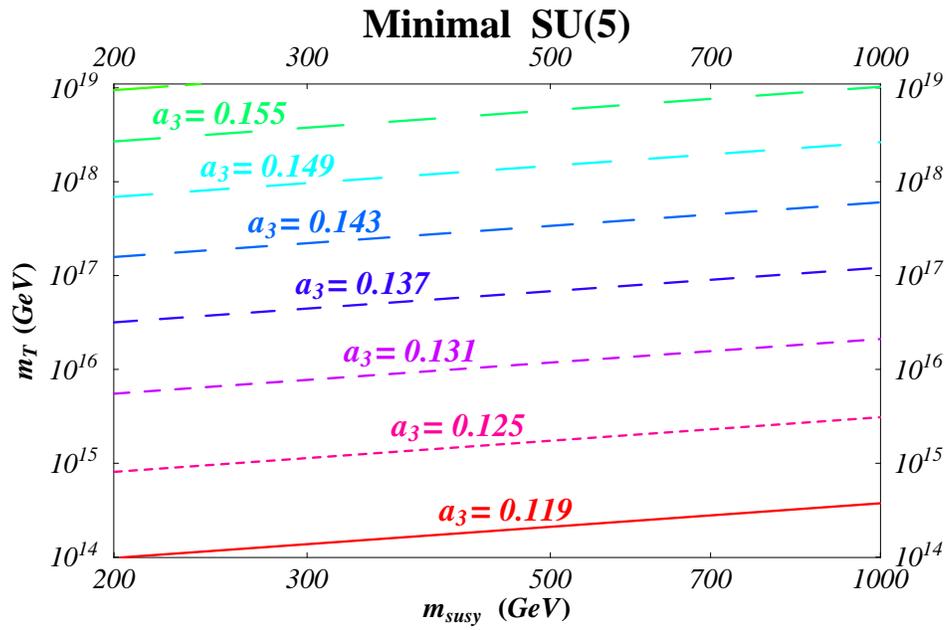,width=1.1\textwidth}}
\caption{Contours of $a_{3}\equiv \alpha_{s}(m_Z)$ in the plane 
$(m_{SUSY},m_T)$, 
for minimal $SU(5)$. The SUSY spectrum is parametrized as in 
Table \ref{tabsusysp}.}
\label{alpha3min}
\end{figure}

The values of $k(2)$ and of $k(SUSY)$ are the same in minimal $SU(5)$ and in
its realistic extensions, because only the heavy spectrum changes. Typical values are 
$k(2)=-0.733$ and $k(SUSY)=-0.510$. 
The value for $k(SUSY)=-0.510$ corresponds to the representative spectrum displayed 
in Table \ref{tabsusysp}.
Note that the value of $k(M_{GUT})$ is practically zero for the $24$ of the minimal model. 
The reason is that in minimal $SU(5)$ all coloured GUT multiplets are degenerate, 
as appears clearly giving a look to Table \ref{gutthmin}.
The result is thus:
\beq
k=-0.733-0.510=-1.243~~~~~~~~\rm{minimal~ model}
%k=-0.733-0.510+1.857&=&0.614~~~~~~~~~~~\rm{realistic~ model}\label{k}
\eeq

In fig. {\ref{alpha3min}} the prediction for $a_3$ in the plane $(m_{SUSY},m_T)$ is
shown for the case of minimal $SU(5)$.
It can be seen that, in this case, we need to take $m_{SUSY}$ as large as possible and 
$m_T$ as small as possible. But the smaller is $m_T$, the faster is proton decay. 
The best compromise is something like $m_{SUSY}\approx 1$ TeV and 
$m_T\approx M_{GUT}^{(0)}$, which leads to $\alpha_3(m_Z)\approx 0.13$ which is still
rather large and proton decay is dangerously fast, in conflict with experimental limits.
In fact, with this value for $m_T$, we already know from the discussion of section 
\ref{numexam}
that proton lifetime for the channel $p \rightarrow K^+ \bar\nu$ is expected
to be around $10^{31}$ yr and that, in the minimal model, there is no way
to push it above the experimental limit of $\sim 2 \cdot 10^{33}$ yr. 
Thus, we are lead to conclude that minimal SUSY $SU(5)$ is ruled out.

As already pointed out, models based on $SO(10)$ are
highly model dependent, because the choice of Higgs fields which are
introduced to realize the breaking of $SO(10)$ affects many predictions,
among which proton decay and coupling unification.
For instance, in the quite economic $SO(10)$ model of ref. \cite{bapawi},
the corrections to $\alpha_{3}(m_Z)$ coming from representations 
like $10, \overline{10}$, $16, \overline{16}$ and $45$ are expected to be 
positive and of order 10\% with respect to the experimental 
value of $\alpha_{3}(m_Z)$, so that one needs to invoke negative
corrections of the same order arising from Planck scale effects or
other thresholds.

%%%%%%%%%%%%%%%%%%%%%%%%%%%%%%%%%%%%%%

\clearemptydoublepage
\chapter{A realistic SUSY $SU(5)$ GUT Model}

\label{real5}

It has emerged from the survey of the previous chapter that, despite the
idea of GUT is really very attractive, its actual realization is not at all precisely defined. 
Consider also that at the Planck scale, $M_{Pl}$, the unification of gravity 
with gauge interactions is a general property of superstring theories.
But in simple GUT models gauge interactions are unified at a distinctly 
lower mass scale $M_{GUT}$ within the context of a
renormalizable gauge theory. 
However, this neat separation between gauge unification and merging with gravity 
is not at all granted. The gap between $M_{GUT}$ and $M_{Pl}$ could be filled 
up by a number of threshold effects and several layers of additional states. 
Also, coupling unification could be realized without gauge unification, 
as suggested in some versions of superstring theory or in flipped $SU(5)$.

The model we aim at should not rely on large
fine tunings and must lead to an acceptable phenomenology. 
This includes coupling unification with an acceptable value of
$\alpha_s(m_Z)$, given $\alpha$ and
$sin^2\theta_W$ at $m_Z$, compatibility with the more and more stringent 
bounds on proton decay \cite{pbounds, lastSK}, agreement with the observed 
fermion mass spectrum, also considering neutrino masses and mixings and so on. 

Prompted by recent neutrino oscillation data some new studies on realistic GUT 
models have appeared in the context of $SO(10)$ or larger groups 
\cite{bapawi,recentso10}. Here we address the question whether the smallest 
SUSY $SU(5)$ symmetry group can still be considered as a basis for a realistic GUT model. 
We indeed present an explicit example of a realistic $SU(5)$ model 
\cite{afm2}, 
which uses a $U(1)$ flavour symmetry as a crucial ingredient. 
In principle the flavour symmetry could be either global or local. 

We tentatively assume here that the flavour symmetry is global. This is more in the spirit of GUT's
in the sense that all gauge symmetries are unified. The associated Goldstone boson receives a mass from the anomaly.
Such a pseudo-Goldstone boson can be phenomenologically acceptable in view of the existing limits on axion-like
particles \cite{gkr}.

In this model the doublet-triplet splitting problem is solved by the MDM \cite{mdm} 
stabilized by the flavour symmetry against the occurrence of doublet mass 
lifting due to non renormalizable operators. Relatively large representations (50,
$\overline{50}$, 75) have to be introduced for this purpose. 
A good effect of this proliferation of states is that the value of $\alpha_s(m_Z)$ 
obtained from coupling unification in the next to the leading order perturbative 
approximation receives important negative corrections from threshold effects 
near the GUT scale. As a result, the central value changes from 
$\alpha_s(m_Z)\approx 0.130$ in minimal SUSY $SU(5)$ down to 
$\alpha_s(m_Z)\approx 0.116$, in better agreement with observation \cite{yam,bag,cla}. 
The same $U(1)$ flavour symmetry that stabilizes the missing partner mechanism 
is used to explain the hierarchical structure of fermion masses. 
In the neutrino sector, mass matrices very similar to the ones
already presented in the first model of section
\ref{ooff} (and proposed in ref. \cite{af}) are obtained. 
The large atmospheric neutrino mixing is due to a large left handed mixing in the lepton
sector that corresponds to a large right handed mixing in the down quark sector. 
In the present particular version maximal mixing also for solar neutrinos is preferred. 
A possibly problematic feature of the model is that, beyond the unification point, 
when all the states participate in the running, the asymptotic freedom of $SU(5)$ is 
destroyed by the large number of matter fields. 
As a consequence, the coupling increases very fast and the theory becomes non
perturbative below $M_{Pl}$. 
In the past models similar to ours have been considered, but were discarded 
just because they contain many additional states and tend to become non 
perturbative between $M_{GUT}$ and $M_{Pl}$. 
We instead argue that these features are not necessarily bad. While the predictivity 
of the theory is reduced because of non renormalizable operators that are only 
suppressed by powers of $M_{GUT}/\Lambda$ with $\Lambda<M_{Pl}$, still
these corrections could explain the distortions of the mass spectrum with respect 
to the minimal model, the suppression of proton decay and so on. However, it is 
certainly true that also in this case, as for any other known realistic model,
the resulting construction is considerably more complicated than in the 
corresponding minimal model.

\section{The Model}

The symmetry of the model is SUSY $SU(5)\otimes U(1)$. We define $Q$ the charge 
associated with $U(1)$ and assume that the global $U(1)$ flavour 
symmetry is a good symmetry of the 
superpotential, that is it is not violated by non-perturbative effects through operators 
appearing in the superpotential $w$.
The superpotential of the model can be thought as composed of three parts:
\beq 
w=w_1+w_2+w_3~~~~, \label{superpot}
\eeq 
where $w_1$ accounts for the breaking of $SU(5)$, $w_2$ realizes
the doublet-triplet splitting by means of the MDM and $w_3$ contains the Yukawa
couplings.

\subsection{$SU(5)$ Breaking}

The $w_1$ term contains only the field $Y$ of the representation $75$ of $SU(5)$ with $Q=0$:
\beq w_1=c_1~Y^{A B}_{C D}~ Y^{C D}_{E F}~ Y^{E F}_{A B}
        +M_Y~Y^{A B}_{C D}~ Y^{C D}_{A B}~~~~~,\label{2}
\eeq 
where $A,B,...=1,2,...,5$.
$w_1$ has the effect of providing $Y$ with a vev 
$\langle Y \rangle \equiv M_Y/(2 c_1) \sim M_{GUT}$ and of giving a mass to all
physical components of $Y$, i.e. those that are not absorbed by the Higgs mechanism. 
The $75$ uniquely 
breaks $SU(5)$ down to $SU(3)\otimes SU(2)\otimes U(1)$ through the following vev: 
\beq
\langle Y_{C D}^{A B} \rangle =- \dd \langle Y \rangle   
          \left[ \dd\left( {\delta}_{\gamma }^{\alpha } {\delta}_{\delta }^{\beta  }
                      +2 {\delta}_{c}^{a } {\delta}_{d}^{b}
                  - \dd\frac{1}{2} {\delta}_{C}^{A} {\delta}_{D}^{B} \dd\right) 
                  - \left(A \leftrightarrow B\right) \dd\right]~~~~~~,
\label{expvevy}
\eeq
where Greek indices assume values from 1 to 3 and Latin indices from 4 to 5. 

\section{Doublet-Triplet Splitting}

The $w_2$ term induces the doublet-triplet splitting:
\bea w_2&=&        c_2~H^{A} Y^{D E}_{B C} {H_{50}}^{B C F G} {\epsilon}_{A D E F G}~
   +~c_3 {\overline H}_{A} Y^{B C}_{D E} {H_{\overline{50}}}_{B C F G} 
                {\epsilon}^{A D E F G}~\nn\\
    &+&~c_4~{H_{50}}^{A B C D} {H_{\overline{50}}}_{A B C D} X~~~~~~~~~~~~~~~~~~~~~~~~
~~~~~~~~~~~~~~~.\label{www2}
\eea 
Here $H$ and $\overline{H}$ are the usual pentaplets of Higgs fields, except that now 
they carry non opposite $Q$ charges. It is not restrictive to take $c_2$, $c_3$ and $c_4$ 
real and positive.
The renormalizable couplings that appear in $w_2$ are the most general allowed by 
the $SU(5)$ and $Q$ assignments of the fields in eq. (\ref{www2}) which are given in Table 
\ref{afm2ch}. The value of $q$ will be specified later.
%\beq
%\begin{array}{ccccccc}
%{\rm field}& Y& H& \bar{H}& H_{50}& {H_{\overline{50}}}& X\\
%SU(5)& 75& 5& \bar{5}& 50 &\bar{50} & 1\\
%Q& 0& -q& q-1& q& 1-q& -1 
%\end{array}
%\label{4}
%\eeq
\begin{table}[h]
\begin{center}
\begin{tabular}{|c|c|c|c|c|c|c|}
\hline & & & & & &\\
 field & $Y$ & $H$ & $\overline H$ & $H_{50}$ & ${H_{\overline{50}}}$ & $X$ \\& & & & & &\\
\hline & & & & & &\\
$SU(5)$ & $75$ & $5$ & $ \overline 5$ & $50$ & $\overline{50}$ & $1$\\& & & & & &\\
$Q$& $0$ & $-q$ & $q-1$ & $q$ & $1-q$ & $-1$ \\ & & & & & &\\ \hline
\end{tabular}
\end{center}
\caption{$Q$ charge assignments.}
\label{afm2ch}
\end{table}

\subsection{Limit of Unbroken SUSY}

At the minimum of the potential, in the limit 
of unbroken SUSY, the vevs of the fields $H$, $\overline H$,
$H_{50}$ and ${H_{\overline{50}}}$ all vanish, while the $X$ vev remains undetermined. 
These can be easily understood by writing symbolically the equations for the minimum 
of the superpotential $w_i=0$, the subscript  i standing for derivation:
\bea
w_{Y}&=&2 M_{Y} Y+3 c_1 Y^2 +c_2 H H_{50} +c_3 \overline{H} H_{\overline{50}}\nn\\
w_{50}&=&c_2 H Y+c_4 X H_{\overline{50}}\nn\\
w_{\overline{50}}&=&c_3 \overline{H} Y +c_4 X  H_{50} \nn\\
w_H&=&c_2 Y H_{50} \\
w_{\overline{H}}&=&c_3 Y H_{\overline{50}}\nn\\
w_X&=&c_4 H_{50} H_{\overline{50}}\nn
\eea
For any $\langle X \rangle$, a minimum is achieved by taking $\langle H \rangle =
\langle  \overline{H} \rangle = \langle H_{50} \rangle = \langle H_{\overline{50}} \rangle=0$
and $\langle Y \rangle$ as shown in eq. (\ref{expvevy}).

As we shall see, when SUSY is softly broken the light doublets in
$H$ and $\overline H$ acquire a small vev while the $X$ vev will be fixed near the cut-off 
$\Lambda$, of the order of the scale between $M_{GUT}$ and $M_{Pl}$ where the 
theory becomes strongly interacting (we shall see that we estimate this scale
at around $10-20~ M_{GUT}$, large enough that the approximation of neglecting 
terms of order $M_{GUT}/\Lambda$ is not unreasonable). 
The MDM occurs as discussed in eqs. (\ref{wmdm}) and (\ref{hatmt}) with the
only difference that the mass term for $H_{50}$ and ${H_{\overline{50}}}$
originates from the vev of $X$.
The $U(1)$ flavour symmetry protects the doublet Higgs to take mass from radiative 
corrections because no $H\overline H$ mass term is allowed. 
Also no non renormalizable terms of the form $H {\overline H} Y^m X^n$ $(m,n\ge 0)$ are 
possible, because $X$ has a negative $Q$ charge. As anticipated in
section \ref{mdmcrit}, this version of the MDM
was introduced in ref. \cite{ber} and overcomes the
observation in ref. \cite{ran} that, in general, non renormalizable interactions 
spoil the mechanism. The Higgs colour triplets mix with the analogous states 
in the $50$ and the resulting mass matrix is of the see-saw form:  
\beq \hat{m}_T= 
\left[\matrix{ 0& - 4 \sqrt{3} c_2 \langle Y\rangle
\cr - 4 \sqrt{3} c_3\langle Y\rangle&c_4\langle X\rangle    } 
\right]~~~~~. 
\label{5}
\eeq
Defining $m_{\phi}=c_4\langle X\rangle$ the eigenvalues of the matrix 
$\hat{m}_T \hat{m}_T^\dagger$ are the squares of:
\beq
m_{T1,2}=\frac{1}{2}\left[
\sqrt{m_{\phi}^2+48~(c_2+c_3)^2\langle Y\rangle^2}\pm \sqrt{m_{\phi}^2+48~(c_2-c_3)^2\langle Y\rangle^2}\right]~~~.
\label{autreal}
\eeq
and $m_{T1}m_{T2}=~48~ c_2~ c_3~ \langle Y\rangle^2$. 
The effective mass that enters in the dimension 5 operators with $|\Delta B=1|$ is 
$1/(\hat{m}_T^{-1})_{1 1}$:
\beq
m_T=\frac{{{m_T}_1} {{m_T}_2}}{m_{\phi}}=\frac{48~ c_2~ c_3~ \langle Y\rangle^2}
{c_4~ \langle X\rangle}\label{mT}~~~.
\eeq

\subsection{Broken Supersymmetry}

The breaking of SUSY fixes a large vev for the field $X$ by removing the corresponding 
flat direction, gives masses to s-partners, provides a small mass to the Higgs doublet 
and introduces a $\mu$ term. 
In the $(50,\overline{50},X)$ sector, the terms that break SUSY softly are:
\beq
-{\cal L}_{soft}~=~m_x^2 |x|^2+m_{50}^2|c|^2~+~m_{\overline {50}}^2|\bar c|^2~+
m_{3/2} c_4 A_4(c \bar c x~+~h.c.)~+.....\label{lsoft}
\eeq
where $c$, $\bar c$ and $x$ denote the scalar components of 
$H_{50}$, ${H_{\overline{50}}}$ and $X$ respectively, 
$m_x$, $m_{50}$, $m_{\overline {50}}$ their soft breaking masses, $m_{3/2}$ the gravitino
mass and $A_4$ a trilinear coupling.  Note that $c_4$ has here the same role 
played in SUSY radiative electroweak breaking by the top Yukawa coupling.
Dots stand for the remaining soft breaking terms, including mass terms for the scalar 
components of the Higgs doublet fields
\footnote{To find the minima of the scalar
potential it is not restrictive to set to zero the imaginary part of $x$, which will be understood
in the remaining part of this section.}.

Consider first the limit $m_x=m_{50}=m_{\overline {50}}=m_{3/2}\equiv m$ and $c_4 A_4=1$.
In the SUSY limit and neglecting the mixing between the $(50,{\overline{50}})$ 
and the $(5,\bar{5})$ sectors, fermions and scalars in the $50$ and 
${\overline{50}}$ have a common squared mass $c^2_4 x^2$. 
When SUSY is broken by the soft terms for each fermion of mass $c_4 x$ there are 
two bosons of squared masses $c^2_4 x^2+m^2\pm mx$. 
The $x$ terms in the scalar potential at one
loop accuracy are given by:
\bea
V&= &m^2 x^2+\frac{50}{64\pi^2}\left\{2(c^2_4 x^2+m^2+mx)^2 \left[\log\left({\frac{c^2_4 x^2+m^2+mx}{\Lambda^2}}\right)
-3/2\right]\right.\nn\\
&+&2(c^2_4 x^2+m^2-mx)^2
\left[\log\left({\frac{c^2_4 x^2+m^2-mx}{\Lambda^2}}\right)-3/2\right]\nn\\
&-&\left.4c^4_4 x^4\left[\log\left({\frac{c^2_4 x^2}
{\Lambda^2}}\right)-3/2\right]\right\}
\label{l15}
\eea
A numerical study of this potential in the limit $m\ll\Lambda$ and for $c_4$ of order one,
shows that the minimum is at $x$ of order $\Lambda$ (somewhat smaller than $\Lambda$ 
but close to it).
The expression for $V$ in eq. (\ref{l15}) provides a good approximation of the scalar 
potential only in a small region around $\Lambda$. 
Outside this region the perturbative approximation will break down. 
We take our numerical analysis as an indication that the minimum occurs near 
$\Lambda$.

\begin{figure}[p]
\centerline{\psfig{file=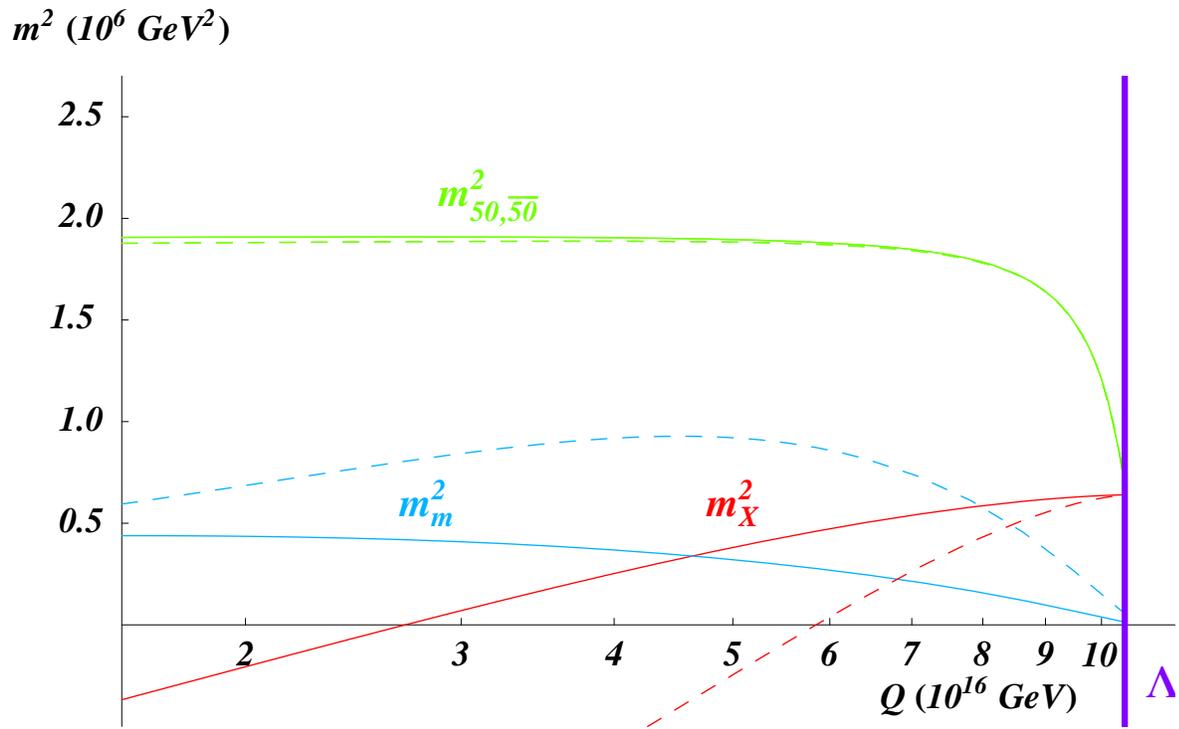,width=1.1\textwidth}}
\caption{Running of 
$m_x^2, m_{50}^2, m_{\overline{50}}^2$ and $m_m^2 \equiv (m_{3/2} c_4 A_4)^2$.
The solid (dashed) lines correspond to: $\Lambda = 1.1 \cdot 10^{17}$ GeV, 
$\alpha (\Lambda) = 0.88$, 
$m_x=m_{50}=m_{\overline{50}}=m_{3/2}=800$ GeV, $\tilde{m}=2$ TeV,
$c_4=0.2$ ($c_4=0.4$) and $A_4=0.7$.
The solid red line gives a 
$ \lambda \equiv \langle X \rangle / \Lambda \sim 0.25$.}
\label{figrunx}
\end{figure}

To analyze in more detail the situation we consider the differential equation
describing the evolution of the $x$ mass and we identify $\langle X\rangle$ 
with the scale at which $m_x^2$ vanishes.
With the soft terms of (\ref{lsoft}), we find that the following set of differential equations
is appropriate to describe the running in range between $M_{GUT}$ and the cut-off:
\bea
\frac{d m_x^2}{d \ln Q}&=& \frac{25}{4 \pi^2} c_4^2 
(m_x^2+m_{50}^2+ m_{\overline {50}}^2+m_{3/2}^2 A_4^2)\nn\\
\frac{d m_{50}^2}{d \ln Q}&=&\frac{d m_{\overline{50}}^2}{d \ln Q}=
-\frac{84}{5 \pi} \alpha \tilde{m}^2+\frac{1}{8 \pi^2} c_4^2 
(m_x^2+m_{50}^2+ m_{\overline {50}}^2+m_{3/2}^2 A_4^2)\nn\\
\frac{d c_4}{d \ln Q}&=&-\frac{42}{5 \pi} \alpha c_4+\frac{13}{4 \pi^2} c_4^3 \\
\frac{d (m_{3/2} A_4)}{d \ln Q}&=&-\frac{84}{5 \pi} \alpha {\tilde m}+\frac{13}{2 \pi^2}
 m_{3/2} c_4^2 A_4\nn\\
\frac{d \tilde{m}}{d \ln Q}&=& \frac{52}{2 \pi} \alpha \tilde{m}\nn\\
\frac{d \alpha}{d \ln Q}&=& \frac{52}{2 \pi} \alpha^2 \nn
\eea 
where $\alpha$ is the running gauge coupling and $\tilde m$ are gaugino masses,
which are supposed to be degenerate.
The value of $\langle X \rangle$ depends on the boundary conditions assigned 
to the relevant parameters at the cut-off scale, in particular $c_4$ and the trilinear 
$c {\bar c} x$ coupling in ${\cal L}_{soft}$. 
To give a numerical example, if $\Lambda = 1.1 \cdot 10^{17}$ GeV and 
$\alpha (\Lambda) = 0.88$, 
then $\langle X \rangle \sim \Lambda /4$ is obtained by  
choosing the following reasonable values of the parameters at the cut-off 
scale: 
$m_x=m_{50}=m_{\overline{50}}=m_{3/2}=800$ GeV, $\tilde{ m} =2 $ TeV,
$c_4=0.2$ and $A_4=0.7$. 
This particular situation is shown in fig. \ref{figrunx}, where the running of 
$m_x^2, m_{50}^2, m_{\overline{50}}^2$ and $m_m^2 \equiv (m_{3/2} c_4 A_4)^2$ 
is displayed. 
In the next section we assume that the true minimum occurs at 
$x=\langle X\rangle=0.25 \Lambda$,
that is, by defining $\lambda \equiv \langle X \rangle / \Lambda$, at 
$\lambda=0.25$. 

We have studied the dependence of $\langle X \rangle$ upon both the boundary values
of $c_4$ and $A_4$.
We find that the dependence on $c_4(\Lambda)$ is logarithmic
and that there is a mild linear dependence on
the trilinear $c {\bar c} x$ term evaluated at $\Lambda$.
The above dependence results mainly from the interplay of
the differential equations describing the evolution of the $x$ mass
and the $c_4$ coupling, the quantities with the faster running
in the region close to the cut-off $\Lambda$. We 
obtain: $\delta \langle X \rangle/\Lambda \approx 0.4~ \delta c_4/c_4$.
This means that, with the same boundaries of the previous numerical example
but with $c_4$ doubled, the value of $\langle X \rangle$ becomes greater by a factor $2.2$, as can be seen from fig. \ref{figrunx}.
The vev of $X$ seems thus to be not dangerously unstable with respect to the
boundaries. This situation differs from the analogous mechanism of radiative electroweak
symmetry breaking, where a doubling of $y_t$ can cause a variation of several
orders of magnitude in the vev of the Higgs doublet.    
We have checked that, in our case, the mild dependence on $c_4$ is due to the running 
of $c_4$ itself which, in the appropriate energy window, is not negligible.

The complex field $x$ describes two physical scalar particles. 
The one associated to the real part of $x$ has a mass of order $m_{3/2}$ 
and couplings to the ordinary fermions suppressed by $1/\Lambda$.
The particle associated to the imaginary part of $x$ is massless only in the tree 
approximation. 
Due to the anomaly of the related $U(1)_Q$ current the particle acquires a mass of order
$f_\pi m_\pi/\langle X\rangle$. Cosmological bounds on $\langle X\rangle$ have been 
recently reconsidered in ref. \cite{gkr} where it has been observed that 
$\langle X\rangle$ of the order of the grand unification scale is not in conflict with 
observational data.

An alternative possibility is to assume that the $U(1)$ symmetry is local \cite{ber}. 
In this case the supersymmetric action contains a Fayet-Iliopoulos term and the 
associated D-term in the scalar potential provides a large vev for $x$, of the order 
of the cut-off scale $\Lambda$.

A $\mu$ term for the fields $H$ and $\overline H$ of the appropriate order of magnitude 
can be generated according to the Giudice-Masiero mechanism \cite{giu}. Assume that 
the breaking of SUSY is induced by the $\theta^2$ component of a chiral 
(effective) superfield $S$, singlet under $SU(5)$ with $Q=0$. 
In the Kahler potential a term of the form
\beq
K~=~\frac{S^\dagger X^\dagger H \overline H}{\Lambda^2}+{\rm h.c.}\label{16}
\eeq
is allowed. The vevs of $S$ and $X$, $\langle S\rangle\sim\theta^2 m_{3/2} M_{Pl}$, 
$\langle X\rangle=\lambda\Lambda$ lead to an equivalent term in the
superpotential of the desired form 
$\mu H \overline H$ with $\mu\sim \lambda m_{3/2} M_{Pl}/\Lambda$ which can 
be considered of the right order. 

\section{Coupling Unification}

As was pointed out previously, it is well known that in the minimal version of 
SUSY $SU(5)$ the central value of $\alpha_3(m_Z)$ required by the constraint of 
coupling unification is somewhat large: $\alpha_3(m_Z)\approx 0.13$ 
\cite{yam,bag,cla,lp,cpp}. 
In the model discussed here, where the doublet-triplet splitting problem is solved by 
introducing the $SU(5)$ representations $50$, ${\overline{50}}$ and $75$, the
central value of $\alpha_3(m_Z)$ is modified by threshold corrections near $M_{GUT}$ 
which bring the central value down by a substantial amount so that the final result can
be in much better agreement with the observed value. As discussed in
refs. \cite{yam,bag}, this remarkable result arises because the $24$ of the minimal 
model is replaced by the $75$. The mass splittings inside these representations 
are dictated by the group embedding of the $75$ singlet under $SU(3)\otimes SU(2)
\otimes U(1)$ that breaks $SU(5)$. 
The difference in the threshold contributions from the $24$ and the $75$ has the right
sign and amount to bring $\alpha_3(m_Z)$ down even below the observed value. 
The right value can then be obtained by moving $m_{SUSY}$ and $m_T$ in a reasonable 
range. 
The difference in the favored value of $m_T$ with respect to the minimal model in order 
to reproduce the observed value of $\alpha_3(m_Z)$ goes in the right
direction to also considerably alleviate the potential problems from the bounds on 
proton decay. 
Note that there are no additional threshold corrections from the $50$ and
${\overline{50}}$ representations because the mass of these states arise from 
the coupling to the field $X$ which is an $SU(5)$ singlet. 
Thus there are no mass splittings inside the $50$ and the
${\overline{50}}$ and no threshold contributions. 
While there is no effect on the value of $\alpha_3(m_Z)$ the presence of the
states in the $50$ and ${\overline{50}}$ representations affects the value of the unified 
coupling at $M_{GUT}$ and also spoils the asymptotic freedom of $SU(5)$ beyond 
$M_{GUT}$. 
We find it suggestive that the solution of the doublet-triplet splitting problem in
terms of the $50$, ${\overline{50}}$ and $75$ representations automatically leads 
to improve the prediction of $\alpha_3(m_Z)$ and at the same time relaxes the 
constraints from proton decay.  We now discuss this issue in more detail.

\subsection{Estimate of $\alpha_3(m_Z)|_{\overline{MS}}$}

The calculation proceeds exactly as in the previous chapter. With the same SUSY 
spectrum, all the differences on the values of $a_5, M_{GUT}, a_3$, with respect to minimal 
SUSY $SU(5)$, are caused by the term $\delta_i (h)$ of (\ref{deltai}). 

Following the steps already encountered,
$\delta_3 (2), \delta_3 (l)$ are those of eqs. (\ref{delta32}) and (\ref{delta3l}), but
$\delta_3 (h)$ is now obtained substituting in eq. (\ref{deltaih}) the $b_i^{(h)}$
reported in Table \ref{htrmdm}.

\begin{table}[h] 
{\begin{center}
\begin{tabular}{|c|c|c|c|c|}   
\hline
& & & & \\                         
state $(SU(3),SU(2),U(1))$ & $ b_1^{(h)} (j) $ & $ b_2^{(h)} (j)$  & $b_3^{(h)} (j) $ & mass  \\ 
& & & &\\
\hline \hline
& & & &\\
$(8,3,0)$ & $0$ & $8$ & $9/2$ & $m_Y$ \\
& & & &\\
$(3,1,5/3) (\bar 3,1,-5/3 )$ & $5$ & $0$ & $1/2$ & $0.8 m_Y$ \\
& & & &\\
$(6,2,5/6) (\bar 6,2,-5/6)$ & $5$ &$3$ &$5$ & $0.4 m_Y$ \\ 
& & & &\\
$(1,1,0)$ & $0$ & $0$ & $0$ & $0.2 m_Y$\\
& & & &\\
$(8,1,0)$ & $0$ & $0$ & $3/2$ & $0.2 m_Y$\\
& & & &\\
\hline 
& & & &\\
$(3,1,-1/3), (\overline 3,1,1/3) $& $1/5$ & $0$ & $1/2$ & ${m_T}_1$\\
& & & &\\
$(3,1,-1/3), (\overline 3,1,1/3) $& $1/5$ & $0$ & $1/2$ & ${m_T}_2$\\
& & & &\\
\hline 
& & & &\\
$50, \overline{50} $& $173/10$ & $35/2$ & $17$ & $m_{\phi}$\\
& & & &\\
\hline
\end{tabular} 
\end{center}}
\caption{GUT thresholds for the MDM. $m_Y \equiv 5 M_{Y}$.}
\label{htrmdm}
\end{table}

For this case one obtains
\beq
\delta_3 (h) =2.047-\frac{9}{14 \pi} \ln{ \left(  \frac{m_T} {M_{GUT}} \right)}
+\frac{3}{14 \pi} \ln{ \left(  \frac{m_Y} {M_{GUT}} \right)}
\eeq
where $m_T$ is the effective triplet mass.
Note that this equation is very similar to eq. (\ref{delta3h}), apart from
the non-logarithmic term. 
We can still use the first of eqs. (\ref{impr}), with the difference that
thresholds from GUT particles within the MDM now give:
\beq
\delta_X (h)=-0.993-\frac{5}{14} \ln{\left(  \frac{m_Y}{M_{GUT}} \right)}
+\frac{1}{14} \ln{\left(  \frac{m_T}{M_{GUT}} \right)}~~.
\eeq

Thus, keeping fixed $m_{SUSY}$, the first of eqs. (\ref{impr}) can be considered 
as a relation between the GUT masses $m_V=M_{GUT}, m_Y, m_T$.
\bea
\ln{\left( \frac{M_{GUT}}{m_Z}\right) }&=&\ln{ \left(    \frac{M_{GUT}^{(0)} }{m_Z} \right) }\nn\\
&+&\left[ 0.469 \right]_{(2)}+\left[ -0.036-\frac{25}{84} \ln{\left(  \frac{m_{SUSY}}{m_Z} \right)}
\right]_{(l)}\nn\\
&+&\left[ -0.993-\frac{5}{14} \ln{\left(  \frac{m_Y}{M_{GUT}} \right)}
+\frac{1}{14} \ln{\left(  \frac{m_T}{M_{GUT}} \right)}\right]_{(h)}
\label{linkmv}
\eea

Substituting it in $\delta_3 (h)$ for $M_{GUT}$, as pointed out in \ref{coupres}, 
we can write the prediction
for $\delta_3$ in the following form, 
\beq
\delta_3 = k+\dd\frac{1}{2 \pi}  \ln{\left(  \frac{m_{SUSY}}{m_Z} \right)} 
-\dd\frac{3}{5 \pi} \ln{\left(  \frac{m_T}{M_{GUT}^{(0)}} \right)}
\eeq
where again $k=k(2)+k(SUSY)+k(M_{GUT})$ 
and it contains the contribution of two-loop diagrams to the running couplings,
$k(2)$, the threshold contribution of states near
$m_{SUSY}$, $k(SUSY)$, and the threshold contribution from states near $M_{GUT}$, 
$k(M_{GUT})$. Remember that
the threshold contributions would vanish if all states had the mass 
$m_{SUSY}$ or $M_{GUT}$ so that only mass splittings contribute to
$k(SUSY)$ and $k(M_{GUT})$. 
The values of $k(2)$ and of $k(SUSY)$ are essentially the same in the minimal and 
the realistic model. As in the previous chapter we take: $k(2)=-0.733$ and $k(SUSY)=-0.510$, 
corresponding to the representative spectrum displayed 
in Table \ref{tabsusysp}.
The value of $k(M_{GUT})$ was zero for the $24$ of the minimal model, but now
we have $k(M_{GUT})=1.857$ for the $75$ of the realistic model. The $75$ contains
in fact non degenerate coloured multiplets. Thus we obtain:
\bea
k=-0.733-0.510&=&-1.243~~~~~~~~\rm{minimal~ model}\nonumber\\
k=-0.733-0.510+1.857&=&0.614~~~~~~~~~~~\rm{realistic~ model}\label{k}
\eea

\begin{figure}[p]
\centerline{\psfig{file=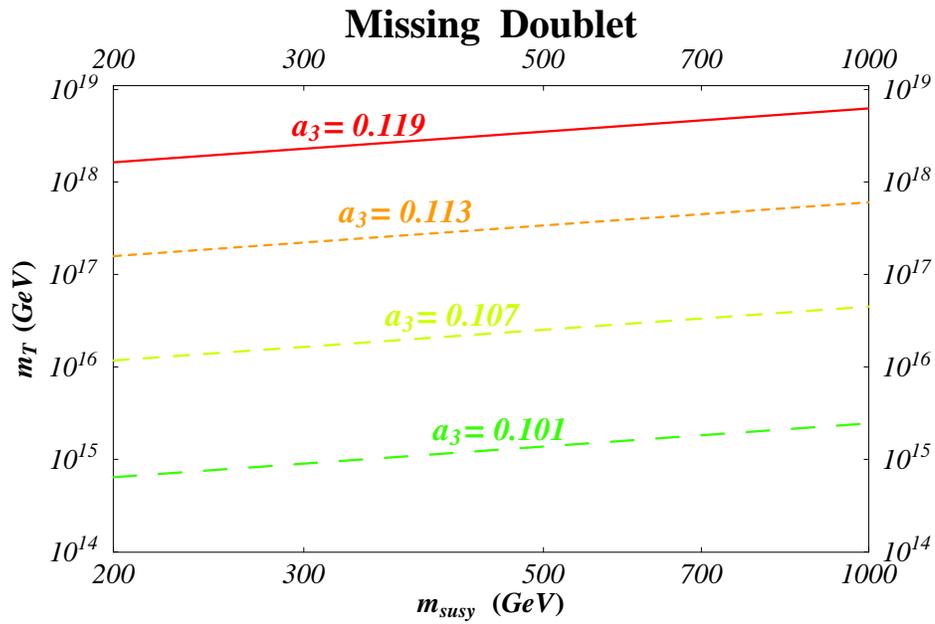,width=1.1\textwidth}}
\caption{Contours of $a_{3}\equiv \alpha_{s}(m_Z)$ in the plane $(m_{SUSY},m_T)$, 
for the missing doublet
model. The SUSY spectrum is parametrized as in Table \ref{tabsusysp}.}
\label{alpha3mdm}
\end{figure}

This difference is very important and makes the comparison with experiment 
of the predicted value of $\alpha_3(m_Z)$ much more favorable in the case of the 
realistic model. In fact for $k$ large and negative as in the
minimal model we need to take $m_{SUSY}$ as large as
possible and $m_T$ as small as possible. But the smaller is $m_T$, the faster is proton 
decay. The best compromise is
something like $m_{SUSY}\approx 1$ TeV and $m_T\approx M_{GUT}^{(0)}$, which 
leads to $\alpha_s(m_Z)\approx 0.13$ which is still
rather large and proton decay is too fast. 
This is to be confronted with the case of the realistic model where $k$
is instead positive and large enough to drag $\alpha_s(m_Z)$ below the observed value. 
We now prefer $m_T$ to be larger than $M_{GUT}^{(0)}$ by typically a factor of 20-30, 
which means a factor of 400-1000 of suppression for the proton decay rate with respect
to the minimal model. For example, for $m_{SUSY}\approx0.25$ TeV and 
$m_T\approx 6\cdot 10^{17}$ GeV we obtain $\alpha_3(m_Z)\approx 0.116$ which is 
acceptable.
The predictions for $a_3$ versus $(m_{SUSY},m_T)$ in the 
minimal and the missing doublet model
are shown in figs. \ref{alpha3min} and \ref{alpha3mdm}.

Due to the large matter content the model is not asymptotically
free and the gauge coupling constant $\alpha_5$ blows up at
the scale
\beq
M_{GUT}~~ {\rm exp}{\dd\left(\frac{\pi}{26 a_5}\right)}~~~~~.
\label{blow}
\eeq
Taking into account that threshold effects modify $a_5$ with respect
the LO value $a_{5}^{(0)}\approx 1/24$, we find that the pole occurs
near $10^{17}$ GeV, the precise value depending on the details
of the heavy spectrum.

\subsection{Theoretical Uncertainties}

Clearly, there is an uncertainty on $k(SUSY)$, related to the possibility of varying
both the parameterization used and 
the relations assumed between the parameters $m_0$, $m_{1/2}$, $\mu$, $m_H$.
An overall uncertainty of $\pm 0.005$ on $a_3$ has been estimated in 
ref. \cite{lp} by
scanning many models consistent with electroweak symmetry breaking, 
a neutral lightest supersymmetric particle and sparticle masses above the experimental 
bound and below $\sim$ 2 TeV. 
%This corresponds to an uncertainty on 
%$k(SUSY)+1/(2\pi)\log(m_{SUSY}/m_Z)$ of about $\pm 0.35$ 

We stress that in the present analysis we are assuming universal soft breaking 
parameters at the cut-off scale. If we relax this assumption, then a larger range for
$a_3$ could be obtained. Indeed $a_3$ is particularly sensitive to
wino and gluino masses, that, at the electroweak scale, 
are approximately in the ratio 1:3 
when a universal boundary condition on gaugino masses is imposed. It has been observed 
that, by inverting this ratio, a negative contribution of about $-0.01$ to
$a_3$ is obtained \cite{rs}.

Another source of uncertainty on $a_3$ is related to the unknown physics above the 
cut-off scale of the grand unified theory. There can be threshold effects due to new 
heavy particles or even non perturbative effects that arise in the underlying 
fundamental theory. These effects can be estimated from the non-renormalizable
operators, suppressed by $\langle Y\rangle/\Lambda$, that split the gauge couplings 
at the scale $M_{GUT}^{(0)}$ \cite{nro}.  
For generic, order one coefficients and barring cancellations among different terms, 
the presence of these operators may affect $a_3$ by additional contributions of 
about $\pm 0.005 M_{Pl}/\Lambda$ \cite{lp}. As we will see
the model under discussion requires $\Lambda\approx 0.1 M_{Pl}$. Therefore, 
to maintain the good agreement
between the experimental and predicted values of $a_3$, we need a suppression 
of this contribution 
by about a factor of 10, which we do not consider too unnatural.

In view of the large theoretical uncertainties on $a_3$ we cannot firmly
conclude that the gauge coupling unification fails in the minimal model
while it is completely successful in the realistic one.
However we find very encouraging that, by solving
the doublet-triplet splitting problem within the missing partner model,
acceptable values of $a_3$ can be easily obtained and that they
are directly related to a proton lifetime potentially
larger than in the minimal model.

\section{Yukawa Couplings}

The $w_3$ term contains the Yukawa interactions of the quark and lepton fields 
$\Psi_{10}$, $\Psi_{\overline{5}}$ and $\Psi_1$, transforming as the representations $10$, 
$\overline 5$ and 1 of $SU(5)$ respectively. 
We assume an exact $R$-parity discrete symmetry under which 
$\Psi_{10}$, $\Psi_{\overline{5}}$ and $\Psi_1$ are odd whereas
$H$, $\overline H$, $H_{50}$ and ${H_{\overline{50}}}$ are even. 
The $w_3$ term is symbolically given by
\bea 
w_3&=&\Psi_{10}G_u(X,Y)\Psi_{10}H~+~\Psi_{10}G_d(X,Y)\Psi_{\overline 5}\overline H~+~
\Psi_{\overline 5}G_{\nu}(X,Y)\Psi_{1}H\nn\\
&+&M\Psi_{1}G_M(X,Y)\Psi_{1}~+~\Psi_{10}{G_{\overline{50}}}(X,Y)\Psi_{10} H_{{\overline{50}}}~~~.
\label{w3}
\eea 
The Yukawa matrices $G_u$, $G_d$, $G_{\nu}$, $G_M$ and ${G_{\overline{50}}}$ depend 
on $X$ and $Y$ and the associated mass matrices on their vevs. 
The last term does not contribute to the mass matrices because of the vanishing 
vev of ${H_{\overline{50}}}$, but is important for proton decay. 
The pattern of fermion masses is determined by the $U(1)$ flavour symmetry that fixes the
powers of $\lambda\equiv\langle X\rangle/\Lambda$ for each entry of the mass matrices.  
In fact $X$ is the only field with non vanishing $Q$ that takes a vev. 
The powers of $\lambda$ in the mass terms are fixed by the $Q$ charges of the 
matter field $\Psi$ and of the Higgs fields $H$ and $\overline H$. 
We can then specify the charge $q$ that appears in Table \ref{afm2ch} and 
the $Q$ charges of the matter fields $\Psi$ in order to obtain realistic textures 
for the fermion masses. We choose $q=2$, so that we have:
\beq Q(H)=-2~~{\rm and}~~Q(\overline H)=1~~~,\label{7}  
\eeq and, for matter fields
\beq 
Q(\Psi_{10})=(4,3,1)~~~,~~~~~~~Q(\Psi_{\bar 5})=(4,2,2)~~~,~~~~~~~Q(\Psi_{1})=(1,-1,0)~~~.\label{8} 
\eeq 
The Yukawa mass matrices are of the form:
\beq G_r(\langle X\rangle,\langle Y\rangle)_{ij}~=~\lambda^{n_{ij}}G_r(\langle Y\rangle)_{ij},~~~~~~~~{r=u, d, \nu, M}
~~~. \label{9}
\eeq 
We expand $G_r(\langle Y\rangle)_{ij}$ in powers of $\langle Y\rangle$ and consider the 
lowest order term at first. Taking $G_r(0)_{ij}$ of order 1 and $n_{ij}$ as dictated 
by the above charge assignments we obtain 
\footnote{In our convention Dirac mass terms are given by $\bar L m R$ and the light 
neutrinos effective mass matrix is $m_\nu m^{-1}_{maj} m_\nu^T$.}: 
\beq
\begin{array}{lr}
m_u~=~ 
\frac{1}{\sqrt{2}}\left[\matrix{
\lambda^6&\lambda^5&\lambda^3\cr
\lambda^5&\lambda^4&\lambda^2\cr
\lambda^3&\lambda^2&1    } 
\right]~v_u~,& 
~~~~~~~~~~m_d~=~m_e^T~=~\frac{1}{\sqrt{2}} 
\left[\matrix{
\lambda^5&\lambda^3&\lambda^3\cr
\lambda^4&\lambda^2&\lambda^2\cr
\lambda^2 & 1 & 1    } 
\right]~v_d\lambda^4~,\\
& \\
m_{\nu}~=~\frac{1}{\sqrt{2}}
\left[\matrix{
\lambda^3&\lambda&\lambda^2\cr
\lambda&0&1\cr
\lambda&0&1    } 
\right]~v_u~,&
m_{maj}~=~ 
\left[\matrix{
\lambda^2&1&\lambda\cr 1&0&0\cr
\lambda&0&1    } 
\right]~M~. 
\label{Mnu}
\end{array}
\eeq
For a correct first approximation of the observed spectrum we need 
$\lambda\approx\lambda_C\approx 0.22$, $\lambda_C$ being the Cabibbo angle. 
These mass matrices are very similar to those already introduced in section
\ref{ooff}, with two important differences.
 
We have here that $\tan\beta=v_u/v_d\approx m_t/m_b\lambda^4$, which is small. 
The factor $\lambda^4$ is obtained as a consequence of the Higgs and matter fields 
charges Q, while in chapter \ref{modneu} the $H$ and $\overline H$ charges were taken 
as zero. 
We recall that a value of $\tan\beta$ near 1 is an advantage for suppressing proton 
decay.  
A small range of $\tan\beta$ around one is currently disfavored by the negative results 
of the SUSY Higgs search at LEP \cite{s2k}. 
Of course we could easily avoid this range, if necessary.

Remember that the zero entries in the mass matrices of the neutrino sector occur 
because the negatively $Q$-charged $X$ field has no counterpart with positive $Q$-charge. 
Neglected small effects could partially fill up the zeroes. 
As explained in ref. \cite{af} these zeroes lead to near maximal mixing also for 
solar neutrinos. At variance of the example of section \ref{ooff}, due to the different
charge assignments for ${\Psi_{\bar 5}}_1$, the LOW solution to the solar neutrino 
problem emerges.

As pointed out, a problematic aspect of this zeroth order approximation to the mass
matrices is the relation $m_d=m_e^T$, which is good as an order of magnitude 
relation because relates large left handed mixings for leptons to large right handed 
mixings for down quarks \cite{af,bralav,berrossi, albrbarr,largem}. 
However the implied equalities $m_b/m_{\tau}=m_s/m_{\mu}=m_d/m_e=1$ are good only 
for the third generation while need to be corrected by factors of 3 for the first two 
generations. As already mentioned in the discussion about the mass matrices presented 
in section \ref{secfmm}, the necessary corrective terms can arise from the neglected 
terms in the expansion in $\langle Y\rangle$ of $G_r(\langle Y\rangle)_{ij}$ \cite{elg}. 
The higher order terms correspond to non renormalizable operators with the 
insertion of $n$ factors of the 75, which break the transposition relation between 
$m_d$ and $m_e$. 
For this purpose we would like the expansion parameter $\langle Y\rangle/\Lambda$ 
to be not too small in order to naturally provide the required factors of 3.
We will present in the following explicit examples of parameter choices that lead to a 
realistic spectrum without unacceptable fine tuning. 

\subsection{A Numerical Example}

To reproduce the fermion mass spectrum we must further 
specialize the Yukawa couplings
introduced in eq. (\ref{w3}):
\bea
w_3 &=& \frac{1}{4}  \Psi_{10}^{A B}~ G_u~ \Psi_{10}^{C D}~ 
H^E \epsilon_{A B C D E}
+ \frac{\sqrt{3}}{2}  \Psi_{10}^{A B}~ G_{\overline{50}}~ \Psi_{10}^{C D}~ 
{H_{\overline{50}}}_{A B C D}
\nn\\
&+& \sqrt{2}~ \Psi_{10}^{A B}~ G_d~ {\Psi_{\bar 5}}_A~ {\overline H}_B
+\frac{1}{\Lambda}~\sqrt{2}~ \Psi_{10}^{A B}~ F_d~ {\Psi_{\bar 5}}_C~ 
{\overline H}_D~ Y^{C D}_{A B}\nn\\
&+& {\Psi_{\bar 5}}_A  ~G_\nu~\Psi_1~  H^A
- \frac{1}{2}~ M~ \Psi_1~ G_M~ \Psi_1+...
\label{w3yuk}
\eea
where $G_r~(r=u,d,\nu,M,{\overline{50}})$ is proportional to $G_r(\langle X\rangle,0)$ of 
eq. (\ref{w3}). We have
explicitly introduced a term linear in $Y$, whose couplings are described
by the $\langle X\rangle$-dependent matrix $F_d$.
There are other terms linear in $Y/\Lambda$, not explicitly
given above. In particular we may insert $Y/\Lambda$ in the renormalizable
term providing masses to the up type quarks. We neglect such a term 
since, on the one hand, the matrix $G_u$ is already sufficient to
correctly reproduce the up quark masses and, on the other hand,
this operator would not significantly modify the results for proton decay. 
As we will show, $Y/\Lambda$ is close to 0.1 in our model and 
higher order terms in the $Y/\Lambda$ expansion can be safely neglected.
The interaction term involving  ${H_{\overline{50}}}_{A B C D}$ does 
not contribute
to the fermion spectrum, but it will be relevant for the proton 
decay amplitudes. 

The term linear in $Y$ in the previous equation is sufficient to 
differentiate the spectra in the charged lepton and down quark sectors.
We get the following Dirac mass matrices:
\beq
m_{u,\nu}=y_{u,\nu}~ \frac{v_u}{\sqrt{2}}~~~,~~~~~m_{d,e}=y_{d,e}~ 
\frac{v_d}{\sqrt{2}}~~~~~,
\eeq
\beq
\begin{array}{lll}
y_u=G_u~~~& &y_\nu=G_\nu~~~~~,\\
& \\
y_d=G_d + \dd\frac{\langle Y\rangle}{\Lambda}F_d~~~& & 
y_e^T=G_d - 3 \dd\frac{\langle Y\rangle}{\Lambda}F_d~~~,
\end{array}
\eeq
where $v_u$, $v_d$ and $\langle Y\rangle$ parameterize the vevs of $H$, ${\overline H}$ and
$Y$ respectively. The fermion spectrum can be easily fitted by appropriately
choosing the numerical values of the matrices $G_u$, $G_d$, $F_d$, $G_\nu$  
and $G_M$. The most general fitting procedure would leave a large number of free 
parameters. Here we limit ourselves to the discussion of one particular example. 
In agreement with eq. (\ref{Mnu}), we take:

\beq
G_u=
\left[\matrix{(-0.51+0.61i) \lambda^6&   (0.42-0.70i) \lambda^5&   (0.27+0.86i) \lambda^3\cr
(0.42-0.70i) \lambda^5&   (-0.39+0.52i) \lambda^4&   (-0.30-1.14i) \lambda^2\cr
(0.27+0.86i) \lambda^3&   (-0.30-1.14i) \lambda^2&           1.39}
\right]~~~,
\label{gu}
\eeq

\beq
G_d=\lambda^4
\left[\matrix{(2.39-1.11i) \lambda^5&   (0.33-0.59i) \lambda^3&   (0.13+0.45i) \lambda^3\cr
(0.87+0.55i) \lambda^4&   (2.76+0.89i) \lambda^2&   (0.69-0.51i) \lambda^2\cr
(-1.50+0.94i) \lambda^2&   (0.45+1.78i)           &            1.94}
\right]~~~,
\label{gd}
\eeq

\beq
\frac{\langle Y\rangle}{\Lambda} F_d=\lambda^4
\left[\matrix{
(0.38-0.18i) \lambda^5&  (-0.16-0.06i) \lambda^3&   (0.07+0.04i)\lambda^3\cr
(-0.08-0.05i) \lambda^4&     (-0.20-0.15i) \lambda^2&    (0.15-0.11i) \lambda^2\cr
(-0.09+0.12i) \lambda^2&     (0.07+0.14i) &               -0.19}        
\right]~~~,
\label{fd}
\eeq 

\beq
G_\nu=
\left[\matrix{
(-0.78-0.19i) \lambda^3&  (0.52-0.34i) \lambda& (1.38+0.39i) \lambda^2\cr 
(-1.23-0.34i)\lambda&      0&          (1.04+1.31i)\cr
(0.45+1.18i)\lambda&  0 &        0.8+1.2i} 
\right]~~~,
\eeq

\beq
G_M=
\left[\matrix{
(1.50+0.55i) \lambda^2&    (1.41+1.19i)&    (0.35-1.53i) \lambda\cr
(1.41+1.19i)&             0&           0\cr            
(0.35-1.53i) \lambda&    0&      1.26+1.48i}
\right]~~~.
\eeq

While the generic pattern of $G_u$, $G_d$, $F_d$, $G_\nu$ and $G_M$ is dictated by
the $U(1)$ flavour symmetry, the precise values of the coefficients multiplying
the powers of $\lambda$ are chosen to reproduce the data. 
We take $\lambda\equiv\langle X\rangle/\Lambda=0.25$, $\tan\beta=1.5$ and $M=0.9\cdot 10^{15}$ GeV.    
In SU(5) the matrix $G_u$ contains two additional phases \cite{ncafasi},
$\phi_1$ and $\phi_2$ that have been set to zero in eq. (\ref{gu}).
These phases do not affect the fermion spectrum but enter the
proton decay amplitude. When discussing the proton decay we will
analyze also the dependence on $\phi_1$ and $\phi_2$.

From the above matrices we obtain, at the unification scale:

\beq
\begin{array}{lll}
m_t=200~ {\rm GeV}& m_c=0.27~ {\rm GeV}& m_u=0.9~ {\rm MeV}\\
   m_b=1.0~ {\rm GeV}& m_s=26~   {\rm MeV}& m_d=1.1~ {\rm MeV}\\
m_\tau=1.1~ {\rm GeV}& m_\mu=71~  {\rm MeV}& m_e=0.34~ {\rm MeV}
\end{array}
\eeq
\beq
|V_{us}|=0.22~~~~~|V_{ub}|=0.0022~~~~~ |V_{cb}|=0.052~~~~~J=1.9\cdot 10^{-5}~~~,
\eeq 
where $J$ is the CP-violating Jarlskog invariant.
In the neutrino sector, we find:
\beq
m_1=0.81\cdot 10^{-3} {\rm eV}~,~~~~m_2= 0.88\cdot 10^{-3} {\rm eV}~,
~~~~m_3=0.061~ {\rm eV}~~~.
\eeq
More precisely:
\beq
\Delta m^2_{sol}\equiv m_2^2-m_1^2=1.1\cdot 10^{-7}~ {\rm eV}^2~~,~~~~~
\Delta m^2_{atm}\equiv m_3^2-m_2^2=3.7\cdot 10^{-3}~ {\rm eV}^2~~~. 
\eeq
The neutrino mixing angles are
\beq
\theta_{12}\sim \frac{\pi}{4}~~,~~~~~\theta_{23}\sim \frac{\pi}{4}~~,~~~~~\theta_{13}=0.06~~~.
\eeq
%We have exploited here the well known fact that the operators obtained via an insertion
%of $Y$ differentiate between the down quarks and the charged leptons and may be arranged
%to break in the desired way the rigid mass relation of minimal $SU(5)$.
Until now we have not specified the value of $\langle Y\rangle/\Lambda$. We know 
that $\langle Y\rangle$
should be
around $M_{GUT}^{(0)}$. The cut-off $\Lambda$ cannot be too close
to $\langle Y\rangle$, otherwise most of the spectrum of the model would lie beyond the cut-off.
At the same time $\Lambda$ cannot be too large: it is bounded from above
by the scale at which the $SU(5)$ gauge coupling blows up, which, as we 
see from eq. (\ref{blow}), occurs more or less one order of magnitude below the Planck mass. This is
welcome. If, for instance, we take $\langle Y\rangle/\Lambda=0.05-0.1$, it is reasonable to neglect 
multiple insertions of $Y$ in the Yukawa operators. At the same time, in the example given in 
eq. (\ref{fd}), 
the coefficients of the powers of $\lambda$ in $F_d$ remain of order one, even if  
$\langle Y\rangle/\Lambda$ is as small as 0.05.
A too large cut-off would have been ineffective in separating down quarks from charged
leptons unless we had chosen unnaturally large coefficients in the allowed operators.   
What is usually considered a bad feature of the missing partner mechanism -
the lack of perturbativity before the Planck mass - turns out here to be an advantage
to provide a correct description of the fermion spectrum.

In the neutrino sector, we obtain a bimaximal neutrino mixing with the so-called 
LOW solution to the solar
neutrino problem  \footnote{As pointed out, in chapter 1,
latest preliminary results from Super-Kamiokande
\cite{lastSK}, 
including constraints from day-night spectra, seem in fact 
to prefer bimaximal neutrino mixing and to revamp interest 
in the LOW solution.}. Within the same $U(1)$ flavour symmetry 
considered here we could as well
reproduce the quasi-vacuum oscillation solution, by appropriately tuning the 
order one
coefficients in $G_\nu$ and $G_M$. We recall that the value 
of $M$ required to fit
the observed atmospheric oscillations is probably 
somewhat small in the context of
$SU(5)$, where a larger scale, closer to the cut-off 
$\Lambda$, is expected.
This feature might be improved by embedding the model in 
$SO(10)$ where $M$
is directly related to the $B-L$ breaking scale.

In conclusion, the known fermion spectrum can be reproduced starting from a superpotential
with order one dimensionless coefficients. Mass matrix elements for charged leptons and down quarks
match only within $10-20\%$, due to $\langle Y\rangle/\Lambda\approx 0.1$, and this produces the required 
difference between the two sectors. The neutrino mixing is necessarily bimaximal in our model,
with either the LOW or the vacuum oscillation solution to the solar neutrino problem.

\section{Proton Decay}
\label{secpdec}

Similarly to the case of minimal $SU(5)$, we expect that the 
main contribution to proton decay comes from  
the dimension five operators \cite{five} originating at the 
grand unified scale when the colour triplet superfields
are integrated out \cite{enr, susygut, hmy}. 
We denote the colour triplets contained in $H$, ${\overline H}$, $H_{50}$, 
${H_{\overline{50}}}$ 
by $H_{3u}$, $H_{3d}$, $H'_{3d}$ and $H'_{3u}$, respectively. 
The part of the superpotential depending
on these superfields reads:
\bea
w&=&H_{3u}\left[-\frac{1}{2} Q G_u Q + U^c G_u E^c\right]\nn\\
&+&H_{3d}\left[Q\hat{C} L 
+ U^c \hat{D} D^c\right]\nn\\
&+&H'_{3u}\left[\frac{1}{4} Q G_{\overline{50}} Q + U^c G_{\overline{50}} E^c\right]\nn\\
&-&4 \sqrt{3} c_2 \langle Y\rangle~ H'_{3d} H_{3u} 
- 4 \sqrt{3} c_3 \langle Y\rangle~ H_{3d} H'_{3u} + c_4 \langle X\rangle~ H'_{3d} H'_{3u} +...
\eea
where
\beq
\hat{C}=-G_d-\frac{\langle Y\rangle}{\Lambda} F_d~~,~~~~~~~~
\hat{D}=G_d-\frac{\langle Y\rangle}{\Lambda} F_d
\eeq
and $Q$, $L$, $U^c$, $D^c$ and $E^c$ denote as usual the chiral multiplets associated to 
the three fermion generations.
Notice that, at variance with the minimal $SU(5)$ model, an additional interaction term
depending on $H'_{3u}$ is present, which originates from the interaction of $\overline{50}$
with the $10$ matter fields. In Appendix \ref{appgroup} we give a short discussion
on its properties and justify the choice of the numerical coefficient $\sqrt{3}/2$ in eq
(\ref{w3yuk}).
%The coupling constants in $G_{\overline{50}}$ are completely independent from 
%those in $G_u$, even though both are constrained by the same $U(1)$ flavour symmetry.
By integrating out the colour triplets we obtain the following effective superpotential:
\beq
w_{eff}=\frac{1}{m_T}\left[Q \hat{A} Q~
Q \hat{C} L + U^c \hat{B} E^c~
U^c \hat{D} D^c\right]+...
\eeq
where $m_T$ has been defined in eq. (\ref{mT}),
\beq
\hat{B}=\left(G_u+\frac{4 \sqrt{3} c_2\langle Y\rangle}{c_4 \langle X\rangle} 
G_{\overline{50}}\right)~~~,~~~~~
-2 \hat{A}=
\left(G_u-\frac{2 \sqrt{3} c_2 \langle Y\rangle}{c_4 \langle X\rangle} G_{\overline{50}}\right)
\eeq
and dots stand for terms that do not violate baryon or lepton number.
Minimal $SU(5)$ is recovered by setting $G_{\overline{50}}=F_d=0$. 
As discussed, in that case the matrices 
$\hat{A}$, $\hat{B}$, $\hat{C}$ and $\hat{D}$ 
are determined by $G_u$ (in which now also $\phi_1$ and $\phi_2$ play a role) 
and $G_d$ and therefore strictly related 
to the fermionic spectrum \cite{enr}.
In our case we have a distortion due to the terms proportional to $G_{\overline{50}}$ 
and $F_d$.
These distortions have different physical origins. 
On the one hand the terms containing $F_d$ are 
required to avoid the rigid mass relation of minimal $SU(5)$. They are suppressed by
$\langle Y\rangle/\Lambda$ and we expect a mild effect from them \cite{inserz}.
On the other hand the terms containing $G_{\overline{50}}$ are a consequence 
of the missing doublet mechanism and they might be as important as those 
of the minimal model.

%At lower scales the dimension five operators give rise to the four fermion operators
%relevant to proton decay, via a ``dressing'' mainly due to 
%chargino exchange \cite{five,enr,dress}. 
%When considering the operators $QQQL$ the main contribution, 
%here called $L_1$,
%comes from the exchange of a wino 
%\footnote{Gluino dressing contributions cancel among each other in case of
%degeneracy between first two generations of squarks \cite{ehnt,gluino}.}. 
%Charged higgsino exchange provides instead
%the most important dressing of the operators $U^c U^c D^c E^c$~\cite{goto}. We term $L_2$ the 
%leading four-fermion operator in this case. Beyond $L_1$ and $L_2$ other 6 operators are generated
%by chargino exchange, 3 from $QQQL$ and 3 from $U^c U^c D^c E^c$. The contributions
%to the proton decay amplitudes from these operators are suppressed by at least 
%$\lambda^2$ with respect to those associated to $L_1$ and $L_2$, and can be safely neglected
%in the present estimate. $L_1$ and $L_2$ are given by:
%\bea
%L_1&=&\frac{K_1}{m_D}\frac{g^2}{8 \pi^2}~[f(\tilde{u},\tilde{d})+f(\tilde{u},\tilde{e})]~ 
%(\nu_l d_i)~(d_j u_k)~ P_{ij}~ Q_{kl}+{\rm h.c.}\nn\\
%L_2&=&-\frac{K_2}{m_D}\frac{1}{8 \pi^2}~f(\tilde{u^c},\tilde{e^c})~ 
%(\bar{d}_k \bar{\nu}_l)~(u^c_i d^c_j)~ S_{kl}~ T_{ij}+{\rm h.c.}~~~,
%\eea

\begin{figure} [ht]
\centerline{
\psfig{file=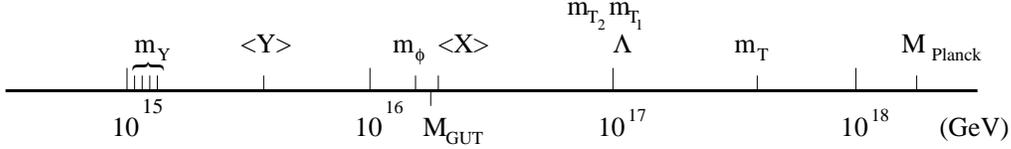,width=0.90\textwidth}
}
\caption{Heavy sector of the spectrum.}
\label{spectrum}
\end{figure}

To estimate the proton decay rates we should specify some important parameters.
First of all the mass $m_T$. From the discussion of the threshold 
corrections we know that a large value for $m_T$ is preferred in our model.
We should however check that this value can be obtained with reasonable choices
of the parameters at our disposal. The unification conditions constrain $\langle Y\rangle$
in a small range around $10^{16}$ GeV. More precisely, after the inclusion
of threshold corrections and two loop effects, eq. (\ref{linkmv})
becomes a relation among the masses of the super-heavy particles and the 
leading order quantity $M_{GUT}^{(0)}$. This relation limits the allowed range
for the heavy gauge vector bosons and indirectly, through eq. (\ref{mvy}), 
pushes the $SU(5)$ breaking vev $\langle Y\rangle$ close to $10^{16}$ GeV. 
Then the cut-off scale and the $X$ vev are fixed by the phenomenological requirements
$\langle Y\rangle\approx 0.05~ \Lambda$ and $\lambda\equiv \langle X\rangle/\Lambda=0.25$.
Large values of $m_T$ could be obtained either by taking $m_{T_1} m_{T_2}$ large
or by choosing a small $m_\phi$. This last possibility is however not practicable,
since the gauge coupling $\alpha_5$ blows up at approximately 20 $m_\phi$:
it is not reasonable to push $m_\phi$ below $5\cdot 10^{15}$ GeV. 
The only remaining freedom to obtain the desired large value for $m_T$
is represented by the coefficients $c_2$ and $c_3$ that, however,
cannot be taken arbitrarily large. We should also check that
all the heavy spectrum remains below $\Lambda$ and this requirement
imposes a further constraint on our parameters.
A choice that respects all these requirements
is provided by
\footnote{The values of the coefficients 
$c_2$ and $c_3$ here adopted raise doubts on the validity of the perturbative approach that has been 
exploited in several aspects of the present analysis.
We adhere to this choice also to show the difficulty met to obtain acceptable
phenomenological results within a not too complicated scheme.}:
\beq
c_1=0.035 ~~~,~~~~~c_2=c_3=2.8,~~~~~c_4=0.7~~~,
\eeq
\beq
\langle X\rangle=3.0\cdot 10^{16}~{\rm GeV},~~~~~\langle Y\rangle=5.7\cdot 10^{15}~{\rm GeV}~~~~~
(\Lambda=1.2\cdot 10^{17}~{\rm GeV})~~~,\eeq
which leads to:
\beq
M_{GUT}=2.9\cdot 10^{16}~{\rm GeV}~~~,~~~~~m_\phi=2.0\cdot 10^{16}~{\rm GeV}~~~,
\eeq
\beq
%m_\phi=2.0\cdot 10^{16}~{\rm GeV}~~~,~~~~~~
m_{T_1}=1.2\cdot 10^{17}~{\rm GeV}~~~,~~~~~
m_{T_2}=1.0\cdot 10^{17}~{\rm GeV}~~~,~~~~~m_T=6\cdot 10^{17}~{\rm GeV}~~~.
\eeq
The heavy sector of the particle spectrum is displayed in fig. \ref{spectrum}. 

To evaluate the loop function $f$
we also need the spectrum of the supersymmetric particles. As an example
we take here the same spectrum considered in Table \ref{tabsusysp}, 
with $m_{SUSY}=250$ GeV.
This leads to a squark mass of about $800$ GeV, a slepton mass of approximately $350$ GeV
a wino mass of $250$ GeV and a charged higgsino mass of $125$ GeV.

The matrix ${G_{\overline{50}}}$ is not directly related to any accessible observable 
quantity and the only constraint we have on it comes from the $U(1)$ flavour symmetry
that requires the following general pattern:
\beq
{G_{\overline{50}}}=
\left[\matrix{\lambda^7&   \lambda^6&   \lambda^4\cr
\lambda^6&   \lambda^5&   \lambda^3\cr
\lambda^4&   \lambda^3&   \lambda}
\right]~~~.
\label{g50b}
\eeq
With the above numerical values we also obtain:
\beq
\left| \frac{2 ~\sqrt{3}~ c_2~ \langle Y\rangle}{c_4~ \langle X\rangle} \right| \simeq 2.7~~~.
\eeq
The texture for ${G_{\overline{50}}}$ has an overall suppression factor $\lambda$ 
compared to $G_u$. Therefore the contribution of ${G_{\overline{50}}}$ to the matrices
$\hat{A}$ and $\hat{B}$ is comparable or even slightly larger than the minimal
contribution provided by $G_u$. 
The dominant operators are then the ones already described and the calculation can 
be carried out following Table \ref{chir}.
%the results of a chiral Lagrangian computation \cite{chiral}.
%The rates for the dominant channels are given in .

At variance of the minimal case, here the amplitude with $G_u$ and the one 
with ${G_{\overline{50}}}$ interfere. 
This interference can be either constructive or destructive, depending
on the relative phases between the two terms.
We have scanned several examples for ${G_{\overline{50}}}$, obtained by
generating random coefficients for the order one variables in eq. (\ref{g50b}).
By keeping fixed all the remaining parameters we obtain a proton decay 
rate 
% o lifetime?%
in the range $2\cdot 10^{32} \div 4\cdot 10^{34}$ yr for the channel 
$p\to K^+ \overline{\nu}$ and a rate
between $3\cdot 10^{32}$ yr and $5\cdot 10^{34}$ yr for the channel 
$ p\to\pi^+ \overline{\nu}$.
For comparison, considering the same choice of parameters but setting 
${G_{\overline{50}}}=F_d=0$, 
we obtain $9\cdot 10^{32}$ yr and $2 \cdot 10^{33}$ yr respectively, for the above 
channels. The present 90\% CL bound on $\tau/BR(p\to K^+\bar{\nu})$ is 
$1.9~10^{33}$ yr \cite{lastSK}.
These estimates have been obtained by setting to zero the two physical 
phases $\phi_1$, $\phi_2$ contained in the matrix $G_u$. 
These additional parameters may increase the uncertainty
on the proton lifetime. For instance, in minimal $SU(5)$, the proton decay rates for the
channels considered above, change by about one order of magnitude when $\phi_1$ and $\phi_2$
are freely varied between 0 and $2 \pi$. 
Even when the inverse decay rates for the channels $K^+ \bar{\nu}$ and $\pi^+ \bar{\nu}$
are as large as $10^{34}$ yr, they remain the dominant contribution to the proton lifetime.
Indeed, since the heavy vector boson mass, $M_{GUT}$, is equal to $2.9~ 10^{16}$ GeV in 
our model, the dimension 6 operators provide an inverse decay rate for the 
channel $e^+ \pi^0$ larger than $10^{36}$ yr. 

The effective theory considered here breaks down at the cut-off 
scale $\Lambda$. We expect 
additional non-renormalizable operators contributing 
to proton decay amplitudes
from the physics above the cut-off. 
We have already stressed that, by assuming dimensionless coupling constants of 
order one, in unified models without flavour symmetries the proton lifetime induced by 
these operators is unacceptably short, even when $\Lambda=M_{Pl}$ \cite{ehnt,nroforp}. 
In our case these contributions are adequately suppressed 
by the $U(1)$ symmetry. 
If we compare the amplitude $A_{nr}$ induced by the new non-renormalizable 
operators with the amplitude $A_3$ coming from the triplet exchange, we obtain, 
for the generic decay channel,
\beq
\frac{A_{nr}}{A_3}\approx \frac{48~ c_2~ c_3}{c_4} \left(\frac{\langle Y\rangle}{\Lambda}\right)^2\approx 1~~~~~.
\eeq
This supports the conclusion that a proton lifetime range considerably 
larger than the one estimated in minimal models is expected in our case.

\clearemptydoublepage
%\noindent
%{\bf Acknowledgements}

%\noindent
%We would like to thank Zurab Berezhiani, Andrea Brignole, Franco Buccella, Antonio Masiero, 
%Jogesh Pati, Massimo Porrati, Antonio Riotto, Anna Rossi, Carlos Savoy, Mario Tonin and Fabio Zwirner 
%for useful discussions. 
\addcontentsline{toc}{chapter}{Conclusions}
\pagestyle{plain} 

\chapter*{Conclusions}

We have considered the problem of obtaining 
the large atmospheric neutrino mixing  
in the framework of GUT models.

First, we have analyzed the most interesting mechanisms which provide
this large mixing. 

Usually the large atmospheric mixing arises because
of the presence of large mixings in the Dirac and/or Majorana right mass
matrices. In GUT models this situation is naturally realized by exploiting
the possible presence of large right mixings for down quarks, which correspond
to large left mixing for charged leptons. 

We have pointed out an alternative mechanism, which does not require
the presence of any large mixings. 
In fact, even when both the Dirac and Majorana
right matrices possess very small mixings, the large atmospheric mixing 
entirely arises
from the see-saw mechanism if two conditions apply, that is if
the spectrum of right Majorana masses is sufficiently splitted 
and if the left handed $\nu_{\mu}$ and $\nu_{\tau}$ couple 
with comparable strength to the right handed $\nu^c_{\mu}$.
As opposite to the first mechanism,
this situation can be realized even if all left and right mixings of quarks 
and leptons are very small. 
This feature is particularly attractive from the theoretical point of
view, because it implies that this mechanism is well compatible with those
GUT models where right and left mixings are similar.

Despite the fact that GUT has been 
studied since a long time, still a lot of its aspects 
require a better understanding. It would be really improbable
for this idea to survive if inside it there where not something true.
But it is clear that minimal versions of 
SUSY GUTs are already ruled out because unrealistic. 
We have addressed the question whether it is possible to construct
a not too complicated, relatively complete and realistic model which
can serve as benchmark to be compared with experiment.
This model must also correctly reproduce neutrino masses and
mixings.

We have then constructed an 
example of SUSY $SU(5)$ GUT model, with an additional
$U(1)$ flavour symmetry, which is not plagued by the
need for large amounts of fine tunings, like those associated with
doublet-triplet splitting in the minimal model, and leads
to an acceptable phenomenology. This includes coupling unification with a
value of $\alpha_3(m_Z)$ in much better agreement
with the data than in the minimal version, an acceptable pattern for
fermion masses and mixing angles, also including
neutrino masses and mixings, and the possibility of a slower proton decay
than in the minimal version, compatible with the
present limits (in particular the limit from Super-Kamiokande of about
$2\cdot 10^{33}$ yr for the channel $p\rightarrow K^+\bar{\nu}$). 
In the neutrino sector the
preferred solution is one with nearly maximal mixing both for
atmospheric and solar neutrinos. 

The $U(1)$ flavour symmetry plays a crucial role by protecting 
the light doublet Higgs mass from receiving large mass contributions 
from higher dimension operators, by determining the 
observed hierarchy of fermion masses and
mixings and by providing an adequate suppression for non renormalizable
operators inducing proton decay. 
Of course, the $U(1)$ symmetry can only
reproduce the order of magnitude of masses and mixings, while more
quantitative relations among masses and mixings can only
arise from a non abelian flavour symmetry.  

A remarkable feature of the model is that the presence of the
representations $50$, $\overline{50}$ and $75$, demanded by the 
Missing Doublet Mechanism for the solution of the
doublet-triplet splitting problem, directly
produces, through threshold corrections at $M_{GUT}$ from the $75$, a
decrease of the value of $\alpha_3(m_Z)$ that  
corresponds to coupling unification and an increase of the effective mass
that mediates proton decay by a factor of
typically 20-30.  As a consequence the value of the strong coupling is in
better agreement with the experimental value and
the proton decay rate is smaller by a factor 400-1000 than in the minimal
model. The presence of these large representations
also has the consequence that the asymptotic freedom of $SU(5)$ is spoiled
and the associated gauge coupling becomes non
perturbative below 
$M_{Pl}$. We argue that this property far from being unacceptable can
actually be useful to obtain better results for fermion
masses and proton decay.

Clearly such a model is not unique: our version is the simplest realistic
model that we could construct. We think it is
interesting because it proves that SUSY $SU(5)$ GU is not excluded and
offers a benchmark for comparison with experiment.
For example, even including all possible uncertainties, it is difficult in
this class of models to avoid the conclusion that
proton decay must occur with a rate which is only a factor 10-50 from the
present bounds. Failure to observe such a signal
would require some additional specific mechanism in order to further
suppress the decay rate.

Finally, it is a generic
feature of realistic models that the region between $M_{GUT}$ and $M_{Pl}$
becomes populated by many states with
different thresholds and also non perturbative phenomena occur. This
suggests that the reality can be more complicated than
the neat separation between the Grand Unification and the string regime which is
postulated in the simplest toy models of GUTs.

There is then the need for further ideas on the theoretical side.
It would be of crucial importance to actually understand the origin of 
the flavour symmetry. The simple $U(1)$ used in the present model 
is satisfactory at the level of orders of magnitude, but is not
able to really predict the observed values of masses and mixing.
Maybe the flavour has a deeper origin, in a 
fundamental theory of particles that
includes also gravity beyond the strong and electroweak interactions.

\addcontentsline{toc}{chapter}{Acknowledgements}

\chapter*{Acknowledgements}

I warmly thank Guido Altarelli and Ferruccio Feruglio for
their invaluable and pleasant collaboration. 
I would like to thank also Giovanni Costa, 
Gianfranco Sartori, Gianpaolo Valente and Fabio Zwirner
for useful discussions.

\clearemptydoublepage
\pagestyle{headings}

\clearemptydoublepage

\appendix
\chapter[The Calculation of Proton Lifetime]
{Formulae for the Calculation of Proton Lifetime}

\label{appprot}
\section{d=5 B Violating Operators}

The starting point is the most general renormalizable superpotential 
which could come from $SU(5)$. 
Yukawa couplings of doublets and triplets will be considered 
uncorrelated as well as down quarks and charged lepton mass matrices. 
The discussion can thus be adapted for any $SU(5)$ based model.
Omitting colour and weak isospin indices, in the interaction basis the 
relevant terms read:
\bea
w &\ni & Q^T y_u U^c H_{2u} +Q^T y_d D^c H_{2d} + L^T y_e E^c H_{2d} \nn\\
&+& Q^T \hat{A} Q H_{3u}+ {U^c}^T \hat B E^c H_{3u} +Q^T \hat C L H_{3d} +
{U^c}^T \hat D D^c H_{3d}\\
&+& m_T H_{3u} H_{3d}\nn~~~~~,
\eea
where $Q$, $L$, $U^c$, $D^c$ and $E^c$ denote as usual the chiral multiplets 
associated to the three fermion generations, $y_{u,d,e}, \hat A, \hat B, ...$ 
are matrices in generation space and $m_T$ is the triplets mass. 
By integrating out the colour triplets one obtains the following effective 
superpotential
\beq
w_{eff}=\frac{1}{m_T}\left[Q \hat{A} Q~
Q \hat{C} L + U^c \hat{B} E^c~
U^c \hat{D} D^c\right]+...~~~~~,
\eeq
dots standing for terms that do not violate baryon or lepton number.

\begin{table}[!h] 
\begin{center}
\begin{tabular}{|c|}
\hline
\\
$O^L_1 = \dd\frac{g^2}{m_T}    
                      ( \nu_l d_i) (d_j u_k) {[ L_d^T (\hat A+{\hat A}^T) L_d] }_{i j} {[ L_u^T \hat C L_e] }_{k l}
                      (f_{\tilde{w}}({\tilde{u}},{\tilde{e}})+f_{\tilde{w}}({\tilde{u}},{\tilde{d}}))$\\
\\
$O^L_2 = \dd\frac{g^2}{m_T}    
                      ( u_i d_j) (d_k \nu_l) {[ L_u^T (\hat A+{\hat A}^T) L_d] }_{i j} {[ L_d^T \hat C L_e] }_{k l}
                      (f_{\tilde{w}}({\tilde{u}},{\tilde{e}})+f_{\tilde{w}}({\tilde{u}},{\tilde{d}}))$\\
\\
$O^L_3 = - \dd\frac{g^2}{m_T}    
                      ( u_i d_j) (u_k e_l) {[ L_u^T (\hat A+{\hat A}^T) L_d] }_{i j} {[ L_u^T \hat C L_e] }_{k l}
                      (f_{\tilde{w}}({\tilde{d}},{\tilde{\nu}})+f_{\tilde{w}}({\tilde{u}},{\tilde{d}}))$\\
\\
$O^L_4 = \dd\frac{g^2}{m_T}    
                      ( u_j d_k) (u_i e_l) {[ L_u^T (\hat A+{\hat A}^T) L_u] }_{i j} {[ L_d^T \hat C L_e] }_{k l}
                      (f_{\tilde{w}}({\tilde{d}},{\tilde{u}})+f_{\tilde{w}}({\tilde{d}},{\tilde{\nu}}))$\\
\\
\hline
\\
$O^R_1 = - \dd\frac{1}{m_T}    
                      ( \bar d_k \bar \nu_l) (u^c_i d^c_j) {[ L_d^+ y_u^* \hat B y_e^+ L_e^*] }_{k l} 
                                                   {[ R_u^* \hat D R_d^+] }_{i j}
                                        f_{\tilde{h}}({\tilde{u}}^c,{\tilde{e}}^c)$\\
\\
$O^R_2 = - \dd\frac{1}{m_T}    
                      ( {\bar \nu}_l {\bar d}_i) (d^c_j u^c_k) {[ R_u^* \hat B y_e^+ L_e^*] }_{k l} 
                                          {[ L_d^+ y_u^* \hat D R_d^+] }_{i j}
                                      f_{\tilde{h}}({\tilde{u}}^c,{\tilde{e}}^c)$\\
\\
$O^R_3 = - \dd\frac{1}{m_T}    
                      ( {\bar d}_i {\bar u}_j) (u^c_k e^c_l) {[ R_u^*  \hat B  R_e^+] }_{k l} 
                             {[ L_d^+ y_u^*  \hat D y_d^+ L_u^*] }_{i j}
                                      f_{\tilde{h}}({\tilde{u}}^c,{\tilde{d}}^c)$\\
\\
$O^R_4 = - \dd\frac{1}{m_T}    
                      ( \bar u_j \bar d_k) (e^c_l u^c_i) {[ L_d^+  y_u^* \hat B  R_e^+] }_{k l} 
                             {[ R_u^* \hat D y_d^+ L_u^*] }_{i j}
                                f_{\tilde{h}}({\tilde{d}}^c,{\tilde{u}}^c)$\\
\\
\hline
\end{tabular}
\end{center}
\caption{Four-fermion operators arising from dimension 5 operators.
For each operator it is understood to also add the h.c.}
\label{operatgen}
\end{table}

At lower scales these dimension five operators give rise to the 
four-fermion operators relevant to proton decay, via a "dressing'' process.
When considering the operator $QQQL$, the most important dressing 
comes from wino exchange 
\footnote{Gluino dressing contributions cancel among each other in case of
degeneracy between first two generations of squarks \cite{ehnt,gluino}.}. 
Charged higgsino exchange provides instead the most important dressing 
of 
$U^c E^c U^c D^c$ \cite{goto}. 

The integration over the loop has the effect of producing the multiplicative factor
$g^2 f_{\tilde{w}}({\tilde{a}},{\tilde{b}})$ for wino dressing ($g$ being the $SU(2)$ gauge 
coupling) and $f_{\tilde{h}}({\tilde{a}},{\tilde{b}})$ for higgsino dressing.
$f$ is a function coming from the loop integration:
\bea
f_{\tilde{c}}({\tilde{a}},{\tilde{b}})&=& \int{ \frac{d^4 k}{(2 \pi)^4} \frac{-i}{k^2-m^2_{\tilde{a}}}
 \frac{-i}{k^2-m^2_{\tilde{b}}} \frac{-i } { \sls{k} - m_{\tilde{c}} }}
\nn\\
&=&\frac{1}{16 \pi^2} \frac{m_{\tilde{c}}}{(m^2_{\tilde{a}}-m^2_{\tilde{b}})}\left[
\frac{m^2_{\tilde{a}}}{m^2_{\tilde{a}}-m^2_{\tilde{c}}}\log
\frac{m^2_{\tilde{a}}}{m^2_{\tilde{c}}}-
\frac{m^2_{\tilde{b}}}{m^2_{\tilde{b}}-m^2_{\tilde{c}}}\log
\frac{m^2_{\tilde{b}}}{m^2_{\tilde{c}}}\right]~~~.
\label{floop}
\eea

The operators are explicitly given in Table \ref{operatgen}, where we
made the not too bad assumption of degeneracy between the generations 
of each sparticle (the dependence on sparticle masses is logarithmic), 
so that the final expression is more compact.
We also
turned to the mass eigenstate basis by
indicating with $L_{u,d,e}$ and $R_{u,d,e}$ the unitary matrices that 
diagonalize the 
fermion mass matrices:
\beq
L_u^T    y_u R_u^\dagger = y_u^{(d)}~~~~~~~~- L_d^T y_d R_d^\dagger = y_d^{(d)}~~~~~~~~
- L_e^T y_e R_e^\dagger = y_e^{(d)}~~~,
\eeq
with $(y_f)_{(d)}~~(f=u,d,e)$ diagonal and positive. The quark mixing matrix
is then $V_{CKM}=L_u^\dagger L_d$. 
% si fa anche l'ipotesi di f non troppo differenti per d di 1 e 2 gen, entro lambda2.!

\section{The Relevant Operators}

As discussed in chapter \ref{pfsg}, the most important operators are
$O^L_1$ and $O^R_1$, which are explicitly given by:
\bea
O^L_1&=&C^L_{ijkl}~ (\nu_l d_i)~(d_j u_k)+{\rm h.c.}\nn\\
O^R_1&=&C^R_{ijkl}~ (d_k \nu_l)~(\overline{u^c}_i \overline{d^c}_j)+{\rm h.c.}~~~,
\eea
where $i,j,k,l$ are generation indices and
\bea
C^L_{ijkl}&=&\frac{K_L}{m_T} g^2~[f_{\tilde{w}}(\tilde{u},\tilde{d})
+f_{\tilde{w}}(\tilde{u},\tilde{e})]~P_{ij}~ Q_{kl}\nn\\
C^R_{ijkl}&=&-\frac{K_R}{m_T} f_{\tilde{h}}(\tilde{u^c},\tilde{e^c})
S_{kl}^*~ T_{ij}^*
\eea
\beq
\begin{array}{lll} 
P=2~ L_d^T \hat{A}~ L_d & &Q=L_u^T \hat{C}~ L_e\\
& \\
S=L_d^\dagger~ y_u^*~ \hat{B}~ y_e^\dagger~ L_e^*& &T=R_u^*~ \hat{D}~ R_d^\dagger~~~.
\end{array}
\eeq

Considering $O^L_1$, for the channel $p \rightarrow K^+ \bar\nu$ we have to analyze
the contributions from both $(\nu_l d)~(s u)$, which corresponds to $C^L_{121l}$,
and $(\nu_l s)~(d u)$, which corresponds to $C^L_{211l}$.
Since $C^L_{ijkl}$ is proportional to $P_{ij} Q_{kl}$, it is possible to identify
immediately the important $P,Q$ elements. For instance $C^L_{121l}$ is linked to
$P_{12}$ and $Q_{1l}$. Note that $P$ is symmetric, hence $C^L_{121l}=C^L_{211l}$.     
For minimal $SU(5)$, $P, Q$ greatly simplify, as can be seen from Table \ref{opmin}.
In this case 
\beq
P_{12}=[V^T y_u^{(d)} K^+ V]_{12}\sim y_{top} \lambda^5 ~, \eeq
and $Q_{1l}=[V^* y_d^{(d)}]_{1l}$, 
so that \beq Q_{12}\sim Q_{13} \sim y_{bot} 
\lambda^3 \gg Q_{11}~.\eeq 
Thus, since 
\beq y_{top} y_{bot}\sim \lambda^2 (\tan\beta+1/ {\tan\beta})~~,\eeq 
one obtains 
\beq P_{12} Q_{12(3)} \sim \lambda^{10} (\tan\beta+1/ {\tan\beta})~~,\eeq 
as shown in Table \ref{pow}.  
For the channel $p \rightarrow \pi^+ \bar\nu$, we have only to consider
$(\nu_l d)(d u)$, that is $P_{11} Q_{1l}$. Since $P_{11}\sim \lambda P_{12}$,
then \beq
P_{11} Q_{12(3)} \sim \lambda^{11} (\tan\beta+1/ {\tan\beta})~.\eeq

In the case of $O^R_1$, the channel with $K^+$ is obtained for 
$(d \nu_l)~(\overline{u^c} \overline{s^c})$, corresponding to $S_{1l} T_{12}$, and  
$(s \nu_j)~(\overline{u^c} \overline{d^c})$, corresponding to $S_{2l} T_{11}$.
In the minimal case, from 
\beq S=V^+ {y_u^{(d)}}^2 V y_d^{(d)}~~,\eeq 
it turns out that the dominant contribution corresponds to $l=3$, 
that is emission of $\bar\nu_{\tau}$,
and 
$S_{13}\sim y_{top}^2 y_{bot}\lambda^3$, $S_{23}\sim y_{top}^2 y_{bot}\lambda^2$.
From 
\beq T=K V^* y_d^{(d)} \eeq 
it follows $T_{12} \sim y_{bot} \lambda^3$, $T_{11} \sim  y_{bot} \lambda^4$. 
Thus, 
\beq S_{13} T_{12} \sim y_{top}^2 y_{bot}^2 \lambda^6 \sim \lambda^{10}
(\tan\beta+1/ {\tan\beta})^2 \sim S_{23} T_{11}~~.\eeq
For the channel with $\pi^+$, we have to consider 
$(d \nu_l)~(\overline{u^c} \overline{d^c})$,
that is \beq S_{13} T_{11} \sim \lambda^{11}
(\tan\beta+1/ {\tan\beta})^2~~.\eeq
These results are given in Table \ref{pow}.
  
Note that we have also introduced
$K_{L,R}$, which are constants accounting for the renormalization of the operators from 
the grand unification scale down to 1 GeV \cite{enr, hmy, ren}.
In our estimates we take $K_1=K_2=10$.

The next step is to parameterize the partial rates 
according to the results of a chiral Lagrangian computation \cite{chiral}.
The rates for the dominant channels are given in Table ~\ref{chir}.

\begin{table} [ht]
%\title{Proton decay rates}
{\begin{center}
%\footnotesize
\begin{tabular}{|c|c|}   
\hline
& \\                         
channel & rate\\ 
& \\
\hline
& \\
$K^+ \bar{\nu}_\mu$ & $X_K \left\vert \beta \left[\frac{2}{3}\frac{m_p}{m_B} D C^L_{1212}+
\left(1+\frac{m_p}{3 m_B}(D+3F)\right)C^L_{2112}\right]\right\vert ^2$\\ 
& \\
\hline  
& \\
$K^+ \bar{\nu}_\tau$ & $X_K \left\vert \beta \left[\frac{2}{3}\frac{m_p}{m_B} D C^L_{1213}+
\left(1+\frac{m_p}{3 m_B}(D+3F)\right)C^L_{2113}\right]\right.$\\
& \\
&$\left.+\alpha \left[\frac{2}{3}\frac{m_p}{m_B} D C^R_{1312}+
\left(1+\frac{m_p}{3 m_B}(D+3F)\right)C^R_{2311}\right]\right\vert^2$\\ 
& \\
\hline
& \\
$\pi^+ \bar{\nu}_\mu$ & $X_\pi\left\vert\beta \left(1+D+F\right) 2 C^L_{1112}\right\vert^2$\\
& \\
\hline
& \\
$\pi^+ \bar{\nu}_\tau$ & $X_\pi\left\vert\beta \left(1+D+F\right) 2 C^L_{1113}
+\alpha \left(1+D+F\right) 2 C^R_{1311}\right\vert^2$\\
& \\
\hline
\end{tabular} 
\end{center}}
\caption{Proton decay rates. We define $X_{K,\pi}=(m^2_p-m^2_{K,\pi})^2/(32\pi m_p^3 f_\pi^2)$; 
$m_p$, $m_K$, $m_\pi$ are the proton, $K^+$ and $\pi^+$ masses;
$m_B$ is an average baryon mass, $f_\pi$ is the pion decay constant;
$D$ and $F$ are coupling constants between baryons and mesons in the relevant
chiral Lagrangian; $\beta$ and $\alpha$ parameterize the hadronic matrix element. 
In our estimates we take $m_p=0.938$ GeV, $m_B=1.150$ GeV, $m_K=0.494$ GeV, $m_\pi=0.140$ GeV, 
$f_\pi=0.139$ GeV, $D=0.8$, $F=0.45$ \cite{DF}, $\beta=-\alpha=0.003$ GeV$^3$ \cite{albe}. }
\label{chir}
\end{table}
\clearemptydoublepage

\chapter{Representative SUSY Spectrum}

\begin{table}[h]
{\begin{center}
\begin{tabular}{|c||c|}   
\hline
& \\                         
sparticle & mass$^2$\\ 
& \\
\hline \hline & \\
gluinos & $(2.7 m_{1/2})^2$ \\& \\
\hline& \\
winos & $(0.8 m_{1/2})^2$\\& \\ 
\hline& \\
higgsinos & $\mu^2$\\& \\
\hline& \\
extra Higgses & $m_H^2$\\& \\
\hline& \\
squarks & $m_0^2+6 m^2_{1/2}$\\& \\
\hline& \\
(sleptons)$_L$ & $m_0^2+0.5 m^2_{1/2}$\\& \\
\hline& \\
(sleptons)$_R$ & $m_0^2+0.15 m^2_{1/2}$\\& \\
\hline
\end{tabular} 
\end{center}}
\caption{Representative SUSY spectrum.}
\label{tabsusysp}
\end{table}

$SU(2)\otimes U(1)$ breaking effects are neglected.
Numerical coefficients appearing in the previous Table are due
to renormalization. 

The additional freedom related to the parameters $m_0$, $m_{1/2}$, $\mu$, $m_H$ is 
here fixed by choosing $0.8 m_0=0.8 m_{1/2}=2 \mu=m_H$ and taking as 
a definition $m_{SUSY}\equiv m_H$, so that all particle masses can be 
expressed in term of $m_{SUSY}$. This parameterization
leads to $k(SUSY)=-0.510$.

\clearemptydoublepage
\chapter{Group Theory for the MDM}

\label{appgroup}

In this appendix we summarize the method adopted to deal with 
tensors in $SU(5)$. It is convenient to write large representations 
like $75$, $50$ and $\overline{50}$ as products of smaller representations, 
like $10$ and $\overline{10}$. In this way, it is not difficult to recognize
how the multiplets of the SM are embedded in those large representations,
thus staying in touch with the physical role of the various superpotential
terms. In particular, when considering the interaction terms, it is compelling to 
accurately take into account the numerical coefficients which follow directly from 
the tensorial structure of the representations involved. 
Since these coefficients can be large, it is not possible to omit them 
when performing a quantitative analysis. 

\section{Basic Conventions for the Tensors in $SU(5)$}

In $SU(5)$ the matter fields are contained in just the $10$ and $\overline{5}$ representations.
We write them according to 
\beq
\Psi_{10}=\frac{1}{\sqrt{2}}
\left[\matrix{0&   U^c_3& -U^c_2  & U^1&D^1 \cr
            -U^c_3   &   0        &  U^c_1  & U^2 &  D^2 \cr
           U^c_2  &   -U^c_1 &       0     &   U^3  &  D^3 \cr
            -U^1      &   -U^2    &  -U^3     &     0    &   E^c \cr
            -D^1     &     -D^2    &   -D^3    &  -E^c   &   0     }
\right]~~~,
\label{psi10}
\eeq
\beq
{\Psi}_{\bar 5} = \left[ \matrix{ D^c_1 \cr D^c_2 \cr D^c_3 \cr E \cr -\nu } \right]~~~.
\label{psi5bar}
\eeq
The $10$ representation $\Psi_{10}^{AB}$ is thus represented by a tensor whose two 
indices $A,B=1,...,5$ are antisymmetric. Note that the $SU(3)$ indices correspond
to $A,B=1,2,3$ while the $SU(2)$ ones are selected for $A,B=4,5$. To remember about
this in an explicit way, we will sometimes write $\alpha$ ($a$) instead of 
$A$ if $A=1,2,3$ ($4,5$). Due to the antisymmetry of $\Psi_{10}^{AB}$,
it turns out that it is convenient to deal just with $\Psi_{10}^{AB}~|_{A<B}$ .
The $\overline{5}$ representation is instead
written as a tensor with one low index ${\Psi_{\bar 5}}_A$ .

In our conventions, the Higgs fields ${\bar H}_A, H^A$
belonging respectively to the $\overline{5}$ and $5$ representations are written as
\beq
{\bar H} = \left[ \matrix{ {H_{3d}}_1 \cr {H_{3d}}_2 \cr{H_{3d}}_3  \cr H_{2d}^- \cr 
-H_{2d}^0} \right]~~~~~~~~~~~
H = \left[ \matrix{ H_{3u}^1 \cr H_{3u}^2 \cr H_{3u}^3  \cr H_{2u}^+ \cr 
H_{2u}^0} \right]~~~~,
\eeq
where $\langle H_{2d}^0 \rangle = v_d/\sqrt{2}, \langle H_{2u}^0 \rangle = v_u/\sqrt{2}$.

Let us underline a few points.
\begin{itemize}
\item
The factor $1/\sqrt{2}$ is introduced in the definition of $\Psi_{10}$ in order
to have the kinetic terms in the canonical form. In fact, the Kahler potential 
contains
\beq
K \ni  \Psi_{10}^{A B} {\Psi_{10}^{A B}}^* ~~~,
\eeq
where summation over repeated indices has to be understood.
Due to the antisymmetry of $\Psi_{10}$ under $A \leftrightarrow B$,
one immediately realizes that:
\bea 
\Psi_{10}^{A B} {\Psi_{10}^{A B}}^* &=& 2 ~\Psi_{10}^{A B} {\Psi_{10}^{A B}}^* ~\dd|_{A<B} =
\nn\\ 
&=& U^c_\alpha {U^c_\alpha}^* + U^{\alpha} {U^{\alpha}}^*+
D^{\alpha} {D^{\alpha}}^*+E^c {E^c}^*~~~,
\eea
where $\alpha$ is the colour index.
Thus, the fields $U^c,U,D,E^c$ have a canonical kinetic term.
On the other hand the pentaplet ${\Psi_{\bar 5}}_A$ has no problems of multiplicity,
so that there is no need of introducing any coefficient in the definition of (\ref{psi5bar})
to get the canonical form for the kinetic terms of $D^c,E,\nu$.
\item At the contrary of what happens for all the other ${\Psi_{10}^{AB} ~}_{A<B}$, 
we have introduced a minus sign in $\Psi_{10}^{13}$. The reason is the following.
The mass term for up quarks (considering for simplicity one generation only) 
arises from the superpotential term 
\bea
w &\ni & \frac{1}{4} \Psi_{10}^{AB} \Psi_{10}^{CD} H^E \epsilon_{ABCDE}= \nn\\
  &=& \Psi_{10}^{AB} \Psi_{10}^{CD} H^E \epsilon_{ABCDE} \dd|_{A<B,C<D}
\eea
with $E=5$. When one picks the indices corresponding to $U^c_2 U^2$, that is
$(A,B,C,D)=(1,3,2,4)$, a minus sign arises from $\epsilon_{13245}$, at the contrary
of what happens for $U^c_1 U^1$ and $U^c_3 U^3$. The minus sign in $\Psi_{10}^{13}$
is put to eliminate the minus arising from the $\epsilon$.
Clearly there is nothing restrictive in doing this, since we are only exploiting
the freedom of choosing a convenient phase in the definition of our fields.  
\end{itemize}

\section{The $75$ and $50$ representations}

\subsubsection{How to write the $75$}

The $75$ can be conveniently written through the product
\beq
10 \times \overline{10}=1+24+75 ~~.
\eeq
Let us clarify how to isolate the $75$ from the $24$ and the $1$.
We define $10^{AB} \times\overline{10}_{CD} \equiv T^{A B}_{C D}$,
so that $T^{A B}_{C D}$ is antisymmetric in the upper and lower pair
of indices and thus corresponds to 100 independent fields.
It can be split according to
\beq
T^{A B}_{C D}=\underbrace{S^{A B}_{C D}}_{1}+\underbrace{\Sigma^{A B}_{C D}}_{24}+
\underbrace{Y^{A B}_{C D}}_{75}
\label{tabcd}
\eeq
where we identify the $1$ with $S$, the $24$ with $\Sigma$ and the $75$ with $Y$.

Consider the $SU(5)$ singlet first.
One can easily guess it to be given by the combination
\footnote{Here and in the following summation over repeated indices has to be 
understood.} 
\beq
s\equiv T_{MN}^{MN}~~~.
\eeq 
Thus, $S$ has to be
\beq
S^{AB}_{CD}=\frac{1}{20} (\delta^A_C \delta^B_D -\delta^A_D \delta^B_C) T^{M N}_{M N } ~~.
\label{sab}
\eeq
In this case, in fact, substituting $S^{AB}_{CD}$ into eq. (\ref{tabcd})
it turns out that $S^{AB}_{AB}=T^{AB}_{AB}\equiv s$ and that $\Sigma$ and $Y$
must be such that $\Sigma^{AB}_{AB}=Y^{AB}_{AB}=0$.
This indeed means that the singlet has been completely projected out by $S$.

The $24$ can be written as a tensor $\sigma^A_C$ satisfying the constraint 
$\sigma^A_A=0$ .
One can guess 
\beq
\sigma^A_C \equiv T^{MA}_{MC} -\frac{1}{5} \delta^A_C T^{MN}_{MN}~~
\eeq
and
\bea
\Sigma^{AB}_{CD}&=& \dd\frac{1}{3} \left[ 
                       \delta^A_C (T^{M B}_{M D} -\frac{1}{5}\delta^B_D T^{M N}_{M N })
                      -\delta^A_D (T^{M B}_{M C} -\frac{1}{5}\delta^B_C T^{M N}_{M N }) \right. \nn\\
                     &~ & ~~~-\left.\delta^B_C (T^{M A}_{M D} -\frac{1}{5}\delta^A_D T^{M N}_{M N })
                      +  \delta^B_D  (T^{M A}_{M C} -\frac{1}{5}\delta^A_C T^{M N}_{M N }) \right]~~.
\label{sigmaabcd}
\eea
Notice that this form for $\Sigma$ satisfies $\Sigma^{AB}_{AB}=0$, that is $\Sigma$ doesn't
contain the combination corresponding to the singlet $s$.
Substituting this expression in eq. (\ref{tabcd}), we get $\Sigma^{MA}_{MC}=\sigma^A_C$,
$S^{MA}_{MC}=0$ and the requirement $Y^{MA}_{MC}=0$. This means that the $24$ is completely
projected out by $\Sigma$. 

Finally, with the expressions of $S^{AB}_{CD}$ and $\Sigma^{AB}_{CD}$ at our disposal, we know
how to write the $75$:
\bea
Y^{A B}_{C D} &= & T^{A B}_{C D}\nn\\
                          & - & \dd\frac{1}{3} \left[ 
                       \delta^A_C (T^{M B}_{M D} -\frac{1}{5}\delta^B_D T^{M N}_{M N })
                      -\delta^A_D (T^{M B}_{M C} -\frac{1}{5}\delta^B_C T^{M N}_{M N }) \right. \nn\\
                     &~ & ~~~-\left.\delta^B_C (T^{M A}_{M D} -\frac{1}{5}\delta^A_D T^{M N}_{M N })
                      +  \delta^B_D  (T^{M A}_{M C} -\frac{1}{5}\delta^A_C T^{M N}_{M N }) \right]\nn\\
                       &- & \frac{1}{20} (\delta^A_C \delta^B_D -\delta^A_D \delta^B_C) T^{M N}_{M N }~~~.
\label{yabcd}
\eea
Note that $Y$ is antisymmetric in both the upper and lower indices. This would give
100 independent fields but, since it satisfies the 25 constraints $Y^{MA}_{MC}=0$,
we are correctly left with only 75 independent fields.

It is important to realize that the fields in $Y^{A B}_{C D}$ don't have canonical kinetic terms.
This can be recognized by the following reasoning. Let us suppose to put the 100 elements of 
$Y^{AB}_{CD}|_{A<B,C<D}$ into a column vector, say ${\cal Y}_{\cal A}$, with ${\cal A}=1,2,...,100$.
Clearly only 75 of these 100 elements correspond to independent fields. 
The Kahler potential is then 
\bea 
K & \ni &  {Y^{A B}_{C D}}^* Y^{A B}_{C D} = 4 ~ {Y^{A B}_{C D}}^* Y^{A B}_{C D} ~|_{A<B,C<D}=\nn\\
  &  =  &  4 ~{\cal Y}^+ ~ {\cal Y} ~.
\label{kappay}
\eea
We can always exploit the possibility of rotating the fields by means of a 100 by 100 unitary matrix $U$. 
For instance, we can get rid of the unpleasant redundancy by going into a basis where 
${\cal Y}$ has only 75 non-zero elements.
Then, it is also feasible to find a unitary transformation of the 75 fields which carry us into 
a SM-like basis, that is a basis where the fields are split into multiplets of the SM.  
In a similar basis it is evident that the kinetic terms are not canonical because of the factor $4$
appearing in eq. (\ref{kappay}). The remedy is to define 
${Y}^{A B}_{C D} \equiv  {Y'}^{A B}_{C D}/2$ (and ${\cal Y} \equiv {\cal Y'}/2$), so that 
the fields associated to ${Y'}^{A B}_{C D}$ are those actually possessing the correct coefficient 
1 in the kinetic term.

\subsubsection{How to write the $50, \overline{50}$}

By similar considerations, the $50$ can be written from the product of $10 \times 10$.
In fact
\beq
10 \times 10=\bar 5_s+\bar{45}_a+{50}_s~~~.
\eeq
We denote the $10 \times 10 $ with $T_1^{A B} T_2^{C D}$, where $T_{1,2}$ are antisymmetric under
exchange of the two upper indices.
Again we want to separate $T_1^{A B} T_2^{C D}$ as
\beq
T_1^{A B} T_2^{C D} = \underbrace{C^{ABCD}}_{{\bar 5}_s}+\underbrace{Q^{ABCD}}_{\overline{45}_a}
+\underbrace{F^{ABCD}}_{50_s} ~~~.
\label{t1abt2cd}
\eeq 
It is useful to split $T_1^{A B} T_2^{C D}$ into its symmetric and antisymmetric
parts under the exchange $1 \leftrightarrow 2$, so that 
\bea
T_1^{A B} T_2^{C D} &=& \underbrace{\frac{ T_1^{A B} T_2^{C D} + T_2^{A B} T_1^{C D} }{2}}_{ {\bar 5}_s,50_s} +
 \underbrace{\frac{ T_1^{A B} T_2^{C D} - T_2^{A B} T_1^{C D} }{2} }_{\overline{45}_a}\nn\\
                     &\equiv&S^{ABCD} +A^{ABCD} ~~~.
\eea
It follows that $Q$ is antisymmetric while $C, F$ are symmetric under the exchange 
$(AB) \leftrightarrow (CD)$ (or equivalently $1 \leftrightarrow 2$). 
Thus, clearly the $\overline{45}_{a}$, $q^{ABCD}$, is given by
\beq
q^{ABCD}=A^{ABCD}~~,
\eeq
and the simple identification $Q^{ABCD}=A^{ABCD}$ holds because $(Q^{ABCD}-Q^{CDAB})/2=q^{ABCD}$
while $(C^{ABCD}-C^{CDAB})/2=(F^{ABCD}-F^{CDAB})/2=0$.  
Then, we have only to separate the ${\bar 5}_{s}$ from the $50_{s}$.
One can guess that the ${\bar 5}_s$ is given by 
\beq
c_E=\epsilon_{ABCDE} S^{ABCD}
\eeq
and verify that the following expression for $C^{ABCD}$
\beq
C^{ABCD}=\frac{1}{24} \epsilon^{ABCDE} \epsilon_{FGHIE} S^{FGHI}
\eeq
satisfies $\epsilon_{ABCDE} C^{ABCD}=c_E$. Since $\epsilon_{ABCDE} Q^{ABCD}=0$, 
$C$ completely projects out the $\bar 5$ if also $\epsilon_{ABCDE} F^{ABCD}=0$ holds.
This property is automatically satisfied by
\bea
F^{A B C D} &=& T_1^{A B} T_2^{C D} \nn\\
                      & -& A^{ABCD} \nn\\
                      & -&\frac{1}{24} \epsilon^{A B C D E} \epsilon_{FGHIE} S^{FGHI}~~.
\eea
Note that $F$ is antisymmetric in $[A B]$ and $[C D]$ but is symmetric for the exchange
$(A B) \leftrightarrow (C D)$.

By writing the Kahler potential as
$K \ni  {F^{A B C D}}^* {F^{A B C D}} $, it turns out that
$F^{A B C D} $ hasn't canonical kinetic terms. The proof is very similar
to the one already discussed for the case of the $75$. Also in the present case, 
defining ${F}^{A B C D} \equiv {F'}^{A B C D}/2$, one realizes that the fields ${F'}^{A B C D}$
are those with the kinetic terms in the canonical form. 

The $\overline{50}$ is obtained following the same procedure but starting
from the product of $\overline{10}\times \overline{10}={\overline{T_1}}_{AB} 
{\overline{T_2}}_{CD}$.

\begin{table} [p]
\begin{center}
\bea
SU(5)&\supset & SU(3)\times SU(2)\times U(1)\nn\\
&~& \nn\\
1&=&(1,1,0)  \nn\\ &~& \nn\\
24&=&(1,1,0)+(1,3,0)+(3,2,-5/6)+(\bar 3, 2 ,5/6)+(8,1,0) \nn\\ &~& \nn\\
50&=&(1,1,2)+(\bar 3,1,1/3)+(3,2,7/6)+(6,3,1/3)+(\bar 6,1,-4/3)+\nn\\
&~&+(8,2,-1/2)  \nn\\ &~& \nn\\
\overline{50}&=&(1,1,-2)+(3,1,-1/3)+(\bar 3, 2 -7/6)+(\bar 6, 3,-1/3)+(6,1,4/3)+\nn\\
&~&+(8,2,1/2)  \nn \\&~& \nn\\
75&=&(1,1,0)+(3,1,5/3)+(\bar 3,1,-5/3)+(3,2,-5/6)+(\bar 3,2,5/6)+\nn\\
&~& + (\bar 6,2,-5/6)+(6,2,5/6)+(8,1,0)+(8,3,0)  \nn 
\eea
\end{center}
\caption{The SM multiplets contained in some of the $SU(5)$ representations.}
\label{multiplets}
\end{table}

\section{The Method for Catching SM Multiplets} 

The construction of the $75$, $50$ and $\overline{50}$ in terms of products
of $10$ and/or $\overline{10}$ is particularly useful because one can apply
a simple recipe to recognize how the SM multiplets are included in these large
representations.
The starting point is the identification of the SM multiplets in the $10$ and $\overline{10}$
themselves. Keeping $A<B$:
\bea 
T^{AB}&=&\underbrace{10^{\alpha \beta}}_{(\bar 3, 1, -2/3)} + 
        \underbrace{10^{\alpha b}}_{(3,2,1/6)}+ 
        \underbrace{10^{a b}}_{(1,1,1)}   \nn\\
\overline{T}_{AB}&=&\underbrace{10_{\alpha \beta}}_{(3, 1, 2/3)} + 
        \underbrace{10_{\alpha b}}_{(\bar 3,2,-1/6)}+ 
        \underbrace{10_{a b}}_{(1,1,-1)}~~~~.
\eea
Let us consider the case of the $75$ ($50$) in order to be specific.
One starts by writing the SM multiplets obtained by multiplying each
SM multiplet of $T$ with each SM multiplet of $\overline{T}$ ($T$). 
All the $100$ fields obtained are thus separated in SM multiplets.
%\bea
%T\times \overline{T}=()+()+()+()+() \nn\\
%T \times T =()+()+()+() 
%\eea
Next we have to further correctly separate the various multiplets into the 
$1$, $24$ and $75$ (${\bar 5}_s$, $\overline{45}_a$ and $50_s$).
This is easy to do for such multiplets which are obtained only once, because
from Table \ref{multiplets} we immediately understand in which $SU(5)$ representation
they lie. But for the multiplets
which are obtained more then once, like e.g. the singlet (the color antitriplet), 
we have to understand which linear combination of them is the one actually 
embedded in the $75$ ($50$).   
As an example, we show the procedure which permits to single out the SM singlet
contained in the $75$ and the SM antitriplet contained in the $50$.

\subsubsection{The SM singlet of the $75$}

There are three singlets $(1,1,0)$ arising from the product $T\times \overline{T}$.\\

We call $s_1$ the singlet which is obtained by multiplying   
\bea
T^{\alpha \beta}_{\gamma \delta} \equiv \underbrace{T^{\alpha \beta}}_{(\bar 3, 1, -2/3)} 
\underbrace{\overline{T}_{\gamma \delta}}_{(3, 1, 2/3)} & = & 
\underbrace{\frac{1}{6} (\delta^\alpha_\gamma \delta^\beta_\delta -
\delta^\alpha_\delta \delta^\beta_\gamma) 
T^{\mu \nu}_{\mu \nu} }_{(1,1,0)}  \nn\\ 
& + & \underbrace{T^{\alpha \beta}_{\gamma \delta} -
\frac{1}{6} (\delta^\alpha_\gamma \delta^\beta_\delta -\delta^\alpha_\delta \delta^\beta_\gamma) 
T^{\mu \nu}_{\mu \nu}}_{(8,1,0)}~~~. 
\eea
It is then evident that $s_1 \propto T^{\mu \nu}_{\mu \nu}=
2 T^{\mu \nu}_{\mu \nu}|_{\mu < \nu}$. 
Thus, choosing to normalize to 1 each combination of independent fields:
\beq
s_1= \frac{1}{\sqrt{3}} (T^{12}_{12}+T^{13}_{13}+T^{23}_{23}) 
= \frac{1}{\sqrt{3}} T^{\mu \nu}_{\mu \nu}|_{\mu < \nu}~~~.
\eeq

The second singlet $s_2$ comes from the product
\bea
T^{\alpha b}_{\gamma d} \equiv \underbrace{T^{\alpha b}}_{(3, 2, 1/6)} 
\underbrace{\overline{T}_{\gamma d}}_{(\bar 3, 2, -1/6)} & =&
\underbrace{
\frac{1}{6} \delta^\alpha_\gamma \delta^b_d T^{\mu n}_{\mu n}
}_{(1,1,0)}\nn\\
&+& \underbrace{ 
\frac{1}{3} \delta^\alpha_\gamma T^{\mu b}_{\mu d} - \frac{1}{6} \delta^\alpha_\gamma \delta^b_d 
T^{\mu n}_{\mu n} 
}_{(1,3,0)}\nn\\
&+& \underbrace{
\frac{1}{2} \delta^b_d T^{\alpha n}_{\gamma n} - \frac{1}{6} \delta^\alpha_\gamma 
\delta^b_d T^{\mu n}_{\mu n} 
}_{(8,1,0)} \\
&+& \underbrace{
T^{\alpha b}_{\gamma d}-\frac{1}{3} \delta^\alpha_\gamma T^{\mu b}_{\mu d}- 
\frac{1}{2} \delta^b_d T^{\alpha n}_{\gamma n} +
\frac{1}{6} \delta^\alpha_\gamma \delta^b_d T^{\mu n}_{\mu n}
}_{(8,3,0)} ~~~.\nn
\eea
Thus, $s_2 \propto T^{\mu n}_{\mu n}$. Precisely:
\beq
s_2= \frac{1}{\sqrt{6}} (T^{14}_{14}+T^{24}_{24}+T^{34}_{34}+T^{15}_{15}+T^{25}_{25}+T^{35}_{35}) 
= \frac{1}{\sqrt{6}} T^{\mu n}_{\mu n}~~~.
\eeq

Finally the last singlet $s_3$ is obtained from 
\beq
T^{a b}_{c d} \equiv \underbrace{T^{a b}}_{(1, 1, 1)} 
\underbrace{\overline{T}_{c d}}_{(1, 1, -1)} = \underbrace{\frac{1}{2} (\delta^a_c \delta^b_d-
\delta^a_d \delta^b_c) T^{m n}_{m n}}_{(1,1,0)}~~
\eeq
and is simply 
\beq
s_3=T^{45}_{45}=T^{mn}_{mn}|_{m<n}~~.
\eeq

The next step is to understand which linear combination of $s_{1,2,3}$ is contained in the
$75$. This can be accomplished by using eq. (\ref{tabcd}) and the expressions for 
$S, \Sigma, Y$, that is eqs. (\ref{sab}, \ref{sigmaabcd}, \ref{yabcd}).
Indeed, by constructing on the l.h.s. of eq. (\ref{tabcd}) 
the combination relative to $s_{1,2,3}$, we obtain respectively:
\bea
s_1 &\equiv& \frac{T^{\alpha \beta}_{\alpha \beta}}{ 2 \sqrt{3}}
=\left[ \sqrt{\frac{3}{10}} S_1\right]_{S} 
+ \left[ \sqrt{\frac{8}{15}} S_{24}\right]_{\Sigma} 
+ \left[ \sqrt{\frac{1}{6}} S_{75}\right]_{Y}  \nn\\
s_2 &\equiv& \frac{T^{\alpha b}_{\alpha b}}{ \sqrt{6}}
=\left[ \sqrt{\frac{6}{10}} S_1\right]_{S} 
+ \left[ -\sqrt{\frac{1}{15}} S_{24}\right]_{\Sigma} 
+ \left[-\sqrt{\frac{2}{6}} S_{75}\right]_{Y}  \nn\\
s_3 &\equiv& \frac{T^{a b}_{a b}}{2}
=\left[ \sqrt{\frac{1}{10}} S_1\right]_{S} 
+ \left[ -\sqrt{\frac{6}{15}} S_{24}\right]_{\Sigma} 
+ \left[ \sqrt{\frac{3}{6}} S_{75}\right]_{Y} 
\eea
where 
\bea
S_1 &\equiv & \frac{\sqrt{3} s_1+\sqrt{6} s_2 +s_3}{\sqrt{10}} \nn\\
S_{24} & \equiv & \frac{\sqrt{8} s_1- s_2 -\sqrt{6} s_3}{\sqrt{15}} \nn\\
S_{75} & \equiv & \frac{s_1-\sqrt{2} s_2 +\sqrt{3} s_3}{\sqrt{6}}~~.
\eea
It is clear that by taking in the l.h.s. of eq. (\ref{tabcd}) the combination $S_{75}$,
one obtains $S_{75}=[0]_{S}+[0]_{\Sigma}+[S_{75}]_{Y}$.
The SM siglet lying in the $75$ is then $S_{75}$. 
One can check that the combinations $S_{1}$, $S_{24}$ are instead respectively 
the singlets of the $1$ and $24$ representations.
Note that $S_{75}$ can be written as
\beq
S_{75}=\frac{1}{3 \sqrt{2}} \left[T^{\alpha \beta}_{\alpha \beta} +2 T^{ab}_{ab} 
-\frac{1}{2} T^{AB}_{AB} \right] ~~.
\eeq

\subsubsection{The antitriplet of the $50$}
 
There are three multiplets in the product $10 \times 10$ having the quantum numbers
of the antitriplets, that is $(\bar 3,1,1/3)$. 
Two arise from the products
\beq
\underbrace{T^{ab}_1}_{(1,1,1)} \underbrace{T^{\gamma \delta}_2}_{(\bar 3,1,-2/3)} 
\equiv t_1^{ab\gamma \delta}
\eeq
and 
\beq
\underbrace{T^{\alpha \beta}_1}_{(\bar 3,1,-2/3)} \underbrace{T^{cd}_2}_{(1,1,1)} 
\equiv t_2^{\alpha \beta c d}~~.
\eeq
The other one is contained in 
\beq
\underbrace{T^{\alpha b}_1}_{(3,2,1/6)} \underbrace{T^{\gamma d}_2}_{(3,2,1/6)} 
\eeq
and can be written as
\beq
\frac{1}{2} (T^{\alpha b}_1 T^{\gamma d}_2 +T^{\gamma d}_1 T^{\alpha b}_2 
- T^{\alpha d}_1 T^{\gamma b}_2 -T^{\gamma b}_1 T^{\alpha d}_2 ) \equiv t_3^{\alpha b \gamma d} ~~.
\eeq

Following the same procedure already described in some detail for the extraction
of the SM singlet of the $75$, we take on the l.h.s. of eq. (\ref{t1abt2cd}) the 
combination corresponding to each antitriplet:
\bea
t_1^{ab\gamma \delta}&=&
\left[\sqrt{\frac{1}{6}} T_{(\bar 5)}^{ab\gamma \delta}\right]_{C}
+\left[\sqrt{\frac{1}{2}} T_{(\overline{45})}^{ab\gamma \delta}\right]_{Q}
+\left[\sqrt{\frac{1}{3}} T_{(50)}^{ab\gamma \delta}\right]_{F}\nn\\
t_2^{\gamma \delta a b}&=&
\left[\sqrt{\frac{1}{6}} T_{(\bar 5)}^{ab\gamma \delta}\right]_{C}
+\left[-\sqrt{\frac{1}{2}} T_{(\overline{45})}^{ab\gamma \delta}\right]_{Q}
+\left[\sqrt{\frac{1}{3}} T_{(50)}^{ab\gamma \delta}\right]_{F}\nn\\
t_3^{\gamma a \delta b}&=&
\left[-\sqrt{\frac{2}{3}} T_{(\bar 5)}^{ab\gamma \delta}\right]_{C}
+\left[\sqrt{\frac{1}{3}} T_{(50)}^{ab\gamma \delta}\right]_{F}
\label{t1t2t3}
\eea
where
\bea
T_{(\bar 5)}^{a b \gamma \delta}&=&
\frac{t_1^{ab\gamma \delta} +t_2^{\gamma \delta a b}-2~ t_3^{\gamma a \delta b}}{\sqrt{6}}\nn\\
T_{(\overline{45})}^{a b \gamma \delta}&=&
\frac{t_1^{ab\gamma \delta} -t_2^{\gamma \delta a b}}{\sqrt{2}} \nn\\ 
T_{(50)}^{a b \gamma \delta}&=&
\frac{t_1^{ab\gamma \delta} +t_2^{\gamma \delta a b}+t_3^{\gamma a \delta b}}{\sqrt{3}}  ~~.
\label{t50}
\eea
Thus, one realizes that the linear combination of the above
antitriplets which is contained in the $50$ is $T_{(50)}^{a b \gamma \delta}$.
Note that, as required, it is symmetric under the exchange $T_1 \leftrightarrow T_2$.

To get rid of the indices it is useful to define ${t_{(50)}}_{\alpha}=1/4 ~
\epsilon_{\alpha \gamma \delta} \epsilon_{ab} T_{(50)}^{ab \gamma \delta}$.
Note that, as already discussed, ${t_{(50)}}_{\alpha}$ hasn't canonical kinetic terms.
It is ${H'_{3d}}_\alpha$ (introduced in section \ref{secpdec}), 
defined as ${t_{(50)}}_\alpha \equiv {H'_{3d}}_\alpha /2$,
that possesses the correct coefficient 1:
$K \ni  {F^{A B C D}}^* F^{A B C D} \ni 4 {t_{(50)}}_{\alpha}^* {t_{(50)}}_{\alpha}=
{H'_{3d}}_\alpha^* {H'_{3d}}_\alpha$.

\section{Basic Formulae for the MDM}

\subsubsection{$SU(5)$ breaking}

The part of the superpotential responsible for the breaking of $SU(5)$ reads 
\beq  
w_{Higgs}=c_1~Y^{A B}_{C D}~ Y^{C D}_{E F}~ Y^{E F}_{A B}
        +M_Y~Y^{A B}_{C D}~ Y^{C D}_{A B}~~~~~,
\label{}
\eeq
where, as usual, $A,B,...=1,2,...,5$. This is actually the most general 
renormalizable superpotential that can be written by means of a $75$,
the cubic term picking the combination $(75 \times 75)_{75} \times 75$.
The other cubic term that one could write, namely
$Y^{A C}_{E D}~ Y^{B D}_{F C}~ Y^{E F}_{A B}$, turns out to be proportional to
$Y^{A B}_{C D}~ Y^{C D}_{E F}~ Y^{E F}_{A B}$.   
 
The SUSY vacuum of the $75$ breaks $SU(5)$ uniquely to $SU(3)\otimes SU(2)\otimes U(1)$: 
\beq
\langle Y_{C D}^{A B}\rangle  =- \dd\langle Y \rangle     
    \left[ \dd\left( {\delta}_{\gamma }^{\alpha } {\delta}_{\delta }^{\beta  }
                      +2 {\delta}_{c}^{a } {\delta}_{d}^{b}
                  - \dd\frac{1}{2} {\delta}_{C}^{A} {\delta}_{D}^{B} \dd\right) 
                  - \left(A \leftrightarrow B\right) \dd\right]~~~~~~,
\label{vevy}
\eeq
where, as usual, Greek indices assume values from 1 to 3, Latin indices from 4 to 5 and
\beq
\langle Y \rangle \equiv M_Y/(2 c_1) ~~~.
\eeq  

The above superpotential gives a mass to all
physical components of $Y$, i.e. those that are not absorbed by the Higgs mechanism.
The calculation is not difficult and leads to Table \ref{htrmdm}.

Gauge vector bosons instead acquire a mass $m_V$:
\beq
m_V^2=24 g_5^2 {\langle Y \rangle}^2~~~.
\label{mvy}
\eeq

\subsubsection{The triplet mass matrix}

The doublet-triplet splitting is realized by the following superpotential terms,
\begin{eqnarray}
w_{MDM}=c_2~H^{A} Y^{D E}_{B C} {H_{50}}^{B C F G} {\epsilon}_{A D E F G}~
+~c_3 {\overline H}_{A} Y^{B C}_{D E} {H_{\overline{50}}}_{B C F G} 
                {\epsilon}^{A D E F G}~\nonumber\\
+~M_{50}~{H_{50}}^{A B C D} {H_{\overline{50}}}_{A B C D} ~~~~~~~~~~~~~~~.
\label{3}
\end{eqnarray} 

The multiplet in $Y$ which gets a vev is the SM singlet. 
This means that in the term 
\beq 
c_2~H^{A} Y^{D E}_{B C} {H_{50}}^{B C F G} {\epsilon}_{A D E F G}
\eeq
only the interaction between the triplet contained in $H^A$ and the antitriplet contained 
in $H_{50}$ is allowed. Since $H_{50}^{ABCD}$ can be identified with $F^{ABCD}$,  
using eqs. (\ref{vevy}) and (\ref{t1t2t3}), one obtains
\beq
c_2~H^{A} \langle Y^{D E}_{B C} \rangle {H_{50}}^{B C F G} {\epsilon}_{A D E F G}= 
-4 \sqrt{3} c_2 \langle Y\rangle~ H'_{3d} H_{3u}~~. 
\eeq

Finally,
\beq
{H_{50}}^{A B C D} {H_{\overline{50}}}_{A B C D} \ni H'_{3d} H'_{3u} 
\eeq
so that the coefficients appearing in eq. (\ref{wmdm}) are obtained.

\subsubsection{Interaction of $10 ~10~ \overline{50}$}

We are interested in the interactions of the colour triplet of the $\overline{50}$
with the antitriplet combination of
$10_i \times 10_j$, $10$ being matter fields and $i,j$ family indices.
It is convenient to recognize that
\beq   
I\equiv {\Psi_{10}^{A B}}_i~ {G_{\overline{50}}}_{ij}~ {\Psi_{10}^{C D}}_j~ {H_{\overline{50}}}_{A B C D}
\ni 4~ {G_{\overline{50}}}_{ij} ~
 {\tilde t}_{(50) \alpha ij}~ t_{(\overline{50})}^\alpha~~~,
\label{I}
\eeq
where by ${\tilde t}_{(50) \alpha ij}$ we denote the antitriplet combination of 
$10_i \times 10_j$,
that is ${\tilde t}_{(50) \alpha ij}=
1/4~\epsilon_{\alpha \gamma \delta} \epsilon_{ab} {\tilde T}^{ab \gamma \delta}_{(50) ij}$:
\beq
{\tilde T}^{ab \gamma \delta}_{(50) ij}=
\sqrt{\frac{1}{3}} \left[ 
{\Psi_{10}^{a b}}_i {\Psi_{10}^{\gamma \delta}}_j +
{\Psi_{10}^{\gamma \delta}}_i {\Psi_{10}^{a b}}_j +
\frac{ {\Psi_{10}^{\gamma a}}_i {\Psi_{10}^{\delta b}}_j +
       {\Psi_{10}^{\delta b}}_i {\Psi_{10}^{\gamma a}}_j -
       {\Psi_{10}^{\gamma b}}_i {\Psi_{10}^{\delta a}}_j -
       {\Psi_{10}^{\delta a}}_i {\Psi_{10}^{\gamma b}}_j}{2}
\right]~~~. 
\eeq 
Substituting the above expression in eq. (\ref{I}) and exploiting the symmetry of 
$G_{\overline{50}}$, one obtains
\beq
I = \left[{U^c_\alpha}_i (\frac{2 G_{\overline{50}~ij} }{\sqrt{3}}) E^c_j 
-\frac{1}{2} Q_i^\beta (-\frac{G_{\overline{50}~ij}}{\sqrt{3}}) Q_j^\gamma \epsilon_{\alpha \beta
\gamma} \right] H'_{3u}~~.
\eeq
This is the cause of the introduction of the coefficient $\sqrt{3}/2$ in eq. (\ref{w3yuk}).

\clearemptydoublepage

%\addcontentsline{toc}{chapter}{List of figures}
%\listoffigures
%\clearemptydoublepage

%\addcontentsline{toc}{chapter}{List of tables}
%\listoftables
%\clearemptydoublepage

\end{document}